\documentclass{aa}
\usepackage{url}
\usepackage{graphicx}
\usepackage{natbib}
\usepackage{changebar}
\usepackage{dcolumn}
\usepackage{bm}

\newcommand{\be}{\begin{equation}}
\newcommand{\ee}{\end{equation}}
\begin{document}

\titlerunning{Small-scale turbulent dynamo}

\title{Small-scale turbulent dynamo - a numerical investigation}

\author{R.J. West\inst{1} 
\and 
S. Nazarenko\inst{1} 
\and 
J.P. Laval\inst{2} 
\and 
S. Galtier\inst{3}}

\offprints{S. Nazarenko, \protect\url{snazar@maths.warwick.ac.uk}}

\institute{Mathematics Institute, University of Warwick, Coventry CV4 7AL, 
United Kingdom
\and
Laboratoire de M\'ecanique de Lille, CNRS, UMR 8107, 
Bld. Paul Langevin, 59655 Villeneuve d'Ascq Cedex, France
\and
Institut d'Astrophysique Spatiale, Universit\'e de Paris-Sud -- CNRS, 
UMR 8617, b\^at.~121, 91405 Orsay Cedex, France}

\date{\today}

\abstract{
We present the results of a numerical investigation of the turbulent kinematic dynamo
problem in a high Prandtl number regime. The scales of the magnetic turbulence we 
consider are far smaller than the Kolmogorov dissipative scale, so that the magnetic
wavepackets evolve in a nearly smooth velocity field. Firstly, we consider the
Kraichnan-Kazantsev model (KKM) in which the strain matrix is taken to be independent
of coordinate and Gaussian white in time. To simulate the KKM we use a stochastic
Euler-Maruyama method. We test the theoretical predictions for the growth of rates 
of the magnetic energy and higher order moments \citep{Kazantsev,Kulsrud,Chertkov}, 
the shape of the energy spectrum \citep{Kazantsev,Kulsrud,Schekochihin,NazWesZab} 
and the behaviour of the polarisation and spectral flatness \citep{NazWesZab}. 
In general, the results appear to be in good agreement with the theory,
with the exception that the predicted decay of the polarisation in time is not
reproduced well in the stochastic numerics. Secondly, in order to study the 
sensitivity of the KKM predictions to the choice of strain statistics, we perform 
additional simulations for the case of a Gaussian strain with a finite correlation time 
and also for a strain taken from a DNS data set. These experiments are based on 
non-stochastic schemes, using a timestep that is much smaller than the 
correlation time of the strain. We find that the KKM is generally insensitive
to the choice of strain statistics and most KKM results, including the decay of the
polarisation, are reproduced well. The only exception appears to be 
the flatness whose spectrum is not reproduced in accordance with the KKM predictions
in these simulations. 
\keywords{Galaxies: magnetic fields -- ISM: magnetic fields -- Magnetic fields -- 
Methods: numerical -- MHD -- Turbulence}}

\maketitle

\section{Introduction}

Many astrophysical magnetic fields are believed to be generated via 
a dynamo action \citep{moffatt,parker,childress} 
in systems where the magnetic field diffusivity $\kappa$ 
is much less than that of the kinematic viscosity $\nu$. These systems 
correspond to large magnetic Prandtl number $Pr = \nu / \kappa$ flows. 
Such astrophysical situations are observed in the interstellar medium 
as well as in protogalactic plasma clouds where the magnetic Prandtl number 
varies respectively from $Pr \sim 10^{14}$ to $Pr \sim 10^{22}$ 
\citep{chandran,kulsrud99,Schekochihin}. 
Such high values of $Pr$ lead to an interesting interval of scales 
(from $7$ to $11$ decades in our examples) below the viscous cut-off, 
but above the magnetic diffusive scale, where magnetic fluctuations are 
stretched by a randomly changing {\em smooth} velocity field. 
The small-scale kinematic dynamo problem can thus be formulated as to 
whether a small initially ``seeded'' magnetic field, subject to stretching 
by a prescribed smooth velocity field, will grow in time. 

One can picture the evolution of an ensemble of magnetic wavepackets, each 
traveling together with the fluid particles which are distorted by the local 
strain. 
By assuming a given form of the strain statistics, an idealised homogeneous 
and isotropic strain that is a Gaussian white noise process, one can simplify 
the problem, leading to a productive framework from which analytical results 
can be obtained. This formalism is known as the Kraichnan-Kazantsev model (KKM) 
\citep{Kraichnan,Kazantsev}.
It should be remembered that such a model is a simplification and a real turbulent 
velocity field will exhibit finite correlations of the fluid velocity field, as 
well as being to some degree non-Gaussian, inhomogeneous and anisotropic. 
Therefore, in this numerical investigation we will 
also consider more realistic examples of the velocity field.

The work presented here is based on a kinematic model, making use of the 
linear induction equation of the magnetic field ${\bf B}({\bf x},t)$
\begin{equation}\label{induct}
   D_t {\bf B} = {\bf B}\cdot\nabla{\bf v} + \kappa \nabla^2 {\bf B},
\end{equation}
where $D_t \equiv \partial_t + {\bf v}\cdot\nabla$ is the material derivative, 
${\bf v}$ the velocity, and $\kappa$ the magnetic diffusivity. Physically, we can have 
in mind the following picture for the evolution of our dynamo system.
Initially, for small $\kappa$, the ${\bf B}$-field  will be ``frozen'' into the flow 
behaving as a passive vector field \citep{Soward}. 
After some period of time, and hence growth of the magnetic field, 
diffusion will become important and will reduce this rate of growth. In particular, using 
the KKM the growth of the 
total mean magnetic energy $<{\bf B}^{2}(t)>$ was obtained by \cite{Kazantsev} and that of 
the higher order moments $<{\bf B}^{2n}(t)>$ by \cite{Chertkov}. 
\cite{Schekochihin} have extended this work to include the effects of compressibility. 

The work presented here has two primary aims. The first aim is to numerically verify the 
analytical scaling laws found by \cite{Kazantsev} and \cite{Chertkov} for the even moments 
$<{\bf B}^{2n}(t)>$ (where $n$ is a positive integer) in both the perfect conductor 
(non-diffusive) and diffusive regimes. We will also compute the Fourier space one-point 
correlators which have recently been investigated analytically 
\citep{Schekochihin,NazWesZab}. In the past 
much emphasis has been placed solely on investigations of the magnetic energy spectrum. 
However, higher order correlators are also of significance. 
Indeed, they describe statistics of the fluctuations around the mean energy distributions 
in the Fourier space. 
Also, it has been shown that the new quantities, the mean polarisation of the magnetic turbulence
and the spectral flatness, are of particular interest as they characterise the properties of 
the small-scale intermittency \citep{NazWesZab}. These newer theoretical objects will 
also be investigated numerically.
Our second aim is to study the sensitivity of the dynamo model to the choice of 
strain statistics.
In a quest to test the universality of the analytical results found using the KKM, we will consider 
the case of a finite-correlated Gaussian strain field. We will also consider a more realistic 
strain field generated from a Direct-Numerical-Simulation (DNS) of the Navier-Stokes equations.

The presentation of this paper has been organised into six main sections. 
In the next section we give a brief introduction to the small-scale dynamo problem, 
with a summary of the KKM analytical results as derived by \cite{Chertkov} 
and \cite{Kazantsev}. In the third section we outline some of the new KKM analytical results
in regard to the behaviour of the one-point Fourier space correlators; a more detailed 
discussion of which can be found in \cite{NazWesZab}. The fourth section provides an  
overview of the numerical method used to investigate the KKM and its subsequent results. 
In the fifth section we outline the method and results of the finite-correlated 
Gaussian and DNS based investigations. Finally, we draw our conclusions in the 
sixth section.

\section{Moments of ${\bf B}$}\label{Chert}

Let us briefly summarise the analytical results for the moments of ${\bf B}(t)$ 
obtained within the KKM by \cite{Chertkov}. At scales below the viscous cut-off 
Batchelor \citep{Batch1,Batch2} argued the velocity field would be random and smooth 
(Batchelor regime). 
In MHD a uniform velocity cannot alter a system's magnetic field. The flow 
\begin{equation}\label{straindef}
   v_i = \sigma_{ij} x_j,
\end{equation}
where $\sigma(t)$ is a coordinate independent random strain matrix, represents one of 
the simplest choices of a velocity field that allows for the transfer of kinetic energy 
into magnetic energy \citep{Zeldovich}. The incompressibility condition 
$\nabla \cdot {\bf v} = 0$ in this case corresponds to $Tr[\sigma]=0$ 
for all time. It is natural to reformulate our mathematical description in terms of Lagrangian 
dynamics (namely following particles) in which case $D_t \equiv d/dt$ in the induction equation
(\ref{induct}). \cite{Zeldovich} \citep{Soward} showed that given an initial 
condition, the induction equation (\ref{induct}) can be re-written in Fourier space as
\begin{equation}\label{fourier}
   {\bf B}({\bf k}, t) = J(t)\,{\bf B}({\bf q},0)\, \exp \left( 
                                      -\kappa \int_{0}^{t} k^{2}(t')dt'\right),
\end{equation}
where the Jacobian $J(t)$ satisfies 
\begin{equation}\label{jeq}
   \frac{d}{dt} J = \sigma J,
\end{equation}
with the initial condition $J(0)=I$, with I the unit matrix, and 
${\bf q}={\bf k(0)}$ is the initial wavevector related to ${\bf k}(t)$ by 
\begin{equation}\label{keq}
 {\bf q} = J^{T} {\bf k}.
\end{equation}
Moments of ${\bf B}$ are calculated via
two independent averagings, one over initial statistics, 
and the other over various realisations of the strain matrix \citep{Chertkov}. 
The initial ${\bf B}$ field is assumed to be homogeneous, isotropic 
and Gaussian, with zero mean and the following variance
\begin{equation}\label{initB}
<B_{a}({\bf k},0)B_{b}({\bf k}', 0)> \,\, = \,\, 
\end{equation}
$$\left( \delta_{ab} - \frac{k_{a}k_{b}}{k^{2}} \right) 
k^{2} E_m e^{-k^{2} L^{2}} \delta({\bf k}-{\bf k}'),$$
where $L$ is the length-scale at which the ``seed'' magnetic noise is initially 
concentrated and $E_m$ is a constant which determines the initial turbulence intensity. 
After averaging over the initial statistics \cite{Chertkov} obtained the following 
expression for ${\bf B}^{2}(t)$
\begin{equation}\label{B2t}
{\bf B}^{2}(t) = \int E_m e^{-q^{2} L^{2}}\,e^{-2\kappa{\bf q}\Lambda{\bf q}}\,
q^2\,J_{ia} \left(\delta_{ab} - \frac{q_{a}q_{b}}{q^{2}}\right) J_{ib}\,d{\bf q},
\end{equation}
with summation over repeated indices and where $q = |{\bf q}|$ and $\Lambda(t)$ satisfies
\begin{equation}\label{lameq}
   \Lambda(t) = \int_{0}^{t}\,dt'\,J^{-1}(t')J^{-T}(t').
\end{equation}
The full even moments $<{\bf B}^{2n}(t)>$ where calculated by taking the
value of ${\bf B}^2$ to the correct power and then averaging over the 
realisations of the velocity field. For an isotropic, delta-correlated, 
Gaussian strain field
\begin{equation}\label{ChertStrain}
   <\sigma_{ab}(t)\sigma_{ab}(0)> \,\,=\,\, 10\lambda_1 \delta(t),
\end{equation}
where $\lambda_1$ is the growth rate of a material line element, corresponding to the 
largest Lyapunov exponent of the flow, \cite{Chertkov} derived the 
following expressions for the scaling behaviour of the even moments $<{\bf B}^{2n}(t)>$.
In the perfect conductor regime $t<3t_d/(2n+3)$, we have 
\begin{equation}\label{reg1}
   <{\bf B}^{2n}(t)> \,\,\simeq\,\, e^{2\lambda_1 n (2n+3)t /3},
\end{equation}
where on dimensional grounds the scale at which magnetic diffusion becomes important is
\mbox{$r_d \simeq \sqrt{\kappa/\lambda_1}$} and the corresponding time is 
\mbox{$t_{d} \simeq \lambda_{1}^{-1} \ln(L/r_d)$}. In the diffusive regime $t>3t_d/(n+2)$,
we have 
\begin{equation}\label{reg2}
   <{\bf B}^{2n}(t)> \,\,\simeq\,\, \,e^{\lambda_1 n (n+4)t/4}.
\end{equation}
For $n=1$, this result was derived earlier by \cite{Kazantsev}.

\section{Fourier Space Correlators}\label{Fsec}

Traditionally, Fourier space analysis of the dynamo problem has only really 
involved investigations of the $k$-space correlator corresponding to the energy spectrum 
$E(k,t) = <|{\bf B}({\bf k})|^{2}>$ \citep{Kazantsev,Kulsrud,Schekochihin}. 
\cite{Kazantsev} studied the dynamo system as an eigenvalue problem 
from which he was able to obtain the growth exponents of the total magnetic energy. 
The evolution of the energy spectrum in $k$-space has also been studied both analytically 
and numerically by \cite{Kulsrud}. 

In this KKM analysis, the way in which we define the incompressible strain matrix 
is different to (\ref{ChertStrain}) used in \cite{Chertkov}. We write
\begin{equation}\label{OurStrain}
   \frac{\sigma_{ij}}{\sqrt{\Omega}} = G_{ij} - \frac{Tr[G]}{3}\delta_{ij}
\end{equation}
where $G$ is a $3\times 3$ matrix made up of independent Gaussian $N(0;1)$ random variables, 
normally distributed (i.e. Gaussian) with zero mean and a standard deviation of 1, such that
\begin{equation}\label{Astrain}
   <G_{ij}(t)G_{kl}(0)> = \delta_{ik}\delta_{jl} \delta(t).
\end{equation}
This choice is motivated by its convenience in numerical modelling.
One should note there is a degree of flexibility in how we define our strain statistics,
a more detailed discussion of which can be found in the appendix. The KKM results 
remain the same for different choices of the strain statistics, the only difference
being that time is re-scaled by a constant factor.

\subsection{Energy Spectrum}

\subsubsection{Energy Spectrum : $\kappa=0$ Solution}

In the limit of zero diffusion $\kappa=0$, the 3D energy spectrum has the solution 
\citep{NazWesZab}
\begin{equation}\label{HeatSol}
   E(k,t) = \frac{E_0}{\sqrt{t}} \left(\frac{k}{q}\right)^{-1/2} 
            \exp\left(-\frac{3\ln^{2}(k/q)}{4\Omega t}\right)
            \exp\left(\frac{5\Omega t }{4}\right),
\end{equation}
where $q$ is the wavenumber at which $E(k)$ is initially concentrated at $t=0$, and $E_0$ 
is a constant depending on the initial conditions. 
If we let $t_d$ be the diffusion time as defined in section \ref{Chert} or 
\cite{Chertkov}, then for large time $\Omega^{-1} ln^2(k/q)\ll t\ll t_d$, i.e. 
in the perfect conductor regime, we have $E\sim k^{-1/2}$. 
Alternatively, we can think of the $k^{-1/2}$ slope as being present over an interval of 
wavenumber space 
$k_{c-}\ll k\ll k_{c+}$ where
\begin{equation}\label{kc} 
   k_{c\pm} (t) \sim \exp(\pm\sqrt{4\Omega t/3}).
\end{equation}
The critical wavenumbers $k_{c+}$ and $k_{c-}$ govern when the exponential log squared 
term above becomes important. We see the fronts at $k_{c+}$ and $k_{c-}$ propagate 
exponentially in time, and within this $k^{-1/2}$ range the spectrum also grows 
exponentially. By integrating equation (\ref{HeatSol}), over the whole of wavenumber 
space, one can obtain the growth rate of the total magnetic energy
\begin{equation}\label{OurB2}
   <B^{2}(t)> = \int E(k,t)\,d{\bf k} \simeq \mbox{ const } \exp(\frac{10\Omega t}{3}).
\end{equation}
which agrees with result (\ref{reg1}) 
for $n=1$ when we take into account the slightly different definitions of the strain co-variance.
Indeed, one finds that the $\lambda_1 t$ of (\ref{reg1}) and $\Omega t$ of (\ref{OurB2}) are equivalent
if we re-scale time by a factor of $4/5$. Hence
the scaling laws of \cite{Chertkov}, presented in section \ref{Chert}, 
for the diffusive regime should be re-written for our choice of strain statistics as
\begin{equation}\label{ourchert1}
   <B^{2n}(t)> \,\,\simeq\,\, e^{2n\,\Omega\,(2n+3)t /3}.
\end{equation}
Further, in the diffusive regime we have
\begin{equation}\label{ourchert2}
   <B^{2n}(t)> \,\,\simeq\,\, e^{\Omega\,n(n+4)t /4}.
\end{equation}

\subsubsection{Energy Spectrum : $\kappa\neq0$ Solution}\label{espsec}

If all wavevectors are initially of the same length $|q|$, and we are interested
in large time asymptotics \citep{Schekochihin,NazWesZab}, the solution in the diffusive 
regime takes the form 
\begin{equation}\label{Esoly}
   E(k,t) \simeq \mbox{ const }\,\,  e^{5\Omega t/4} k^{-1/2} t^{-3/2} K_{0}(k/k_\kappa),
\end{equation}
where $K_{0}$ is a MacDonald function of zeroth order and $k_\kappa$ is a constant given by
$k_\kappa = \sqrt{\Omega / 6\kappa}$. In fact this $k_\kappa$ is the wavenumber at which diffusion
becomes important. It is therefore equivalent to the estimated $k_d \sim 1/r_d$ derived on purely
dimensional grounds in section \ref{Chert}.

\subsection{Flatness and Polarisation}

Although the energy spectrum solutions described above provides us with some very 
useful information, they do not provide us with a complete picture of our system. 
In the recent work of \cite{NazWesZab} higher one-point correlators of the magnetic field 
in Fourier space were 
studied\footnote{These were objects of the form $<|{\bf B}|^{2n}|{\bf B}^2|^{2m}>$ which 
represent a basis for all one-point correlators in magnetic turbulence which is isotropic.}. 
In this paper, as well as for the energy spectrum $E(k,t)$, we will also study two other 
correlators
\be
S(k,t) =<|{\bf B}({\bf k})|^{4}>,\label{Sdef}
\ee
\be
T(k,t) =<|{\bf B}^{2}({\bf k})|^{2}>.\label{Tdef}
\ee
Let us briefly review the results from \cite{NazWesZab} regarding these correlators.

\subsubsection{Polarisation Spectrum and Flatness : $\kappa=0$ Solutions}

In the perfect conductor regime, the fourth order correlators of Fourier amplitudes 
\mbox{$S\equiv <|{\bf B}({\bf k})|^{4}>$} and \mbox{$T \equiv <|{\bf B}^{2}({\bf k})|^{2}>$} 
have exact solutions
\be
S(k,t) = \frac{1}{\sqrt{t}} \left(\frac{k}{q}\right)^{1/2} 
\exp\left(-\frac{3\ln^{2}(k/q)}{4\Omega t}\right)
\label{SHeatSol}
\ee
$$
\left[V_0 \exp \left(\frac{21\Omega t}{4} \right) 
+ \frac{3}{7} Q_0 \exp\left(-\frac{3\Omega t}{4}\right) \right],
$$
\be
T(k,t) = \frac{1}{\sqrt{t}} \left(\frac{k}{q}\right)^{1/2} 
\exp\left(-\frac{3\ln^{2}(k/q)}{4\Omega t}\right)
\label{THeatSol}
\ee
$$
\left[V_0 \exp\left(\frac{21\Omega t}{4}\right) 
- \frac{4}{7} Q_0 \exp\left(-\frac{3\Omega t}{4}\right)\right],
$$
where $V_0$ and $Q_0$ are constants. In large time the $Q_0$ terms can be neglected 
and these spectra develop a $k^{1/2}$ scaling between two propagating fronts 
$k_{c-} \ll k \ll k_{c+}$.

Combining these two fourth order correlators, we find an important new quantity 
$P = (S-T)/S$,
the physical significance of which becomes clear when we write $P$ as 
\begin{equation}
P = \frac{4}{<|{\bf B}|^{4}>} \sum_{i\neq j} <( \Im\{B_i B_j^{*}\} )^2> \, = 
\end{equation}
$$
\frac{4}{<|{\bf B}|^{4}>} \sum_{i\neq j} < |B_i|^2 |B_j|^2 \sin^2 (\phi_i - \phi_j)>, 
$$
where $\phi_i$ is the phase of the complex component $B_{i}$ and the operator $\Im\{\cdot\}$ takes the imaginary part
of an expression. In this form we see that $P$ contains information about the phases of the $k$-space 
modes. $P$ can be thought of as the mean normalised polarisation. Indeed, $P=0$ corresponds to the plane polarisation  
of the Fourier modes. In contrast for a Gaussian magnetic field one finds the polarisation $P = 1/3$. 
Combining (\ref{SHeatSol}) and (\ref{THeatSol}), in large time we find the normalised polarisation $P$ evolves as
\begin{equation}\label{PHeatSol}
   P = \frac{Q_0}{V_0} \exp(-6\Omega t).
\end{equation}
That is, the polarisation is independent of wavenumber $k$ and
decays exponentially. This solutions tells us that in the perfect conductor regime, 
the Fourier modes of the magnetic field will eventually become plane polarised. 
In comparison, we should recall a Gaussian field has a finite polarisation, 
thus our turbulent magnetic field is far from being Gaussian. 

Another important measure of turbulent intermittency in a fluid, both in real and Fourier space, is the flatness. 
The spectral flatness is defined as the ratio
\begin{equation}\label{Fdef}
   F = \frac{S}{E^2} = \frac{<|{\bf B}({\bf k})|^{4}>}{<|{\bf B}({\bf k})|^{2}>^{2}}.
\end{equation}
Using (\ref{Fdef}), one finds a Gaussian field has a constant flatness of $3/2$.
On the other hand small-scale intermittency corresponds to a field with a flatness that grows both 
in time and in wavenumber space. Indeed, in the perfect conductor regime, 
the magnetic field displays such small-scale intermittent behaviour \citep{NazWesZab}, with
\begin{equation}\label{Fsoln}
F \sim t^{1/2} \exp \left(\frac{11\Omega t}{4}\right) k^{3/2} 
\exp\left(\frac{3\ln^{2}(k/q)}{4\Omega t}\right).
\end{equation}
For $k_{c-}<k<k_{c+}$, we see there is a range of $k^{3/2}$ scaling, while for larger $k$ 
there is an increase in the flatness arising from the log-squared exponential term. 
This intermittency in Fourier space can be attributed to the presence of coherent structures,
as will be discussed later.

\subsubsection{Polarisation Spectrum and Flatness : $\kappa\ne0$ Solutions}

In the diffusive regime the fourth order correlators $S$ and $T$ have large time 
solutions \citep{NazWesZab}
\be
  S\simeq k^{1/2}t^{-3/2}\left[V_0 e^{21\Omega t /4}+\frac{3}{7}Q_0e^{-3\Omega t/4}\right] K_0\left(\frac{k}{k_{\kappa,2}}\right),
\label{Ssol2}
\ee
\be
  T\simeq k^{1/2}t^{-3/2}\left[V_0 e^{21\Omega t /4}-\frac{4}{7}Q_0e^{-3\Omega t/4}\right] K_0\left(\frac{k}{k_{\kappa,2}}\right),
\label{Tsol2}
\ee
where $k_{\kappa,2} = \sqrt{\Omega/12\kappa}$. This $k_{\kappa,2}$ can again be 
interpreted as the wavenumber 
at which diffusion becomes important. However, in comparison to the energy spectrum 
we should note that the 
spectral cut-off will be at a smaller wavenumber as $ k_{\kappa,2} < k_{\kappa}$.
Importantly we see that the normalised  polarisation (\ref{PHeatSol}) behaves 
identically in both the diffusive and perfect conductor regimes.
In contrast, the behaviour of the flatness is modified
\begin{equation}\label{Fsol2}
F = \frac{S}{E^2} \simeq k^{3/2} t^{3/2} e^{11\Omega t} \frac{K_0(k/k_{\kappa,2})}{(K_0(k/k_{\kappa}))^2},
\end{equation}
when compared to (\ref{Fsoln}). For small $k$, we again have a region of $k^{3/2}$ scaling but now with  
a log correction arising from the MacDonald functions. For large $k$, below the spectral cut-off,
the additional MacDonald functions act to heighten the flatness. That is, the introduction of 
a finite diffusivity actually increases the small-scale intermittency.

\section{KKM Numerical Investigation}

The key to modelling the KKM numerically is the successful integration
of equation (\ref{jeq}) to find $J$. As the strain matrix on the right hand side of
(\ref{jeq}) contains noise, it is evident that this is not a normal ordinary differential
equation (ODE) and should instead be interpreted in the stochastic sense. 
Following the analytical formalism (\ref{OurStrain}) and (\ref{Astrain}), 
the elements of the strain matrix $\sigma$ will be made up from a matrix 
of Gaussian random variables $G$, such that the incompressibility condition 
$Tr[\sigma]=0$ is satisfied. 
As with any stochastic formalism one must decide the form of the integral. Mathematically both the 
Ito and Stratonovich definitions of the stochastic integral are correct. It is only in a few special cases 
that both interpretations produce the same solutions. In the present problem this is not the case.
Indeed, the Stratonovich formalism is the correct interpretation for our problem 
as we are using white noise as an idealisation of what in reality is a coloured noise process
\citep{Kloeden}.

The simplest numerical scheme one can use to solve Stochastic Differential Equations (SDE) 
is called the Euler-Maruyama scheme (EMS), and it can be thought of as generalisation 
of the normal forward 
Euler method \citep{Kloeden}. The stochastic numerical experiments presented 
here can be divided into two separate implementations.
Both these codes require the stochastic integration of the $J$ equation (\ref{jeq}),
which we perform numerically using an EMS. Averaged quantities
are calculated over many realisations of the strain statistics and initial conditions.
An ensemble of particles is chosen as the initial condition, randomly distributed 
on a unit sphere in wavenumber space $|{\bf q}|=1$ (corresponding to an 
isotropic distribution in real space). Each particle is subjected to its own realisation of the 
strain matrix, and its wavevector evolves according to equation (\ref{keq}). 

\subsection{Code 1}

Using the integration process described above for the evolution matrix $J$, Code 1 (C1) 
determines how the magnetic field evolves according to equation (\ref{B2t}).
As each particle is subjected to its own realisation of the strain matrix the integral 
(\ref{B2t}) over initial conditions simplifies to just the value of the integrand.
Higher order quantities, such as $B^{4} \cdots $, can then be calculated 
before performing the final averaging over strain statistics. The initial magnetic noise is
determined by (\ref{initB}) with $L=1$ . In the case of $\kappa\neq 0$ one must also determine 
the value of the matrix $\Lambda (t)$ from equation (\ref{lameq}). This is achieved via a 
forward Euler scheme.
\begin{equation}
   \Lambda_{jk}(t_{n+1}) = \Lambda_{jk}(t_n) 
 + \triangle t \left(\sum_{l=1}^{3} J^{-1}_{jl}(t_n)J^{-T}_{kl}(t_n)\right).
\end{equation}
We see therefore, that we must also determine the value of $J^{-1}$. This can be done in two
different ways. Firstly, we could calculate the inverse of $J$ at each step in the usual 
manner by finding the adjoint and determinant of $J$. However, as some elements of $J$ can become very 
large in time one needs to be careful in finding the determinant as round-off errors can be 
easily introduced. Alternatively, we can integrate the separate 
stochastic equation for the evolution of $J^{-1}$
\begin{equation}\label{jinv}
   \frac{d}{dt} (J^{-1}) = - (J^{-1}) \sigma,
\end{equation}
using an EMS.

The primary aim of this code is to verify the scaling laws for the even moments 
$<{\bf B}^{2n}(t)>$, (\ref{ourchert1}) and (\ref{ourchert2}).
However, by recording the values of the integrand of (\ref{B2t}) and the corresponding 
wavenumbers $k=|{\bf k}|$ for each particle, we can also construct snapshots of the 
energy spectrum in time. 

C1 has also been set up to calculate the eigenvalues of the 
matrix $JJ^{T}$ so that the Lyapunov exponents can be calculated. Indeed, 
one important result for systems with random time-dependent strain is that,
for nearly all realisations of the strain, the matrix $(1/t)\ln (J^T J)$ is found 
to stabilise at large time \citep{Goldhirsch,Falkovich}. 
The eigenvectors ${\bf f}_i$ in this case 
tend to a fixed orthonormal basis for each realisation, where $i=1,2,3$.
As the Jacobian $J$ at $t=0$ is the identity matrix, $J^{T}J$ will also take the form of the 
identity initially. In time the strain $\sigma$ will distort $J$ via (\ref{jeq}). An initial 
sphere will be deformed into an ellipsoid, the volume of which will be 
conserved by the incompressibility condition $Det[J]=1$.
The length of the principal axis of the ellipsoid correspond to the eigenvalues $e^{2\rho_i}$ 
of $J^T J$, while the eigenvectors ${\bf f}_i$ give the directions of these axis. 
It is evident that in considering a time-dependent stochastic strain we have moved to a 
system corresponding
to Lagrangian chaos. In which case, we can define the Lyapunov exponents $\lambda_i$
in terms of the limiting eigenvalues \citep{Goldhirsch,Soward,Falkovich} as
\begin{equation}\label{lyapdef}
   \lambda_i = \lim_{t\rightarrow \infty} \frac{1}{2t}\ln({\bf f}^{T}_i(J^{T}J){\bf f_i}),
\end{equation}
for $i=1,2,3$. It is customary to order the Lyapunov exponents in terms of size, 
$\lambda_1>\lambda_2>\lambda_3$. A material element aligned along 
one of the eigenvectors ${\bf f}_i$ will in the asymptotic limit expand or contract at the 
rate $\exp(\lambda_i t)$. The Lyapunov exponents do not depend on the realisations of the strain,
while in contrast the asymptotic eigenvectors are realisation dependent. Incompressibility 
ensures that $\sum^{3}_{i=1}\lambda_i = 0$ which in turn tells us that
one Lyapunov exponent must be positive if two or more are non-zero. A positive Lyapunov exponent
corresponds to exponential growth of a material line element. 

In the non-degenerate case, if the statistics of the 
strain matrix are symmetric to time reversal, which is true if the strain is 
assumed delta correlated as is the case for the KKM, then $\lambda_2 = 0$ 
\citep{Kraichnan74,Balkovsky} and 
hence the incompressibility condition gives $\lambda_3 = -\lambda_1 < 0$.
\cite{Chertkov} found that it is the largest Lyapunov exponent $\lambda_1$ that is
responsible for the growth of the moments $<{\bf B}^{2n}(t)>$.
\cite{Balkovsky} have also considered the matrix $JJ^T$, which evolves in time according to
\begin{equation}\label{jjt}
  \frac{d}{dt} (JJ^{T}) = \sigma (JJ^{T}) + (JJ^{T}) \sigma^{T}.
\end{equation}
This differs to the matrix $J^{T}J$ in that it does not 
stabilise to some fixed axis at large time. Indeed, the eigenvectors will continue to rotate for 
every realisation of the strain matrix. As in (\ref{lyapdef}), the Lyapunov exponents can also be found from the 
logarithm of the eigenvalues of $JJ^T$ at large time. The eigenvalues of $JJ^T$ and $J^T J$, and hence their
Lyapunov exponents, are the same as they share the same characteristic polynomial.

To determine the Lyapunov exponents of our system numerically we need to determine either 
$JJ^{T}$ or $J^T J$ at each timestep. This can be achieved by either calculating $JJ^T$ 
via its own dynamic equation (\ref{jjt}) and integrate it using an EMS, 
or alternatively calculate $JJ^T$ or $J^T J$ at each timestep from $J$. 
In practice either approach works equally well. Next one must determine the 
eigenvalues of the symmetric matrix $JJ^T$ which must be real. Although 
the matrix is only $3\times 3$ solving the characteristic polynomial using the solution 
to a cubic equation is impractical as the growth/reduction of elements $J$ in time produce
round-off errors numerically. This in turn leads to complex eigenvalues and further inaccuracies.

The method employed in this numerical study is that of Jacobi transformations \citep{numrec}.
The key to this method is a sequence of similarity transformations (rotations).
In the context of ellipsoids, this algorithm can be thought of as a method by which the 
axis of the system are rotated and re-aligned to lie along the principle axis of the ellipsoid, 
the length of these axis corresponding to the eigenvalues, and the directions to the eigenvectors.
Although the Jacobi rotation method is not perfect, it does however perform 
better than the previously mentioned cubic equation approach, in that the eigenvalues 
remain real for all time. Given the set of eigenvalues 
$\exp(2\rho_{i})$ of $JJ^T$, we define the time-dependent Lyapunov exponents to be
\begin{equation}\label{lypnum} 
   \lambda_i(t) = \frac{1}{2t} <\ln(\exp(2\rho_i))> = \frac{1}{t} <\rho_i>,
\end{equation}
where $i=1,2,3$. Analytically the Lyapunov exponents are independent of the strain realisations at large time,
but the averaging above has been performed over all the realisations and initial conditions to improve 
accuracy.

\subsubsection{$\kappa = 0$ simulations}\label{c1resk0}

\begin{figure}[h]
\centering
\includegraphics[width=.49\linewidth]{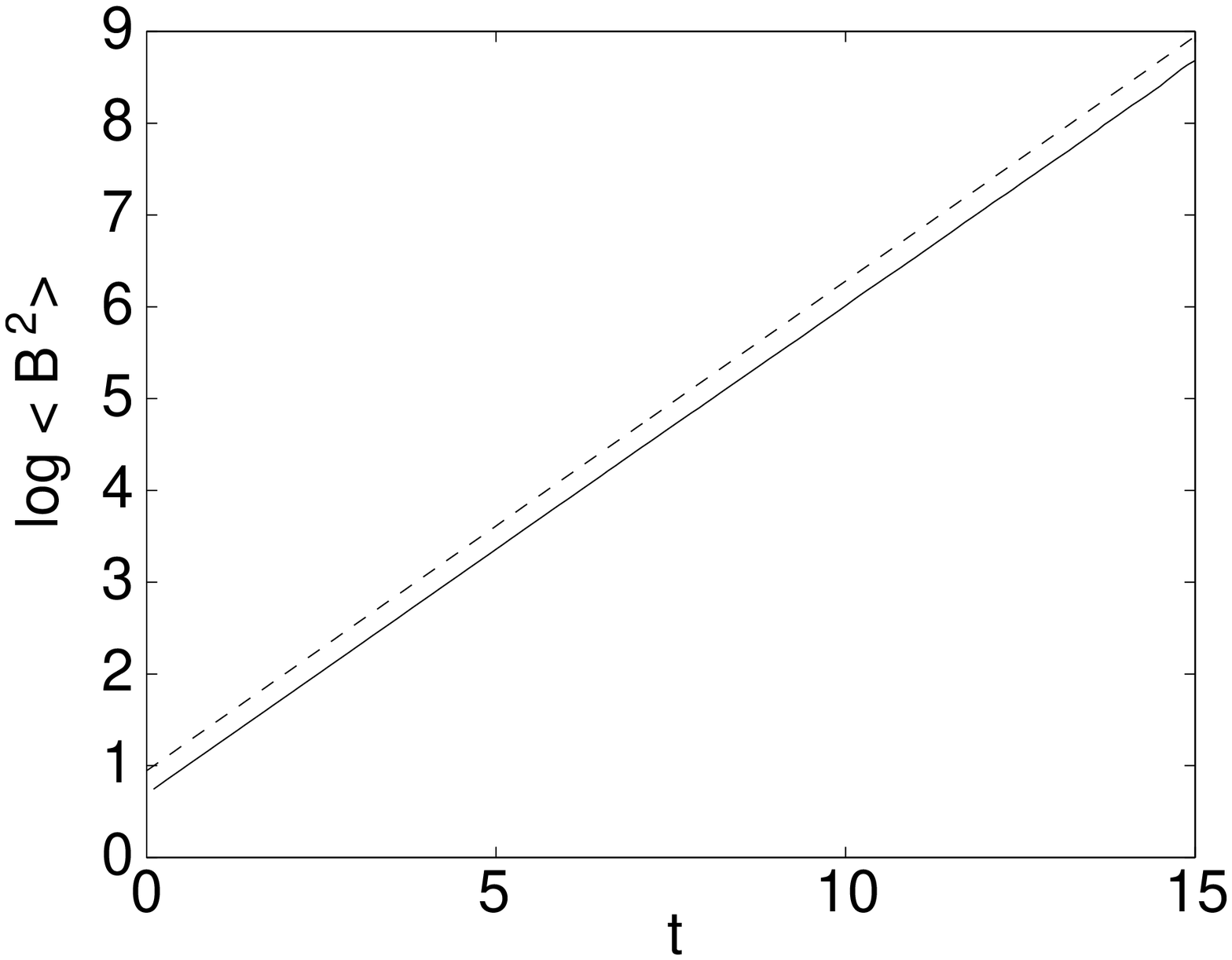}
\includegraphics[width=.49\linewidth]{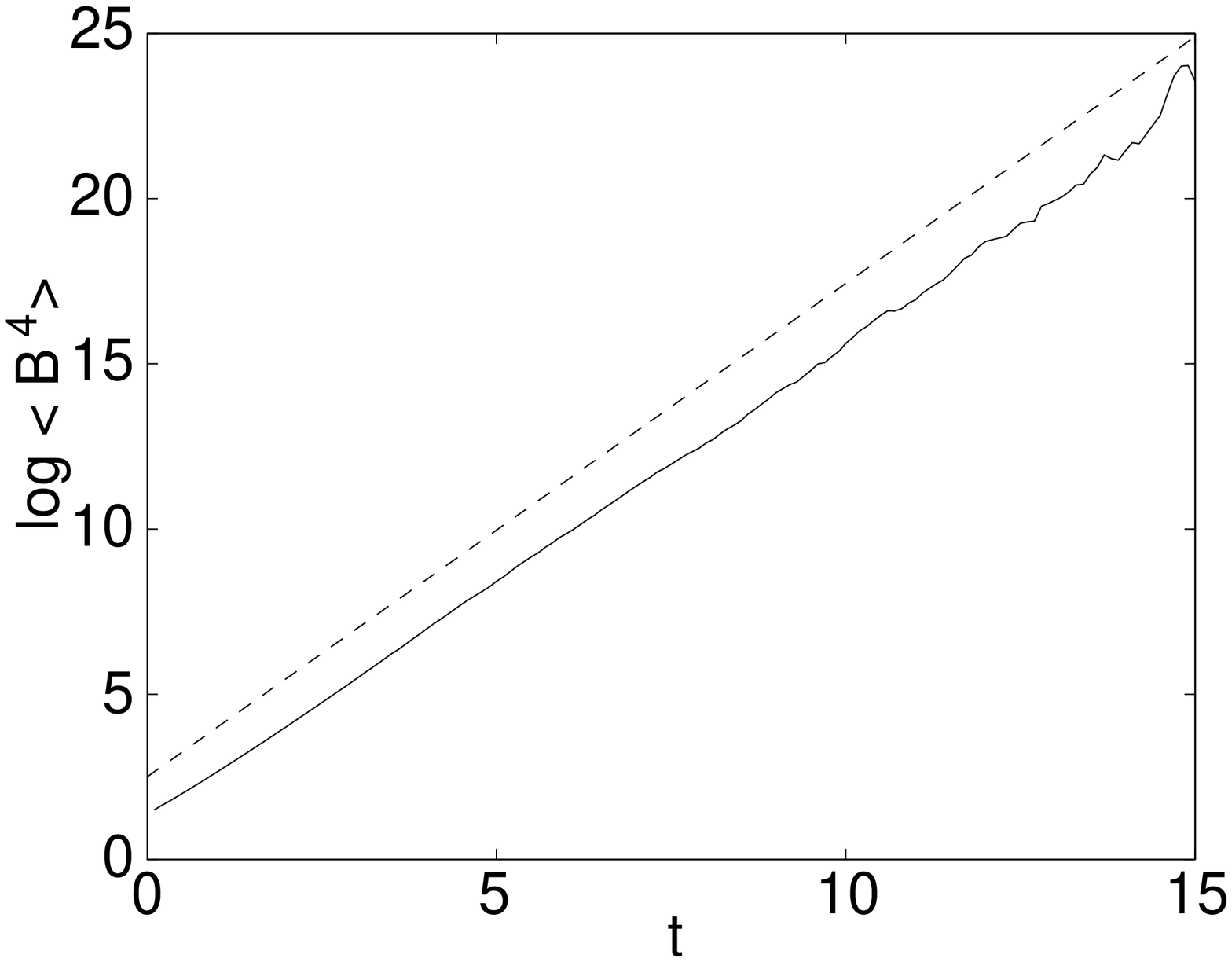}
\caption{
\noindent
$\log < {\bf B}^{2}(t) >$ (left figure) and 
$\log < {\bf B}^4 (t) >$ (right figure) for a 
C1 simulation, with $\Omega = 0.16$ and $\kappa=0$. Averaging 
was performed over $600000$ realisations. The theoretical solutions 
(\ref{ourchert1}) have also been plotted (dashed lines).}
\label{B46s}
\end{figure}

We will start by setting $\kappa=0$ in the numerics so that we can investigate the perfect conductor regime.
Firstly, let us compare the scaling of $<{\bf B}^2(t)>$ with theoretical solution (\ref{ourchert1}).
The left-hand graph of Fig. \ref{B46s} corresponds to the $n=1$ case and we see the 
numerically generated slope agrees nicely with the theoretical scaling (dashed line). 
For the higher order moments $<{\bf B}^{2n}(t)>$, good agreement with the analytical scaling 
(\ref{ourchert1}) can also be seen. The right-hand graph in Fig. \ref{B46s}
shows the slope for $n=2$. As all moments are calculated from the integral (\ref{B2t}), 
it should be noted that any fluctuations in ${\bf B}^2(t)$ will get amplified at higher orders, 
consequently the resulting graphs will become progressively noisier. The only solution to this 
problem is to increase the number of the realisations over which averaging is performed.

Let us now investigate the energy distribution in wavenumber space. 
The energy spectrum \mbox{$<{\bf B}^2(k,t)>$} is obtained at a 
particular instant in time by recording each particles wavenumber $k=|{\bf k}|$ and ``weight'' 
corresponding to the integrand of (\ref{B2t}). One then constructs a histogram by first 
finding the largest and smallest wavenumbers $k_{max}$ and $k_{min}$ of the all the particles, 
and then dividing the interval $\log\, k_{min} < \log\, k < \log\, k_{max}$ into a finite 
number of bins. 
Particles are then sorted into these bins and their corresponding weights are summed. Finally, we normalise
the total weight in each bin, by dividing through by its bin width. One must also divide through by a 
factor of $4\pi k^2$ which originates from the solid angle integration required in converting a 1D spectrum into that of a 3D one (see for example \cite{McComb}). 

\begin{figure}[h]
\centering
\includegraphics[width=.49\linewidth]{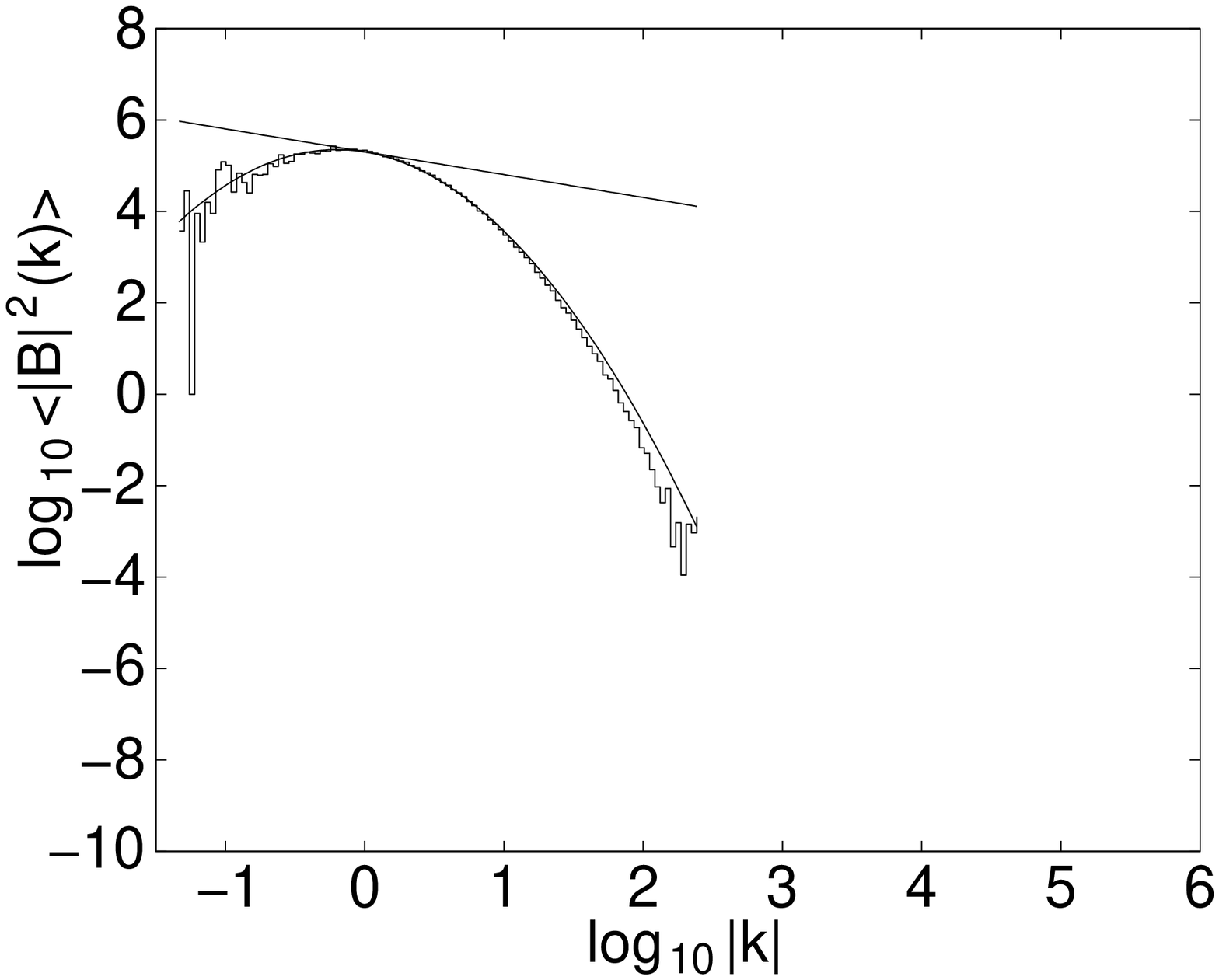}
\includegraphics[width=.49\linewidth]{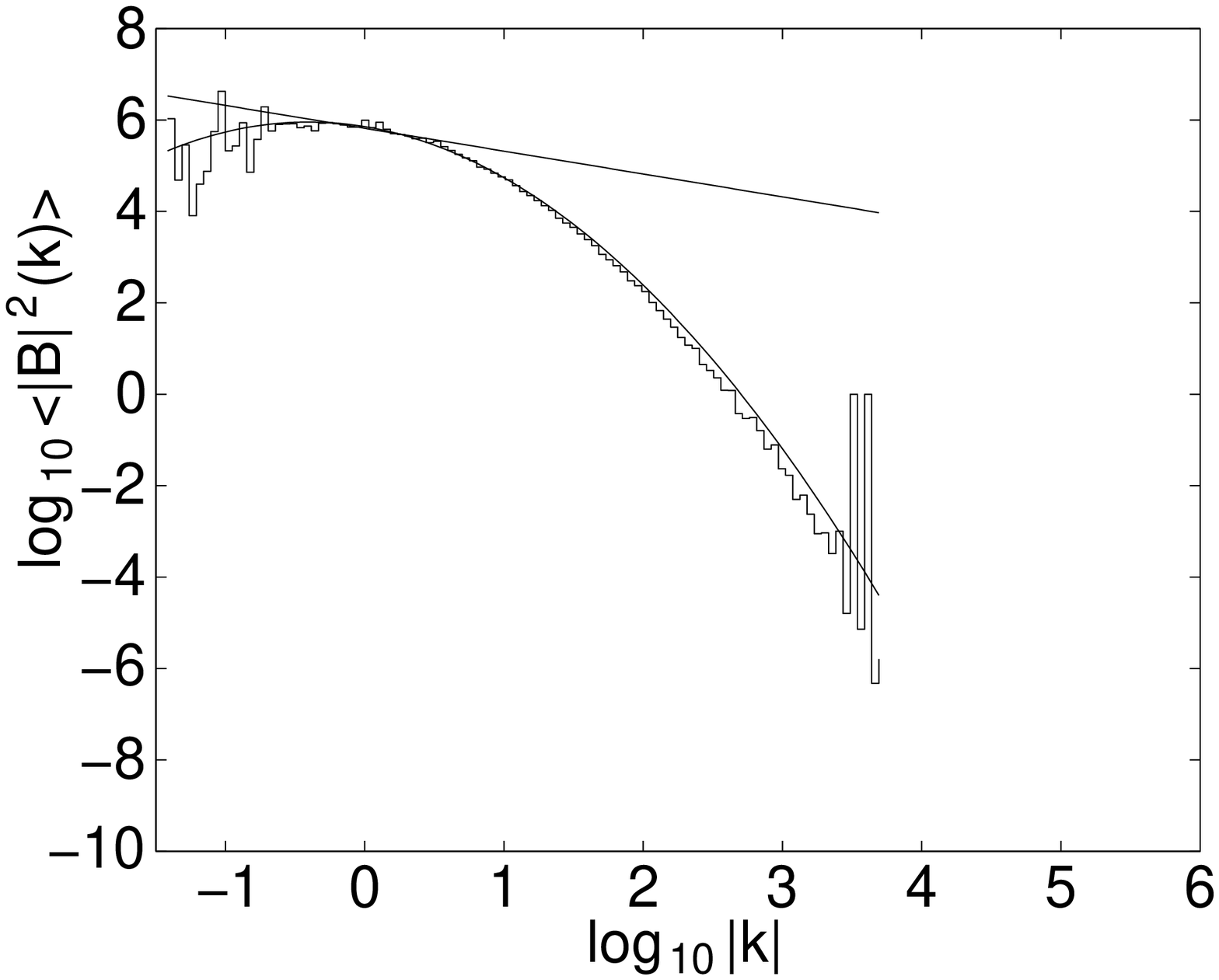}
\caption{
\noindent
$E = < {\bf B}^{2}(k,t) >$ for a C1 simulation with $\Omega = 0.16$ and $\kappa=0$.
Averaging has been performed over 600000 realisations. The time is $t=9$ (left figure)
and $t=18$ (right figure). The particles have been sorted into 100 bins. The theoretical 
solution (\ref{HeatSol}) (curve) and a slope of $k^{-1/2}$ (straight line) have also been 
plotted.}
\label{Eearl}
\end{figure}

\begin{figure}[h]
\centering
\includegraphics[width=.49\linewidth]{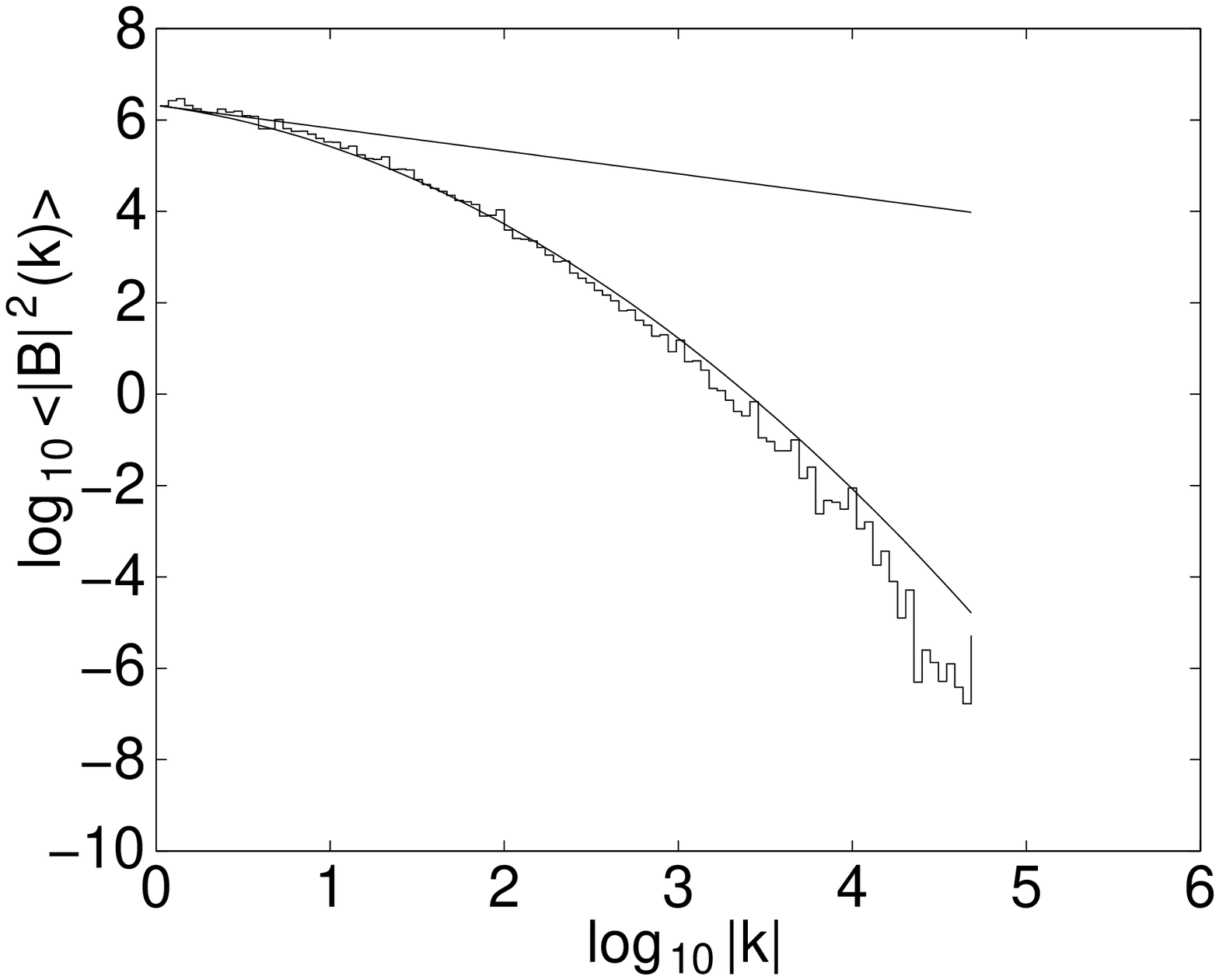}
\includegraphics[width=.49\linewidth]{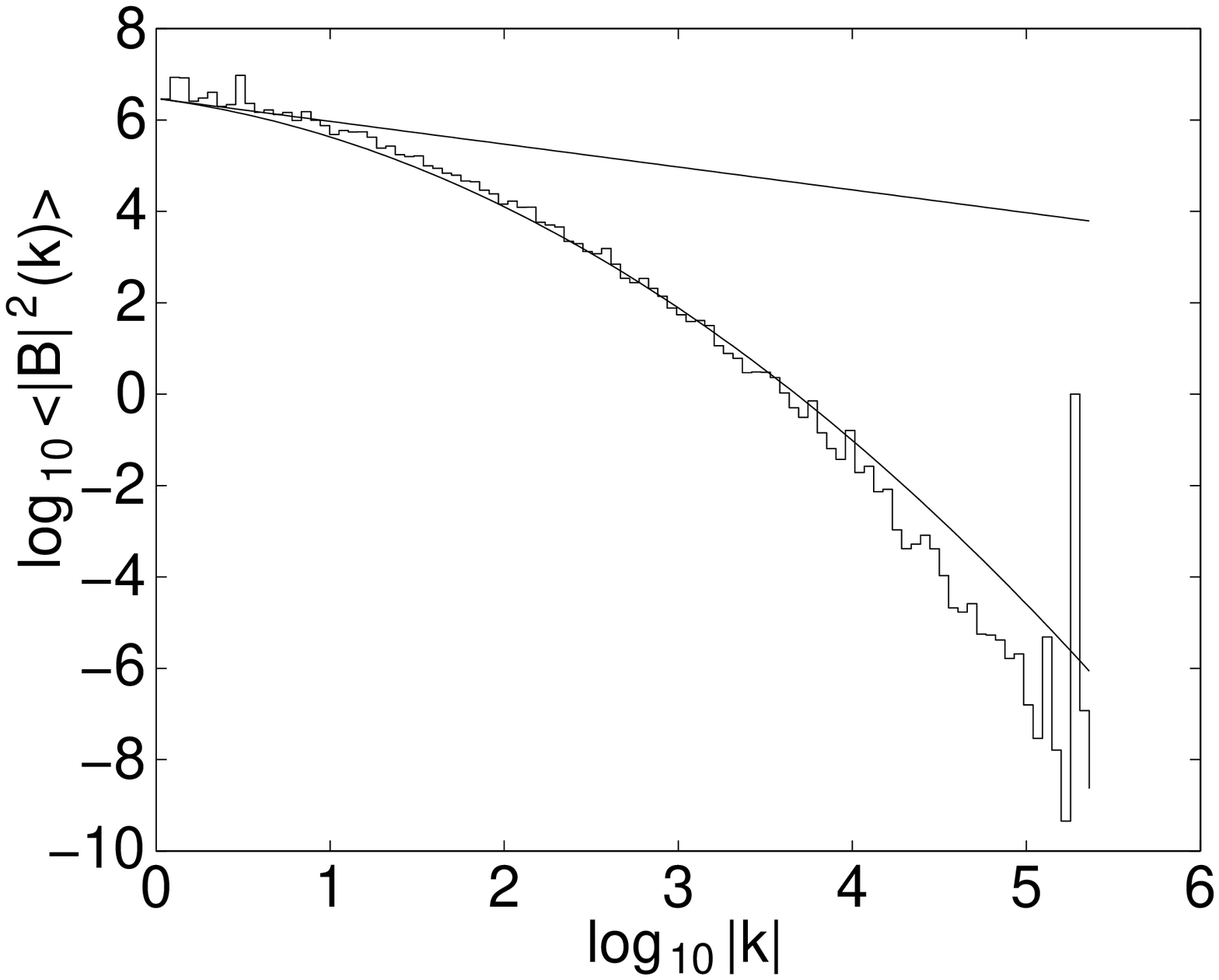}
\caption{
\noindent
$E = <{\bf B}^{2}(k,t)>$ for a C1 simulation with $\Omega = 0.16$ and $\kappa=0$.
Averaging has been performed over 600000 realisations. The time is $t=27$ (left figure)
and $t=31.5$ (right figure). The particles have been sorted into 100 bins. The theoretical 
solution (\ref{HeatSol}) (curve) and a slope of $k^{-1/2}$ (straight line) 
have also been plotted.}
\label{Elat}
\end{figure}

In Fig. \ref{Eearl} the energy spectrum has been plotted 
for two different times $t=9$ (left figure) and $t=18$ (right figure). Here $\Omega=0.16$
and thus using the definition (\ref{kc}) for the critical wavenumber $k_{c+}(t)$ we obtain
$\log_{10} k_{c\pm}(9) = \pm 0.6$ and $\log_{10} k_{c\pm}(18) = \pm 0.85$.
In Fig. \ref{Elat} the same spectrum has been plotted for $t=27$ (left figure) and $t=31.5$ 
(right figure),
correspondingly $\log_{10} k_{c+}(27) = 1.04$ and $\log_{10} k_{c+}(31.5) = 1.13$. 
One should recall that theoretically for wavenumbers $k_{c-}\ll k\ll k_{c+}$ 
we expect a $k^{-1/2}$ scaling range\footnote{Here, we are considering 
a 3D spectrum. \cite{Kulsrud} considered the equivalent 1D spectrum which has a 
corresponding $k^{3/2}$ scaling regime.}, and this is indeed the case here (the straightline in 
each graph corresponding to a slope of $k^{-1/2}$). As a whole we can see that the spectrum 
moves vertically upwards in time. This corresponds to the time dependent increase of the magnetic 
energy at fixed $k$ predicted in the theoretical result (\ref{HeatSol}). This large time asymptotic 
solution has also been plotted on each figure and we see good agreement with the numerically 
generated histograms.

\begin{figure}[h]
\centering
\begin{minipage}[c]{.99 \linewidth}
\includegraphics[width=.49\linewidth]{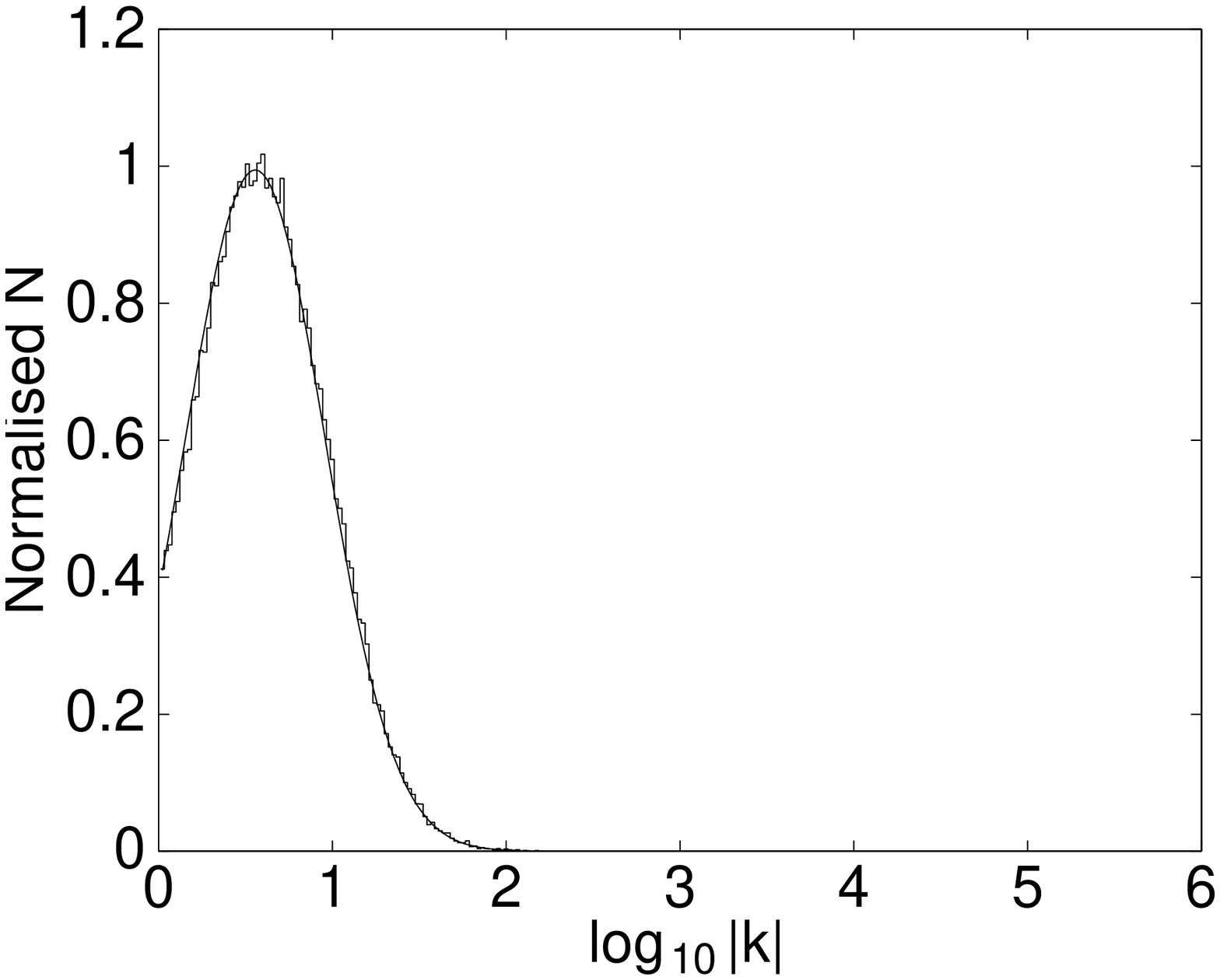}
\hfill
\includegraphics[width=.49\linewidth]{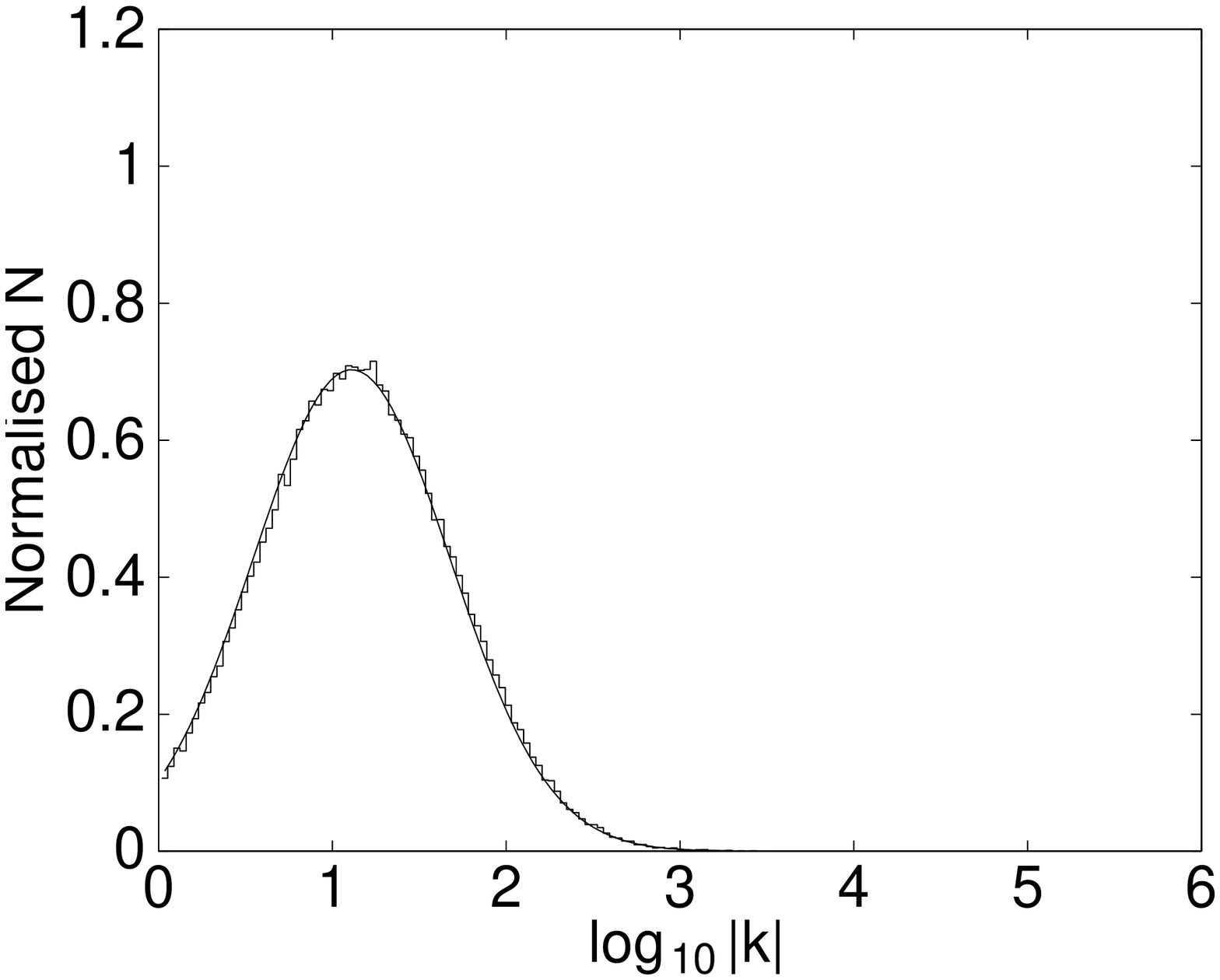}
\end{minipage}\\
\begin{minipage}[c]{.99 \linewidth}
\includegraphics[width=.49\linewidth]{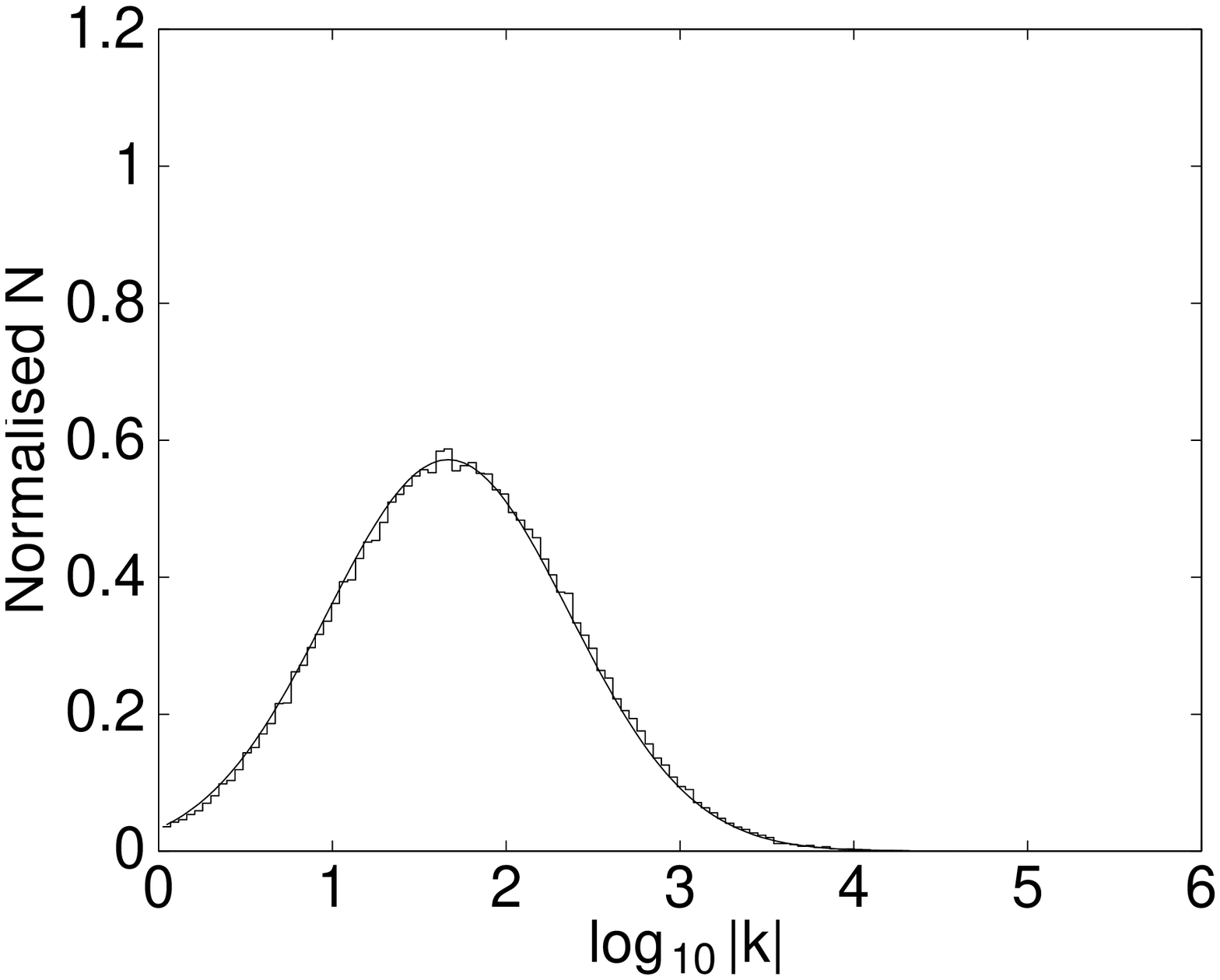}
\hfill
\includegraphics[width=.49\linewidth]{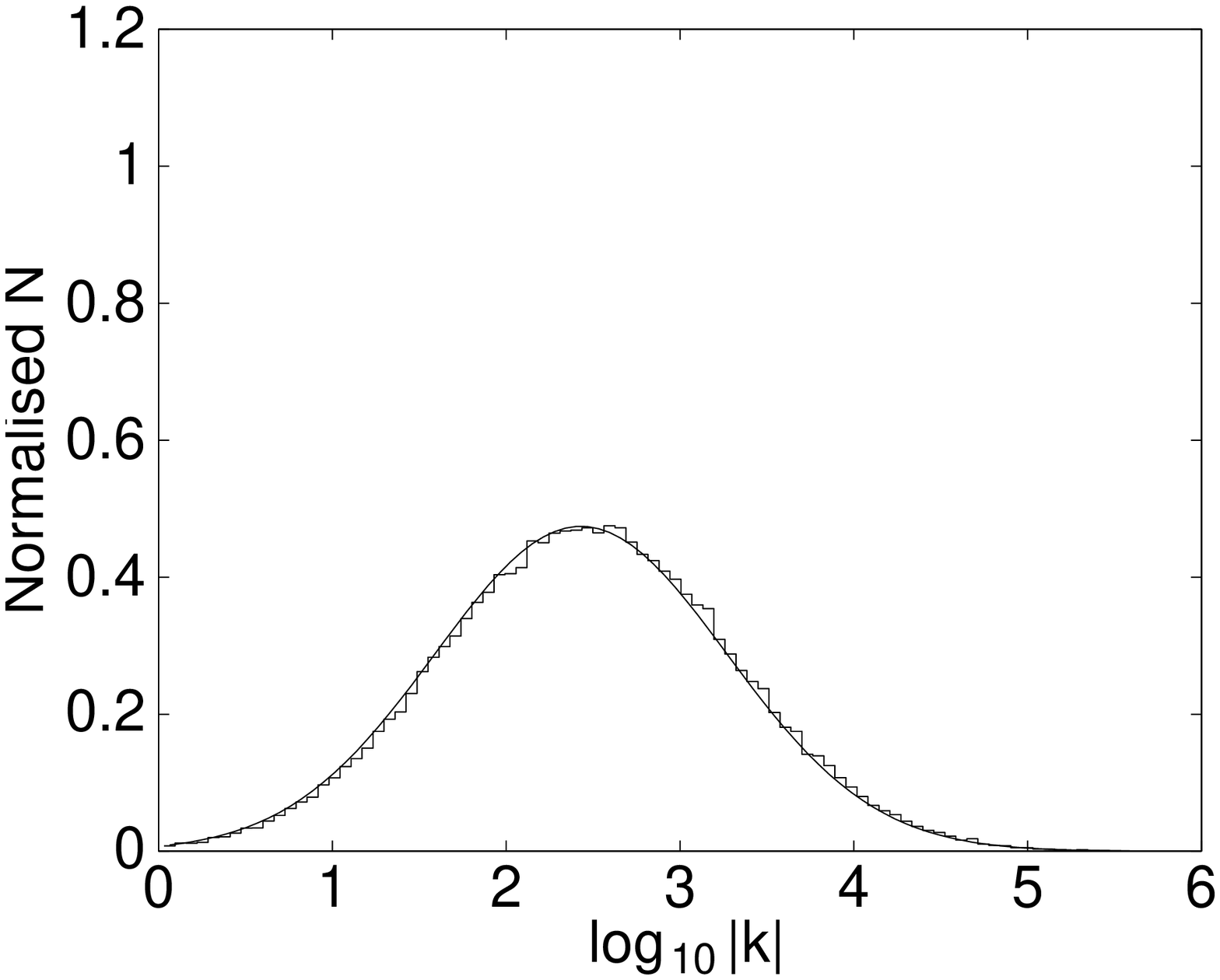}
\end{minipage}
\caption{
\noindent
Normalised particle distribution $N(k,t)$ for a C1 simulation
with $\Omega = 0.16$. Averaging has been performed over 120000 
realisations. Snapshots are shown at $t=9$ (top left), $t=18$ (top right),
$t=27$ (bottom left) and $t=35$ (bottom right). The particles have been sorted 
into 100 bins. A Gaussian curve has also been fitted to the distribution.}
\label{PartDist}
\end{figure}

It should be noted that the later figures show the spectrum for $\log_{10} k > 0$.
This is because the front propagating to small $k$ becomes extremely noisy 
at large time. The reason for this will become apparent shortly.
It is evident from these results that it would be beneficial 
if we could extend the number of decades over which this scaling is observed. 
A possible approach would be to run the simulations for a much 
longer time, remembering that $\log k_{c\pm}(t)\sim \sqrt{\Omega t}$.
In doing this one has to be careful of numerical stability and the generation of round-off errors
as some elements of $J$ will get ever larger. However, this is not the only 
problem, we should remember that our distribution of particles, initially concentrated at 
our initial wavenumber $|{\bf q}|=1$, will broaden in time. The peak of this distribution will 
move towards larger $k$, each individual particle performing its own random walk in wavenumber space.
Fig. \ref{PartDist} shows the distribution of particles at four successive times $t=9,18,27$ and $35$
for a simulation with $\Omega=0.16$. The graph has been normalised so that the area under the curve
is one. We see that with log wavenumber coordinate, the particle distribution fits a moving Gaussian profile.
The mean of this profile moves to higher wavenumbers as $\Omega t$, while the standard deviation
grows as $\sqrt{\Omega t}$. Regrettably, the energy spectrum's $k^{-1/2}$ scaling range of 
interest lies in one of the tails of the particle distribution. This means that there will be 
fewer particles available for averaging when calculating the energy spectrum histogram in this region. 

The only way to improve this averaging is to increase the total number of particles used in a simulation.
Of course, this is not an efficient way of improving the statistics,
as increasing the number of rarer particles located in each tail, will correspond to a much greater
increase in the number of particles located in the centre of the particle distribution.
This is an inherent problem with this type of particle simulation.
Also, as the particle distribution gets more spread out in time, fluctuations at both ends of the 
energy spectrum histogram will increase. This effect can clearly be seen in Fig. \ref{Eearl} and 
{\ref{Elat}}. 
That is why the front $k_{c-}$ that propagates to smaller $k<q<1$ in time becomes so noisy,
as the bulk of the particles are travelling to larger wavenumbers. For this reason in the 
remaining results we will concentrate only on the $\log_{10} k \ge 0$ region of $k$-space.

\subsubsection{$\kappa\neq 0$ simulations}

\begin{figure}[h]
\centering
\includegraphics[width=.49\linewidth]{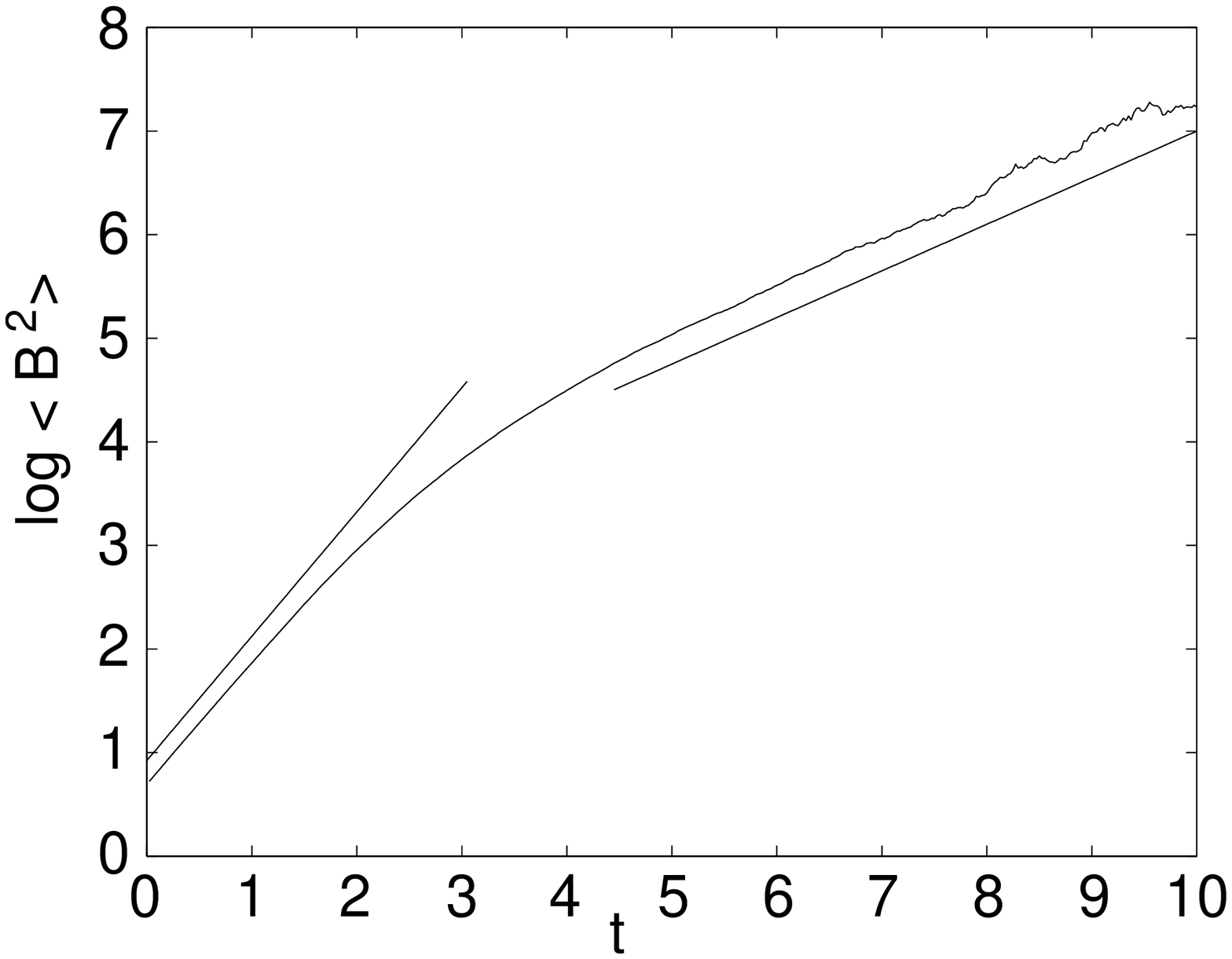}
\includegraphics[width=.49\linewidth]{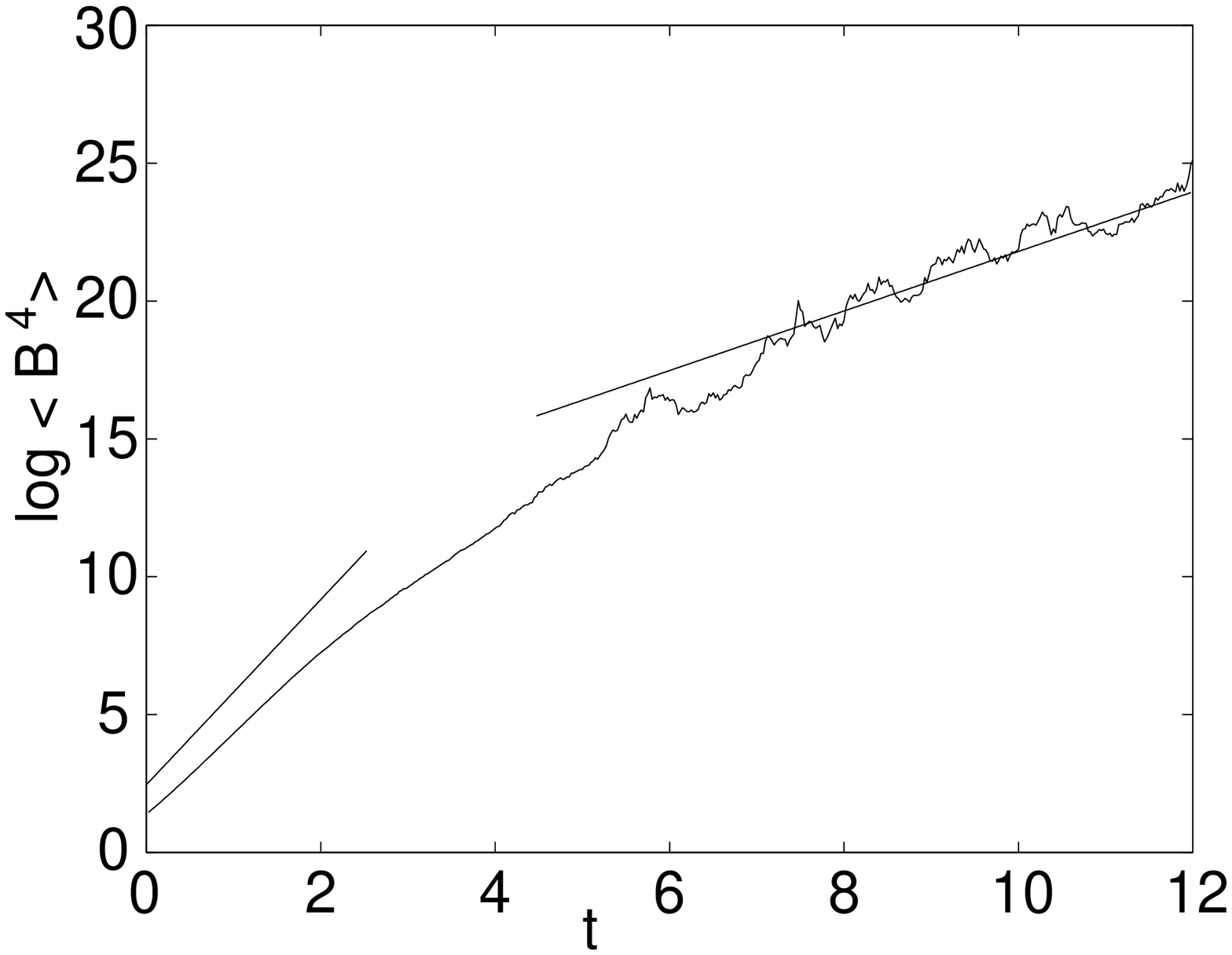}
\caption{
\noindent
$\log <{\bf B}^{2}(t)>$ (left figure) and $\log <{\bf B}^{4}(t)>$ (right figure) 
for a C1 simulation, with $\Omega = 0.36$, $\kappa=0.005$. 
Averaging was performed over $120000$ realisations (left figure) and $480000$ 
realisations (right figure). The theoretical slopes (\ref{ourchert1}) and (\ref{ourchert2})
have also been plotted (dashed lines).}
\label{b2d}
\end{figure}

\begin{figure}[h]
\centering
\includegraphics[width=.49\linewidth]{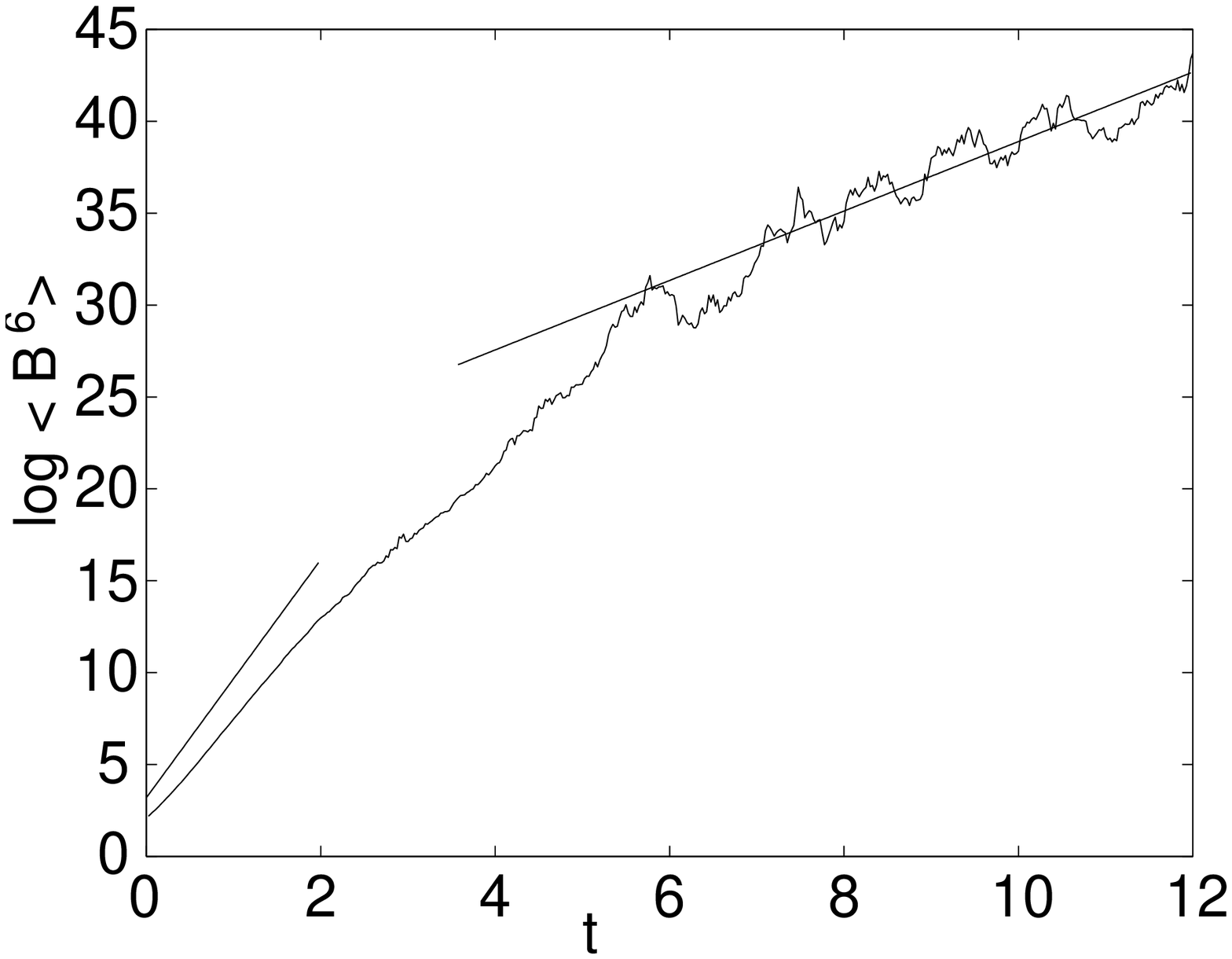}
\includegraphics[width=.49\linewidth]{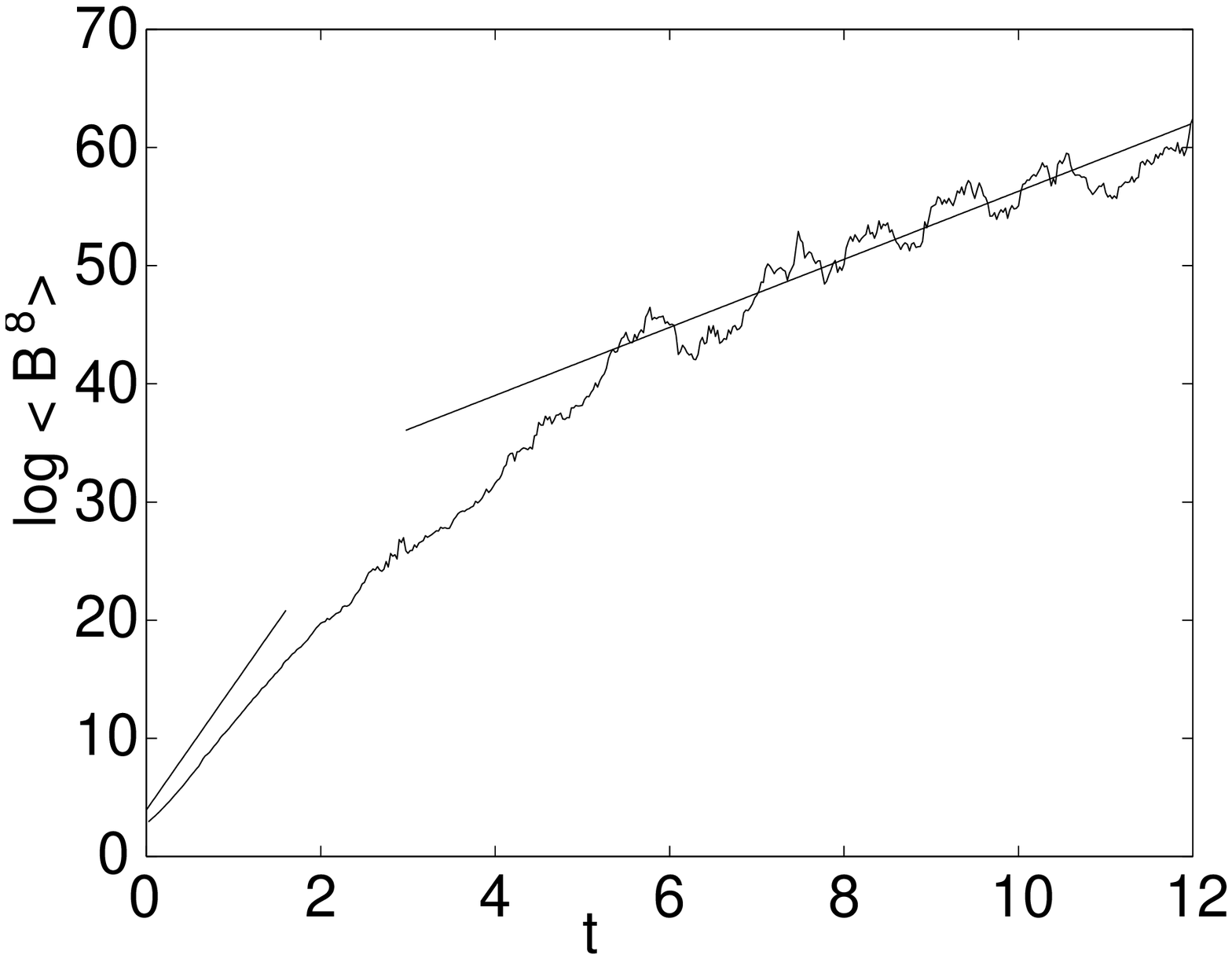}
\caption{
\noindent
$\log <{\bf B}^{6}(t)>$ (left figure) and $\log <{\bf B}^{8}(t)>$ (right figure) 
for a C1 simulation with $\Omega = 0.36$ and $\kappa=0.005$. Averaging has been 
performed over $480000$ realisations. The theoretical slopes (\ref{ourchert1}) and 
(\ref{ourchert2}) have also been plotted (straight lines).}
\label{b46d}
\end{figure}

We will now introduce a non-zero $\kappa$ into the numerical model to investigate
the effects of diffusion. The first thing we will investigate here is the scaling 
behaviour of the total magnetic energy \mbox{$<{\bf B}^2(t)>$} in the diffusive regime. 
The left-hand graph of Fig. \ref{b2d} shows the results of a simulation with 
$\Omega=0.36$ and $\kappa=0.005$. The two separate 
scaling regimes are clearly apparent and the analytical slopes (\ref{ourchert1}) and 
(\ref{ourchert2}), for the perfect conductor and
diffusive regimes respectively, are in good agreement with the numerics. 
It is inevitable that to generate similarly smooth graphs for the higher order moments,
one would have to use a greater number of realisations. Figure \ref{b46d} and the right-hand graph 
of Fig. \ref{b2d} show the scaling of the next three even moments. 
As expected these graphs are noisier, but still in agreement with the
analytical results.

\begin{figure}[h]
\centering
\includegraphics[width=.49\linewidth]{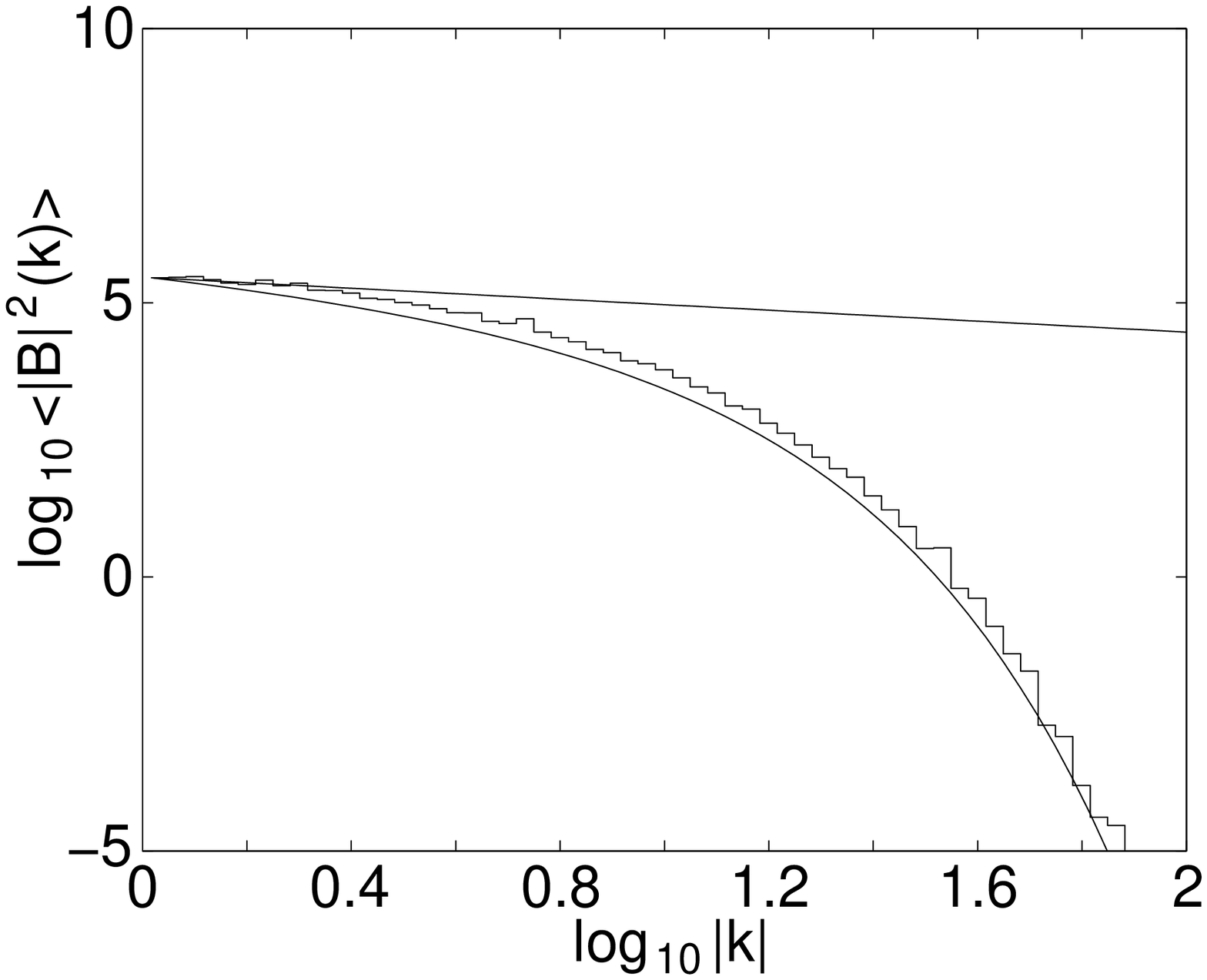}
\includegraphics[width=.49\linewidth]{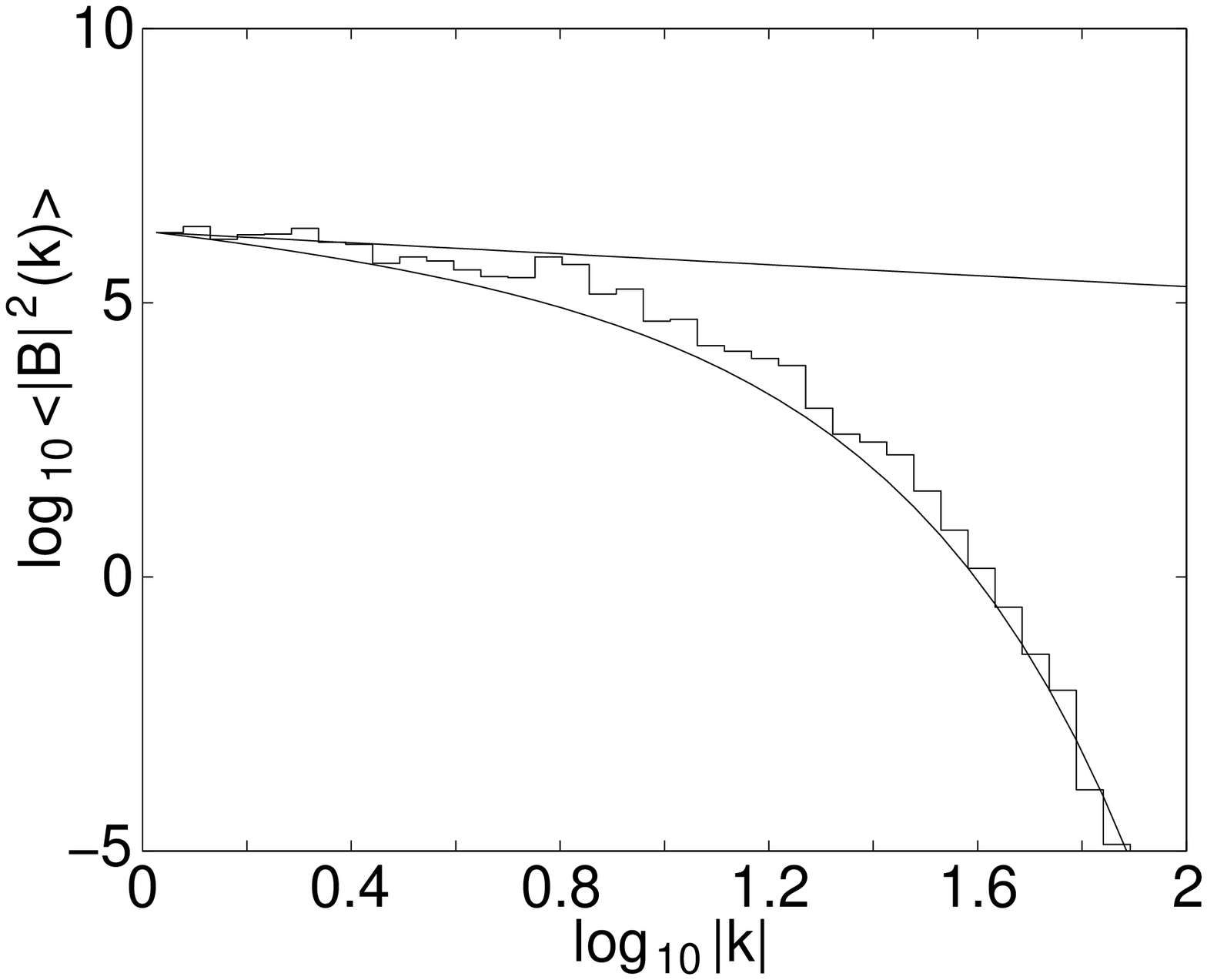}
\caption{
\noindent
$E = <{\bf B}^{2}(k,t)>$ for a C1 simulation with $\Omega = 0.36$ and $\kappa=0.005$.
Averaging has been performed over 480000 realisations. The time here is $t=6$ (left figure)
and $t=12$ (right figure). The particles have been sorted into 100 bins. The theoretical 
solution (\ref{Esoly}) (curve) and a slope of $k^{-1/2}$ (straight line) 
have also been plotted.}
\label{Esp36}
\end{figure}

Next we will consider the energy spectrum. The introduction of a finite diffusivity 
will cut-off the energy spectrum at some finite wavenumber. Figure \ref{Esp36} 
shows two successive snapshots of the energy spectrum at times $t=6$ and $12$ for a simulation 
with $\kappa=0.005$ and $\Omega=0.36$. 
As expected, one observes a spectral cut-off and for the results shown in Fig. \ref{Esp36}, 
the parameter choice gives $k_\kappa = \sqrt{\Omega/6\kappa}$. Theory predicts that for 
$k<k_\kappa$ there should be a region of $k^{-1/2}$ scaling (\ref{Esoly}) 
(with a logarithmic correction). For this particular simulation we have 
$\log_{10}\, k_\kappa \simeq 1.1$.
As with the perfect conductor regime, the spectrum is seen to move vertically upwards
corresponding to a steady increase in the total magnetic energy. The large time theoretical 
solution (\ref{Esoly}) has also been plotted in Fig. \ref{Esp36} and 
good agreement is observed.

\begin{figure}[h]
\centering
\includegraphics[width=.49\linewidth]{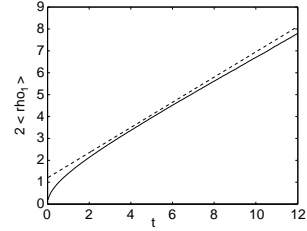}
\caption{
\noindent
$<2\rho_1(t)>$ for a C1 simulation with $\Omega = 0.36$. Averaging has been performed 
over 480000 realisations. A slope of $(8/5)\Omega t$ has also been plotted (dashed line).}
\label{lyp1}
\end{figure}

\begin{figure}[h]
\centering
\includegraphics[width=.49\linewidth]{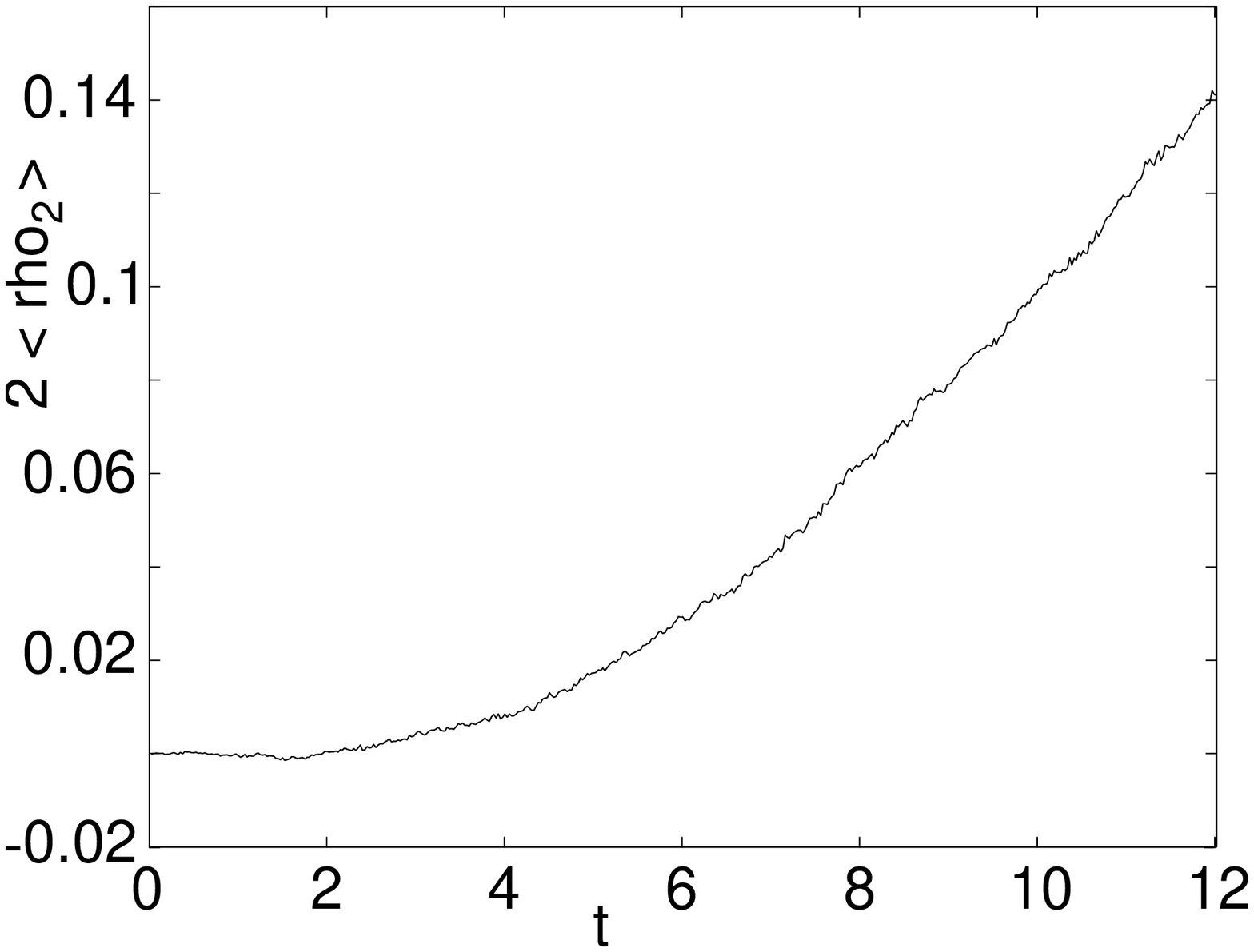}
\includegraphics[width=.49\linewidth]{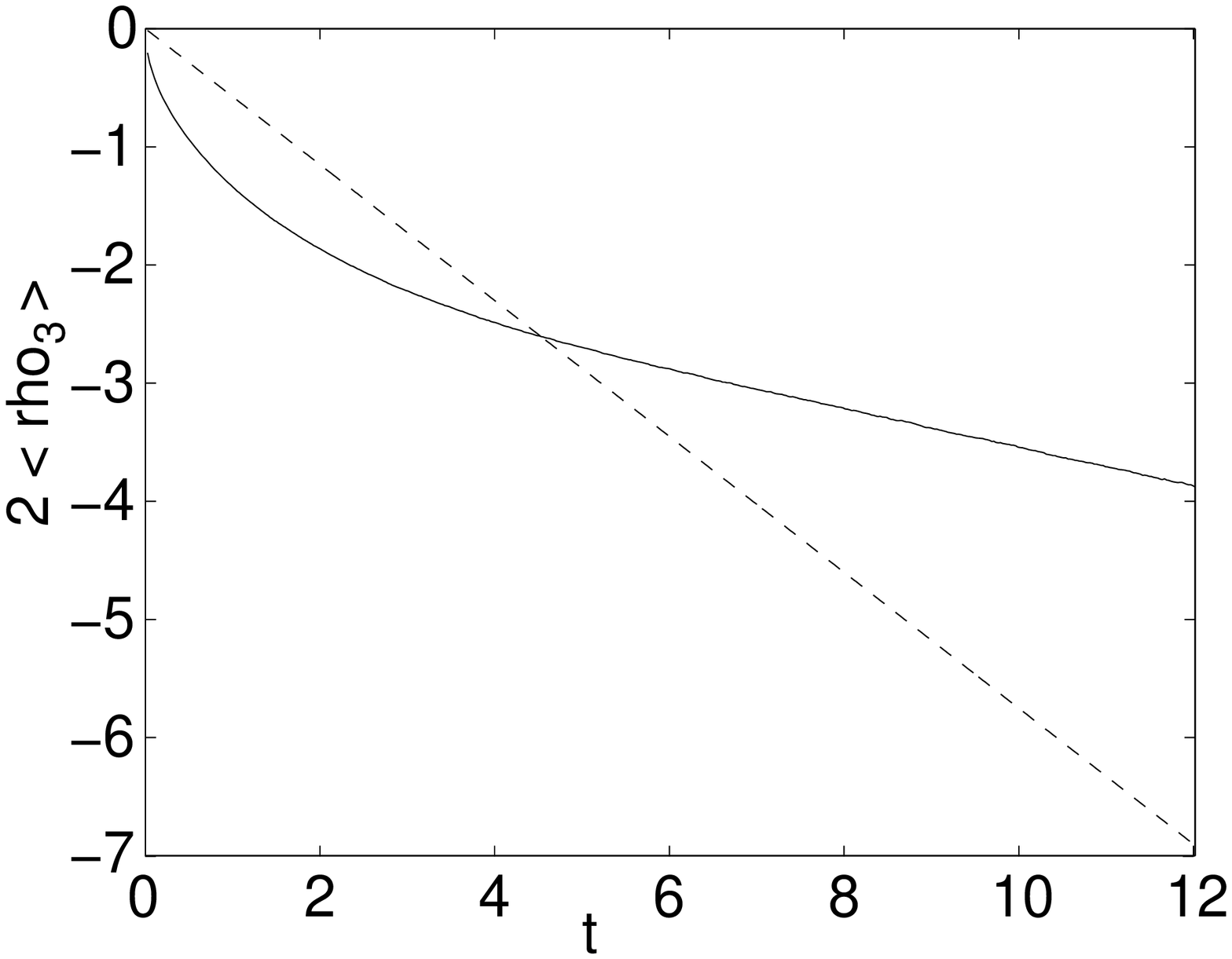}
\caption{
\noindent
$<2\rho_2(t)>$ (left figure) and $<2\rho_3(t)>$ (right figure)  
for a C1 simulation with $\Omega = 0.36$. Averaging has been performed over 480000 
realisations. A slope of $-(8/5)\Omega t$ (right arrow) has also been plotted 
(dashed line).}
\label{lyp23}
\end{figure}

We will briefly review the Lyapunov exponents generated in these C1 numerical experiments
because it help us to understand the origins of some numerical errors in our method. 
Using the Jacobi transform method described earlier, one can calculate the eigenvalues $e^{2\rho_i}$ from either 
$J^T J$ or $JJ^T$, and then order them in terms of size so that $\rho_1>\rho_2>\rho_3$. 
The Lyapunov exponents are then determined via (\ref{lypnum}). Figure \ref{lyp1} shows
$<2\rho_1(t)>$ for a simulation with $\Omega=0.36$. As expected the graph tends to a 
flat line in time, and $\lambda_1>0$. We would expect $\lambda_1 \sim \Omega$, that is
$<\ln e^{2\rho_1} > = 2t \lambda_1  = 2t C \Omega $ where $C$ is the time-scaling constant 
mentioned earlier. In Fig. \ref{lyp1} a line corresponding 
to $C=4/5$ has also been plotted. From theory we would expect
the remaining two Lyapunov exponents to take the values $\lambda_2 = 0$ and $\lambda_3=-\lambda_1$,
with conservation of volume corresponding to  $\sum \lambda_i = 0$. The left and right-hand graphs
of Fig. \ref{lyp23} shows $<2\rho_2(t)>$ and $<2\rho_3(t)>$ respectively. 
Looking at the range of the vertical axis we see $\lambda_2$ does indeed stay around zero 
but there is a gradual deviation from the theoretical value in time. 
On the other hand, $\lambda_3$ is not the same as $-\lambda_1$. 
The line $<\ln e^{2\rho_3} > = -2t \lambda_1  = -2t C \Omega $ with $C=4/5$ has also been 
plotted for comparison. Numerically, we see that $\sum \lambda_i \neq 0$ and hence, 
according to the Lyapunov exponents, the volume is not conserved.
It should also be mentioned that in our simulations, the determinant of the Jacobian $Det[J]$, is held
within $1\%$ of unity, which means that the volume is well conserved. 
We conclude therefore that the deviations observed in the numerical values of $\lambda_i$ from theory 
are mainly due to errors generated in finding the eigenvalues of $JJ^T$ or $J^T J$ using the Jacobi transform
method. This is not overly surprising if we remember that some elements of $JJ^T$ are very large in comparison
to others. Indeed, that is why the alternative method of finding the eigenvalues via solving the
cubic characteristic polynomial of $JJ^T$ runs into difficulties. In the context of the ellipsoids, essentially
the problem arises because each ellipsoid is becoming ever larger in one direction and smaller in another.

It should be noted that the magnetic Reynolds number is relatively small in these simulations. 
The magnetic Reynolds number is by definition
$R_m = \sqrt{L^2\Omega/\kappa}$, giving a range of values $32$ to $72$ in the above simulations, where
we have taken $L$ to be of unity. Thus, it is only the 
relative sizes of $\Omega$ and $\kappa$ that are important. Higher Reynold number simulations
are possible but again increasing $\Omega$ will make the stochastic fluctuations greater and hence
one will need to reduce the timestep in each simulation. Alternatively, decreasing the diffusivity $\kappa$ will
delay the time at which the second regime commences. In which case, the corresponding simulation will have to be 
run over a larger time interval.

\subsection{Code 2}

Code 2 (C2) uses the same foundations as $C1$, an ensemble of particles are initially distributed 
on the surface of a unit sphere in wavenumber space. However, now each particle is given a randomly
orientated magnetic field satisfying ${\bf B}\cdot{\bf q} = 0$. That is, the real and imaginary parts 
of the magnetic field are randomly orientated in a plane which is at right-angles to the particle's
initial wavevector and tangent to the unit sphere in $k$-space. 
Particles are again allowed to evolve in wavenumber space via (\ref{keq}), but in contrast to C1, 
information about the full magnetic field is retained by the integration of equation
(\ref{fourier}). In the case of $\kappa\neq 0$ we need to determine the integral in (\ref{fourier}).
As with C1 this is achieved through the use of a forward Euler scheme.
In this way one can investigate the spectra of Fourier space quantities of interest 
such as the magnetic energy, polarisation and flatness.

\subsubsection{$\kappa = 0$ simulations}

\begin{figure}[h]
\centering
\includegraphics[width=.49 \linewidth]{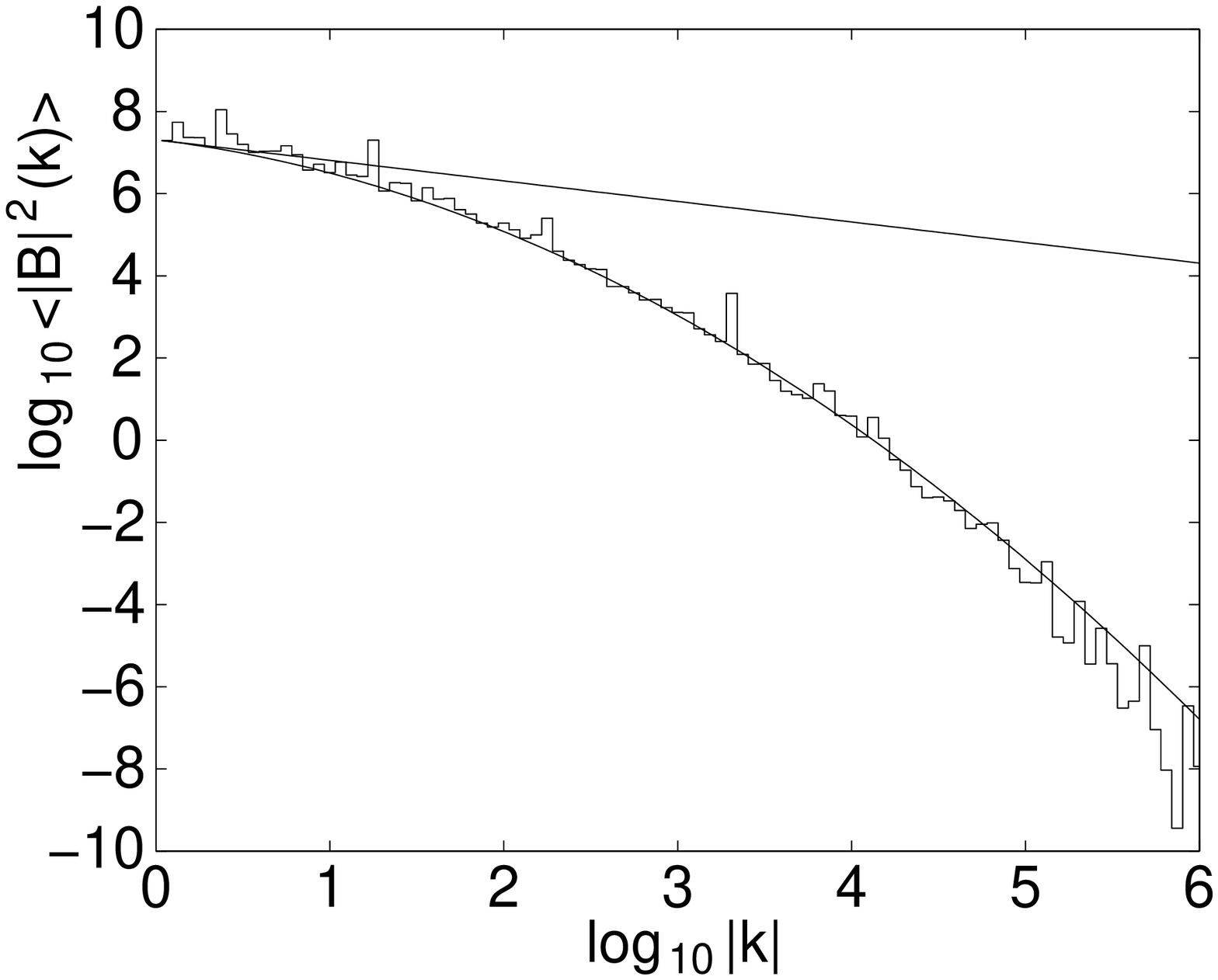}
\includegraphics[width=.49\linewidth]{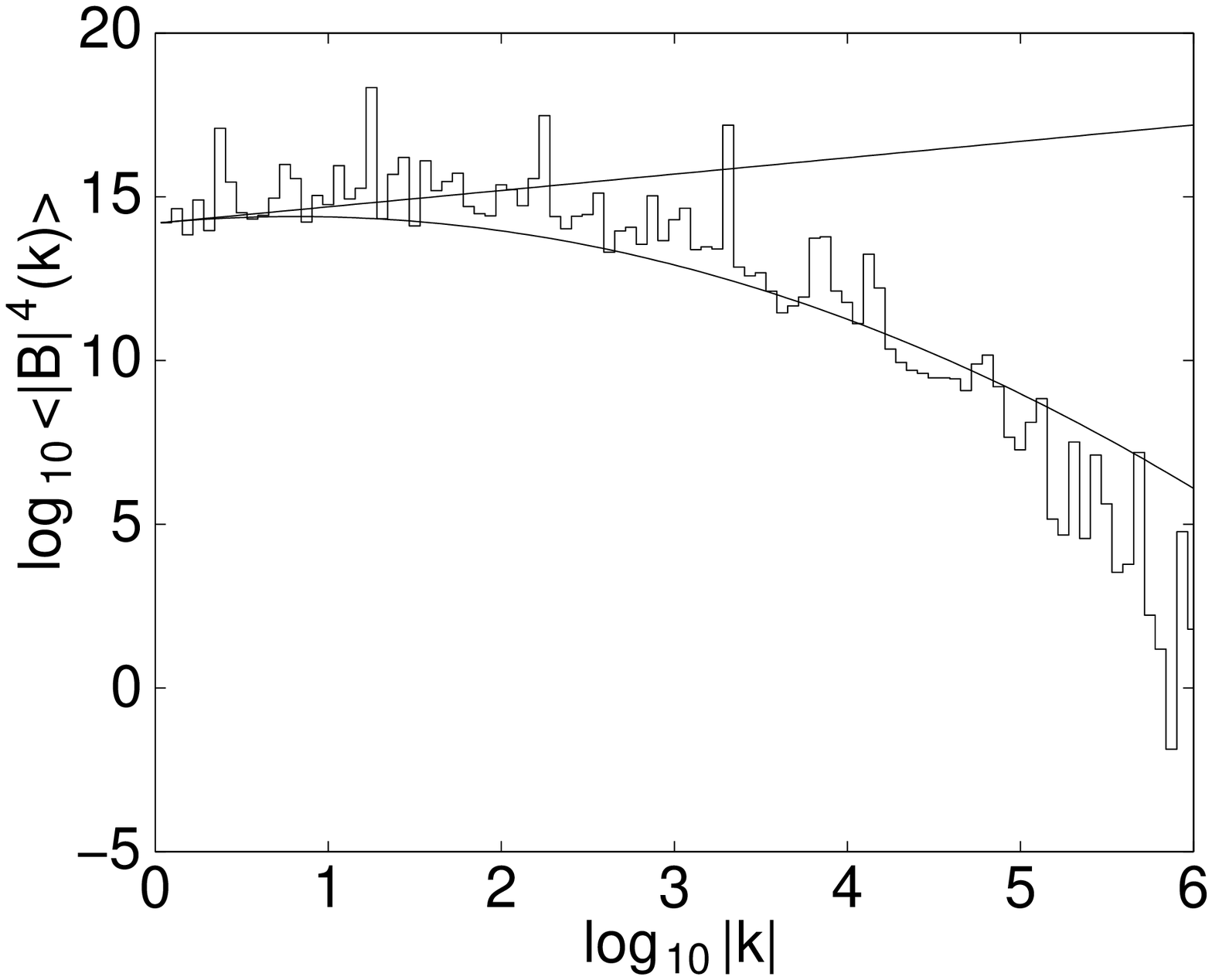}
\caption{
\noindent
$E = <{\bf B}^{2}(k,t)>$ (left figure) and $S = <|{\bf B}|^{4}(k,t)>$ (left figure) 
for a C2 simulation with $\Omega = 0.16$ and $\kappa=0$.
The time here is $t=35$. Averaging has been performed over 600000 realisations.
The particles have been sorted into 100 bins. The theoretical 
solution (\ref{HeatSol}) (curve) and a slope of $k^{-1/2}$ (straight line) 
have also been plotted in the left figure. While the curve (\ref{SHeatSol})
and a slope of $k^{+1/2}$ (straight line) have also been plotted in the right figure.}
\label{Espdiff}
\end{figure}

Starting with the energy spectrum $<{\bf B}^2(k,t)>$. This has been plotted in the left-hand graph of Fig. 
\ref{Espdiff} along with the large time theoretical curve (\ref{HeatSol}) 
and slope $k^{-1/2}$, for the choice of parameters $\Omega=0.16$ and $t=35$. 
As with the C1 experiment Fig. \ref{Elat}, the numerical and analytical 
results support each other well.
As with the investigation of higher order moments $<{\bf B}^2(t)>$ in section \ref{c1resk0},
the higher order correlators, such as  $S=<|{\bf B}|^4 (k,t)>$, will generate noisier histograms
in comparison to their lower order counterparts. Nevertheless, the following numerical results
are still very much of interest. Firstly, we will examine the correlator $S$ which has
been plotted as the right-hand graph of Fig. \ref{Espdiff}. 
The large time asymptotic solutions has also been plotted, and there appears to be good qualitative agreement
with the numerics, although some deviation is evident at large $k$. As with the energy spectrum the 
analytical results tell us that there exists a $k^{+1/2}$ scaling range for $k<k_{c+}$. In this simulation
$\log_{10} k_{c+} = 1.13$. A slope of $k^{+1/2}$ has been included in the graph and again there would seem to be
agreement. However, it must be said that the $k$-interval for this scaling range is perhaps too short and the
histogram too noisy to be conclusive.

\begin{figure}[h]
\centering
\includegraphics[width=.49\linewidth]{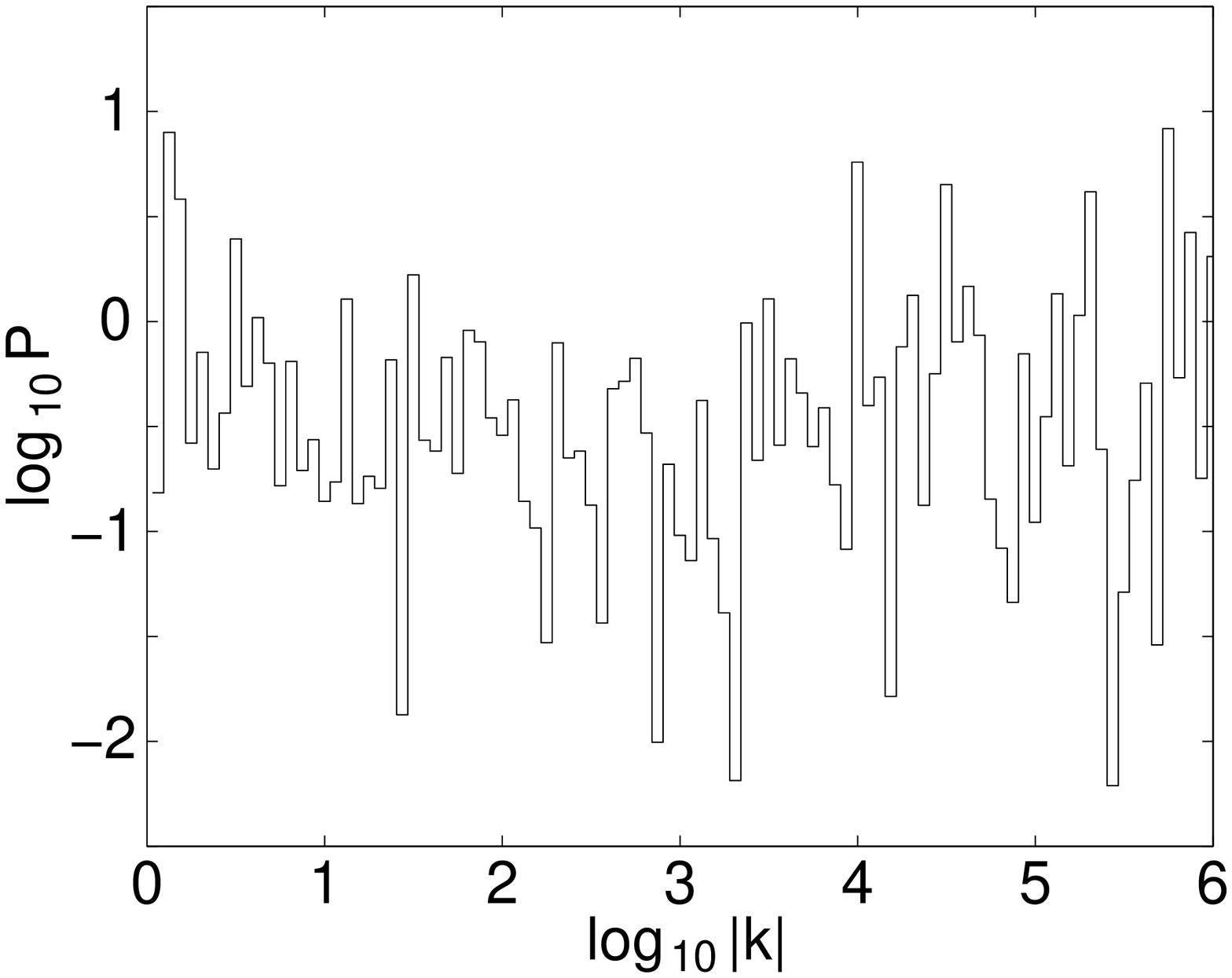}
\includegraphics[width=.49 \linewidth]{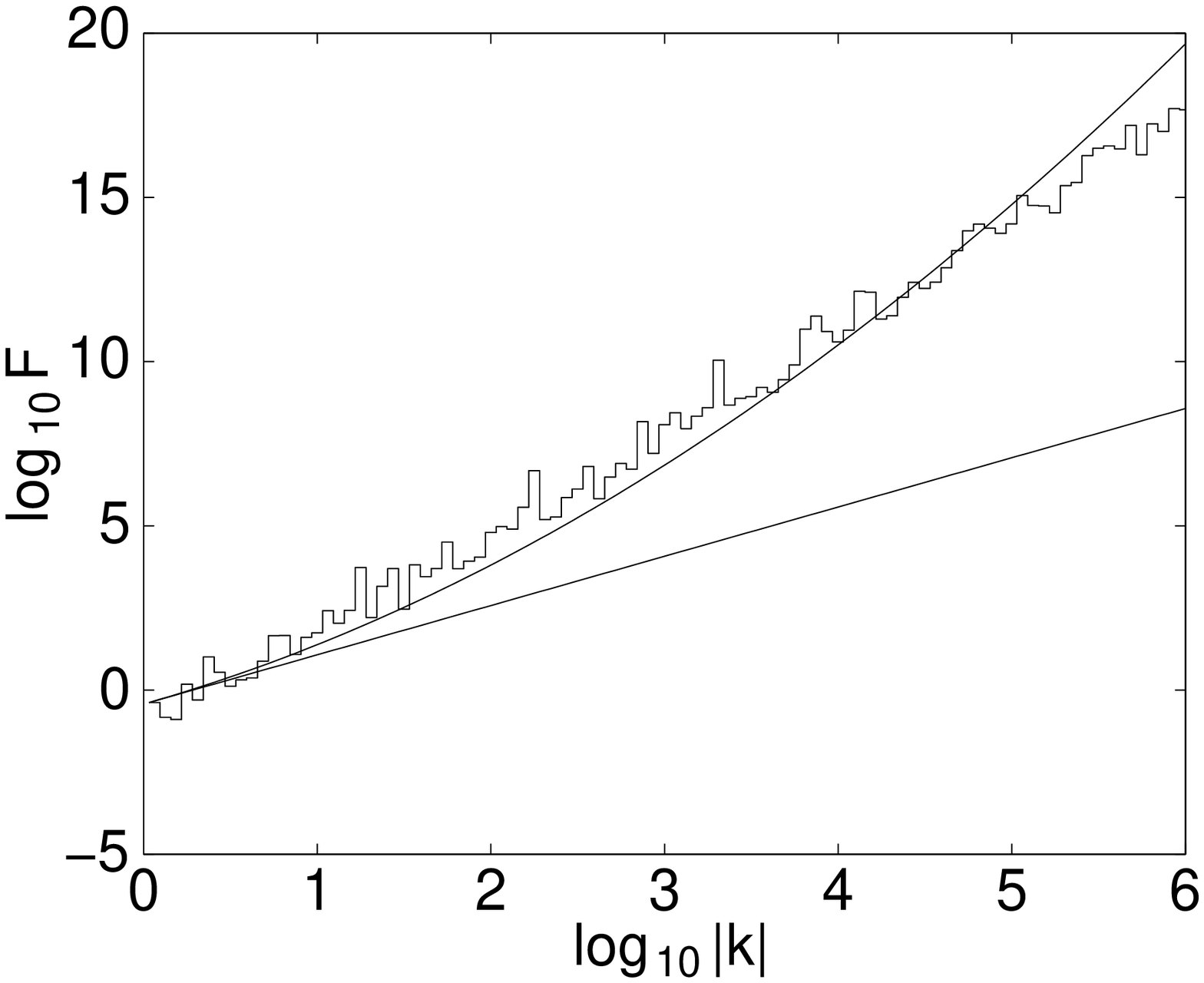}
\caption{
\noindent
$P=(S-T)/S$ (left figure) and $F = S/E^2 $ (right figure) 
for a C2 simulation with $\Omega = 0.16$ and $\kappa=0$. The time here is $t=35$. 
Averaging has been performed over 600000 realisations. The particles have been sorted 
into 100 bins. In the right figure the theoretical curve (\ref{Fsoln}) and slope $k^{+3/2}$ 
(straight line) has also been plotted.}
\label{Polpic}
\end{figure}

For the same simulation, Fig. \ref{Polpic} shows the normalised mean polarisation $P$ (left figure) and 
spectral flatness $F$ (right figure). Let us consider the $P$ spectrum. From the theoretical results 
we expect $P$ to be independent of wavenumber. Although, the histogram is noisy, there does indeed 
appear to be no $k$-dependence, and the resulting spectra is flat. However, for the KKM we also 
have the prediction that the polarisation tends to zero in time (\ref{PHeatSol}). Unfortunately, 
this is not the case numerically. Indeed, taking an average over all $k$ in the histogram we can 
find an approximate value for $P$ from the above spectrum. Repeating this procedure for various 
snapshots in time we find the normalised polarisation remains steady in time at the value of 
$0.32$. One should remember that for a Gaussian field analytically we found $P=1/3$. 
Therefore, in our numerical simulation the polarisation seems to be that of a Gaussian field. 
However, the corresponding spectral flatness is far from being Gaussian. Indeed, the flatness 
$F$ in this perfect conductor simulation can be seen in the right-hand graph of Fig. \ref{Polpic}.
Below $\log_{10} k_{c+}=1.13$ we see agreement with the analytical solution of $k^{+3/2}$. While at 
larger $k$ the theoretical increase in the flatness, due to the log squared term in (\ref{Fsoln}), 
is also well represented.

One may well ask, why is there a discrepancy between theory and the numerically generated polarisation,
but not the flatness? It helps if we can understand the physical significance of these two quantities.
We return to the familiar example of the evolution of an initial ball of isotropic 
magnetic turbulence is wavenumber space. For each realisation this ball is deformed
into an ellipsoid with one large, one short and one neutral direction.
One can visualise this ellipsoid as an elongated flat cactus leaf with thorns aligned to the
magnetic field direction. Note that in this picture one component of the magnetic field
(transverse to the cactus leaf) is dominant which is captured by the fact that the
polarisation $P$ introduced above, tends to zero at large time. 
Another consequence of this picture is that the wavenumber space will be covered by 
the ellipsoids more sparsely at large $k$. This produces large intermittent 
fluctuations of the Fourier transformed magnetic field. These fluctuations are
quantified by the growth of the flatness $F$ as $ k^{3/2}$, a clear indication 
of this small-scale intermittency.

Numerically, it would appear that the flatness is a little more robust than the polarisation, 
arising naturally in the numerics as the ellipsoids become increasingly sparse at large $k$.
In contrast, the apparently Gaussianity of the polarisation can be attributed to the 
mis-alignment of the cactus needles. Indeed, the magnetic needles should align themselves
with the smallest principal axis of the ellipsoid. In this direction the cactus leaf 
is becoming continuously thinner in time, and hence numerically one might expect some 
errors to creep into the alignment of ${\bf B}$. Indeed, this seems to be related to the 
errors observed in calculating the smallest Lyapunov $\lambda_3$.

\begin{figure}[h]
\centering
\includegraphics[width=.49\linewidth]{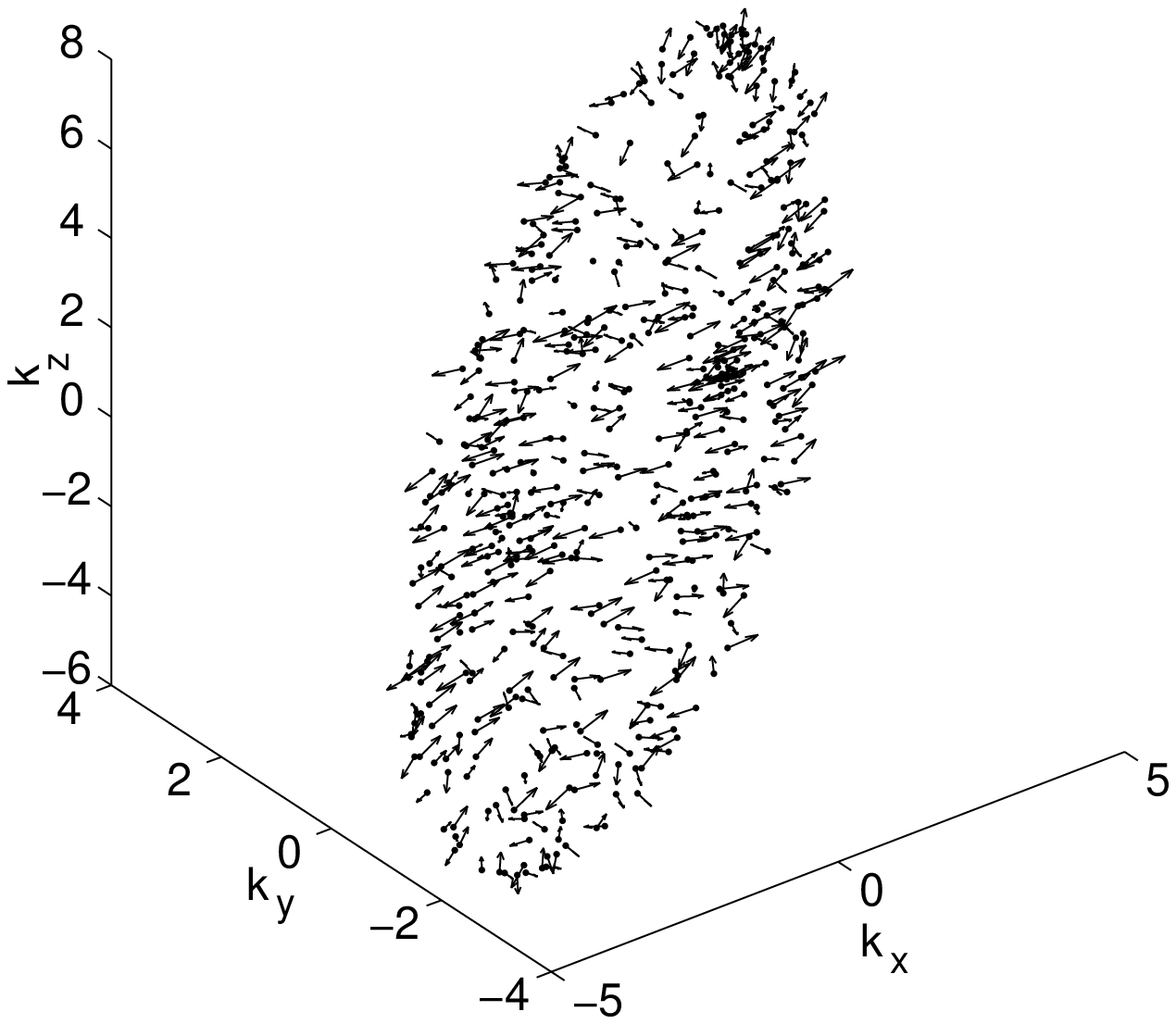}
\includegraphics[width=.49\linewidth]{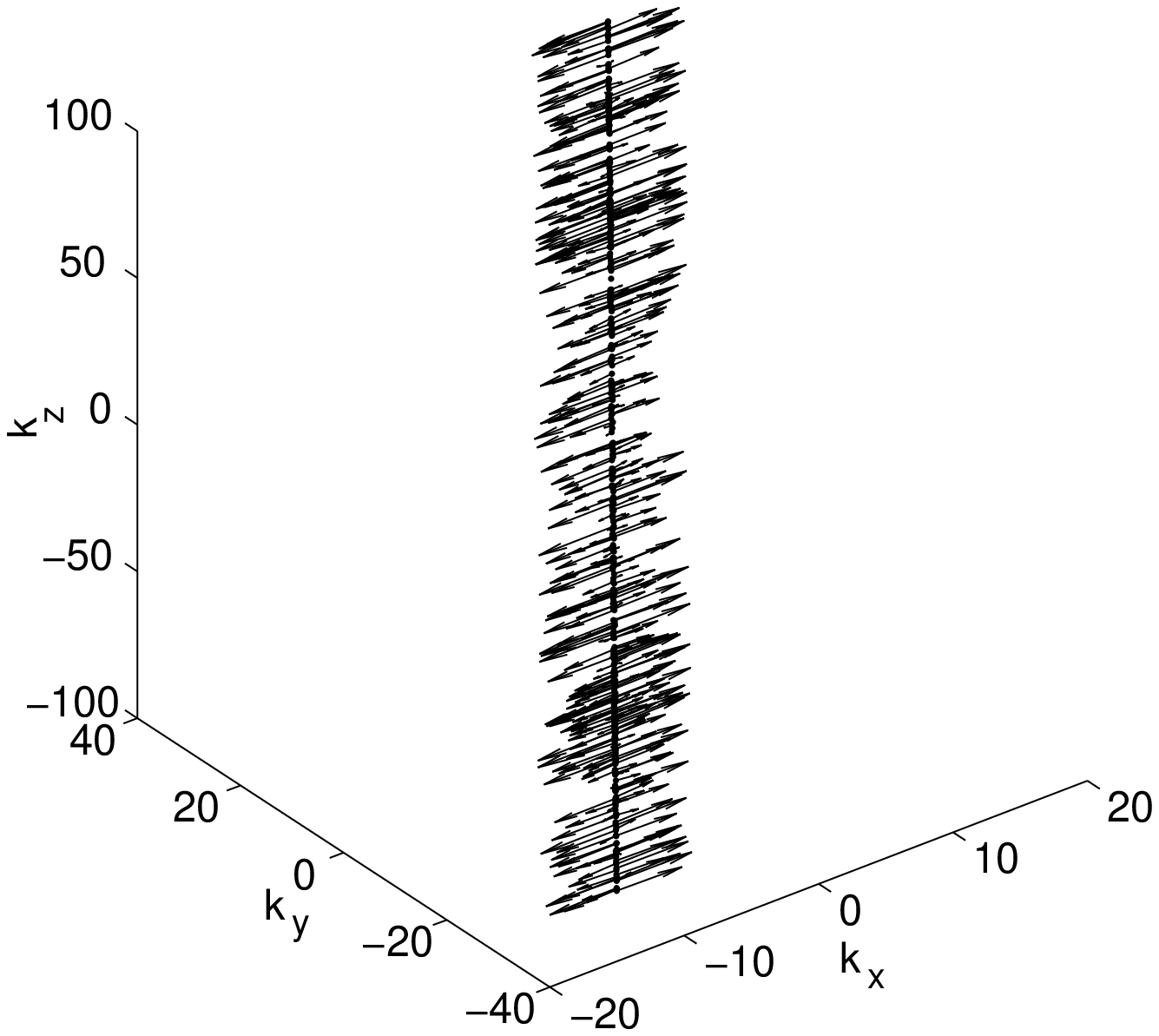}
\caption{
\noindent
The magnetic field of 500 wavepackets in a C2 simulation with $\Omega = 0.36$ 
and $\kappa=0$. The figures show the particle's positions in $k$-space
at $t=6$ and their corresponding real magnetic fields. The
left and right figures correspond to different realisations 
of the strain matrix.}
\label{Ellipsoids1}
\end{figure}

\begin{figure}[h]
\centering
\includegraphics[width=.49\linewidth]{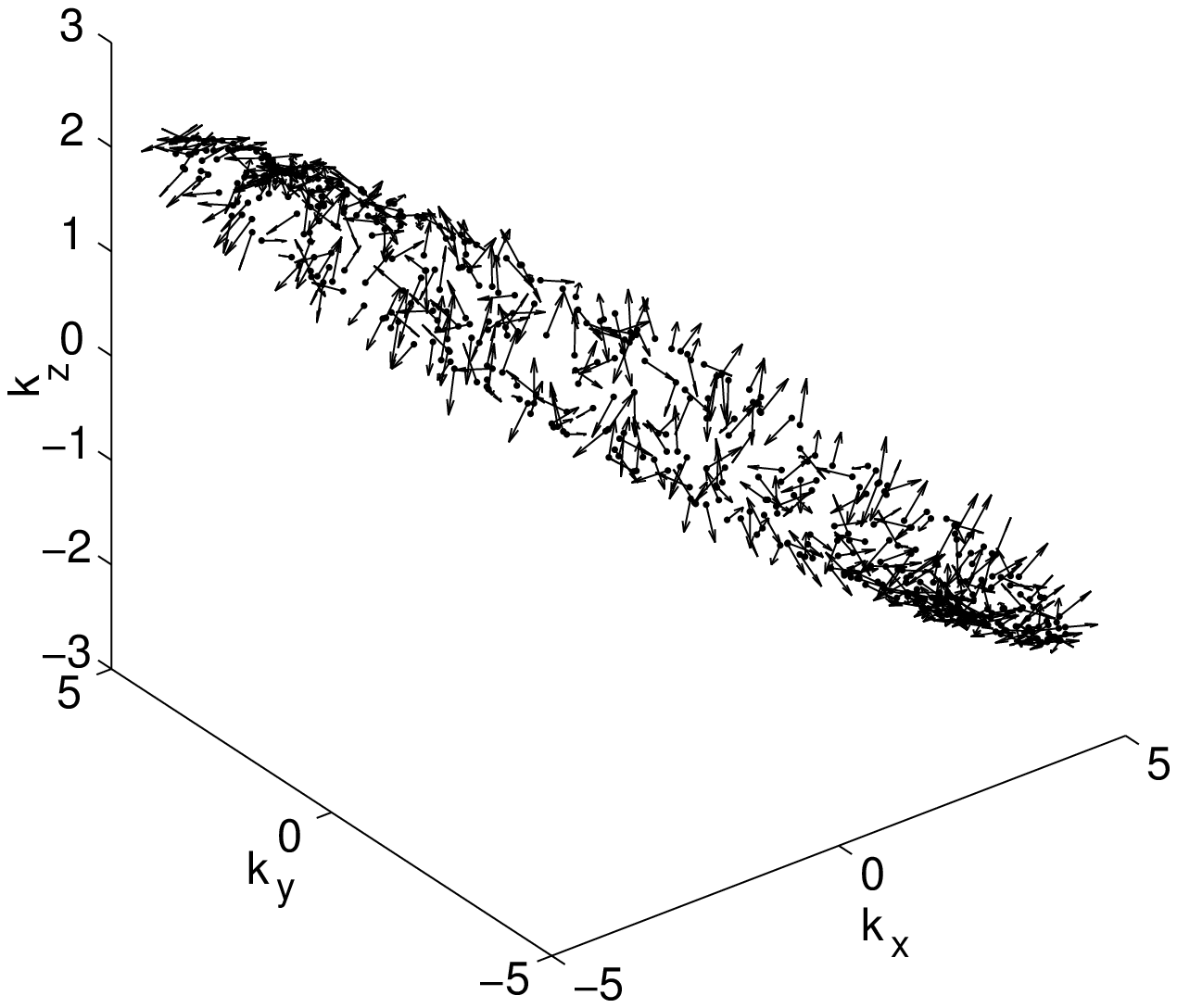}
\includegraphics[width=.49\linewidth]{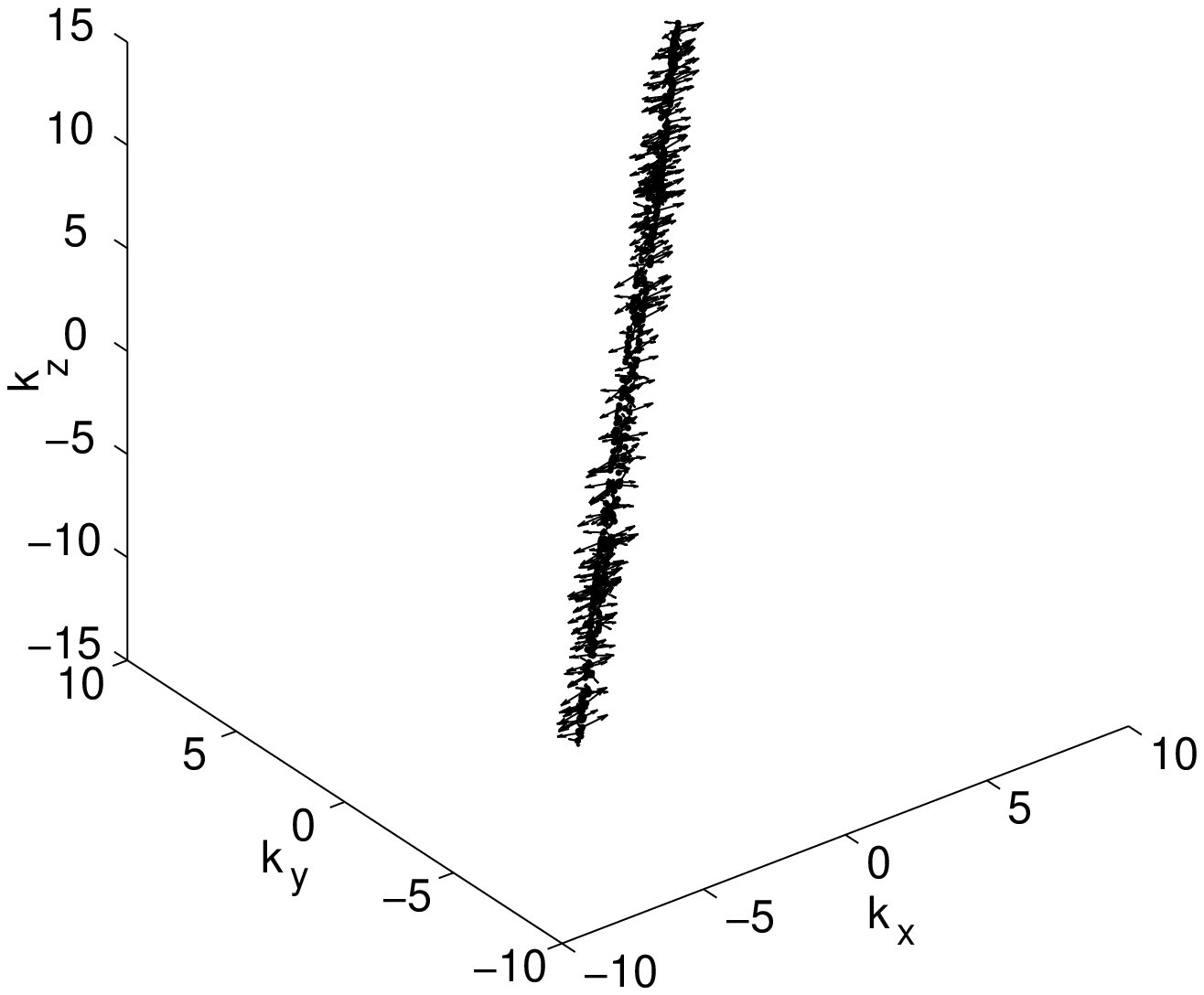}
\caption{
\noindent
The magnetic field of 500 wavepackets in a C2 simulation with $\Omega = 0.36$ and 
$\kappa=0$. The figures show the particle's positions in $k$-space
and their corresponding real magnetic fields at $t=2$ (left figure) and 
$t=5$ (right figure) for one realisation of the strain matrix.
The strain field is the same here as the realisation used in right-hand graph 
figure \ref{Ellipsoids1}.}
\label{EllipEarly}
\end{figure}

\begin{figure}[h]
\centering
\includegraphics[width=.49\linewidth]{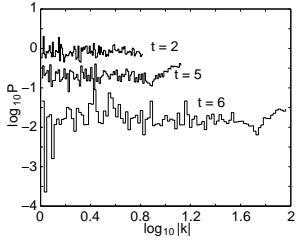}
\caption{
\noindent
The polarisation $P=Q/S$ of 500 wavepackets in a C2 simulation with $\Omega = 0.36$ 
and $\kappa=0$. Here all the particles are subjected to the same 
realisation of the strain matrix. The graph shows the polarisation
at $t=2$ (top line), $t=5$ (middle line) and $t=6$ (bottom line).
The corresponding snapshots of the particle's positions and 
magnetic field can be found in figures \ref{EllipEarly} and 
\ref{Ellipsoids1} respectively.}
\label{EllipPol}
\end{figure}

In fact, although on average our numerical simulations appear to mis-represent the behaviour 
of the polarisation completely, things are not as bad as they would initially seem. Indeed,
further insight can be obtained by considering the behaviour of an ensemble of particles that 
are subjected to the same realisation of the strain. Figure \ref{Ellipsoids1} shows the 
real magnetic fields of a set of $500$ wavepackets that have been subjected 
to two different realisations of the strain matrix\footnote{The imaginary magnetic fields
are qualitatively similar and have therefore not been included.}. 
In the left-hand figure it is clear the magnetic field in this realisation is far from being plane polarised,
the magnetic vectors still being predominantly random in their orientation at this given point in 
time $t=6$. In contrast, in the right-hand figure, which has also been taken at $t=6$, we see the ellipsoid has become very 
elongated and the magnetic field appears plane polarised. Earlier snapshots, at $t=2$ and $5$, 
of the magnetic field for this particular realisation can be found in Fig. \ref{EllipEarly}, 
and the evolution of the corresponding polarisation spectra is shown in Fig. \ref{EllipPol}. 
In this case, we see the polarisation does indeed decay in time and is far from being
Gaussian by $t=6$. In some respects, therefore, our numerical model does get some of 
the polarisation's behaviour at least qualitatively correct. 
 
\begin{figure}[h]
\centering
\begin{minipage}[c]{.99 \linewidth}
\includegraphics[width=.49\linewidth]{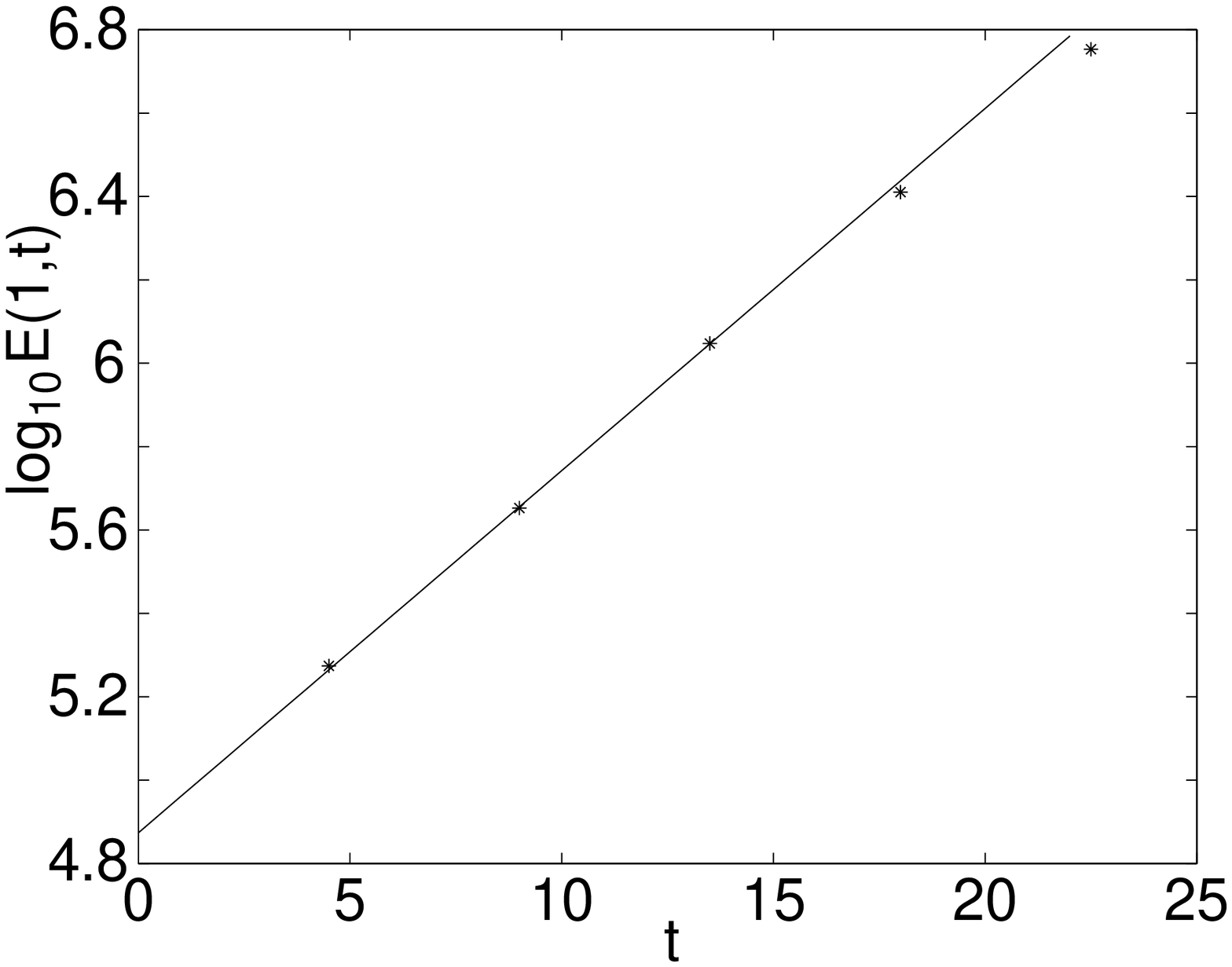}
\hfill
\includegraphics[width=.49\linewidth]{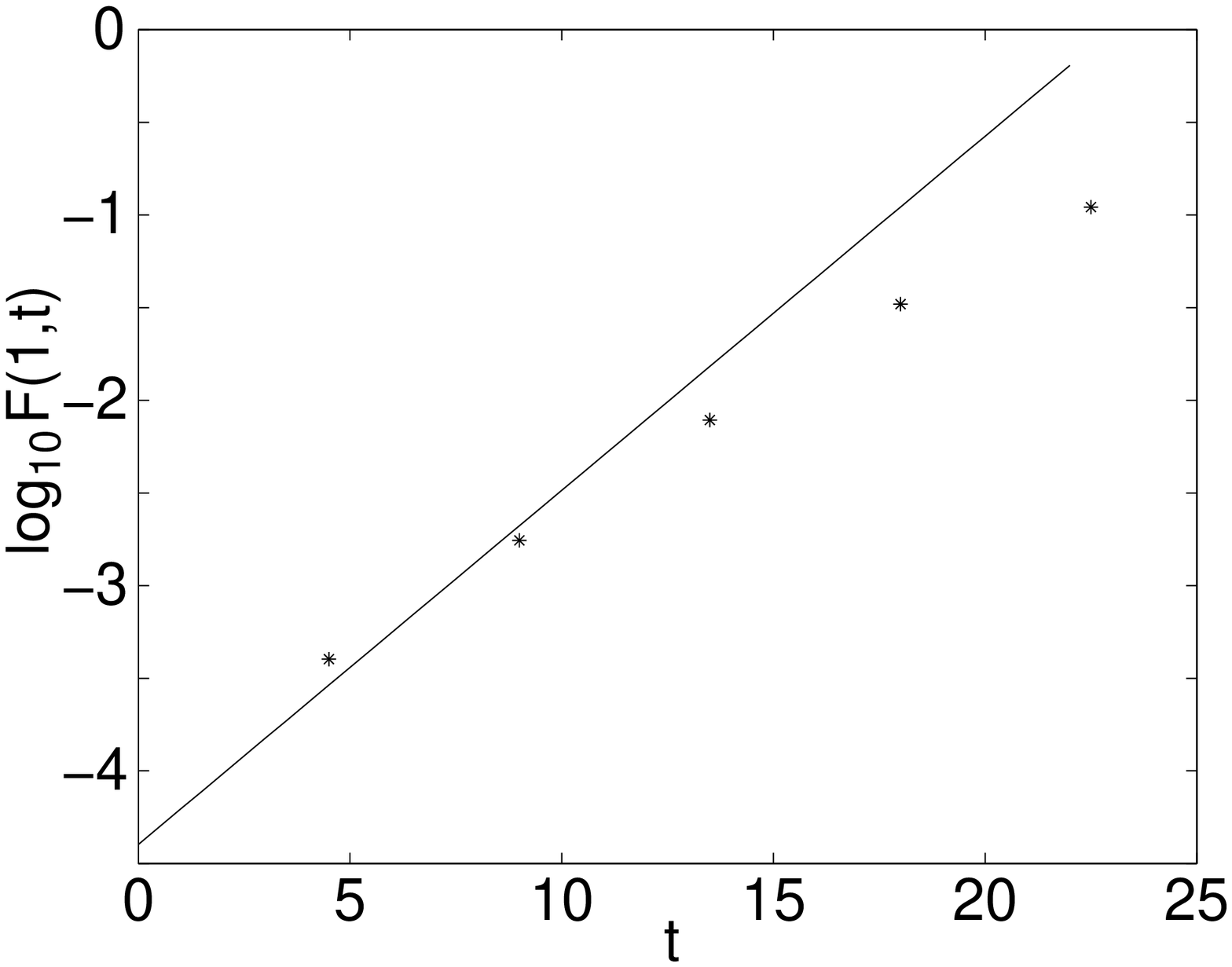}
\end{minipage}\\
\begin{minipage}[c]{.99 \linewidth}
\includegraphics[width=.49\linewidth]{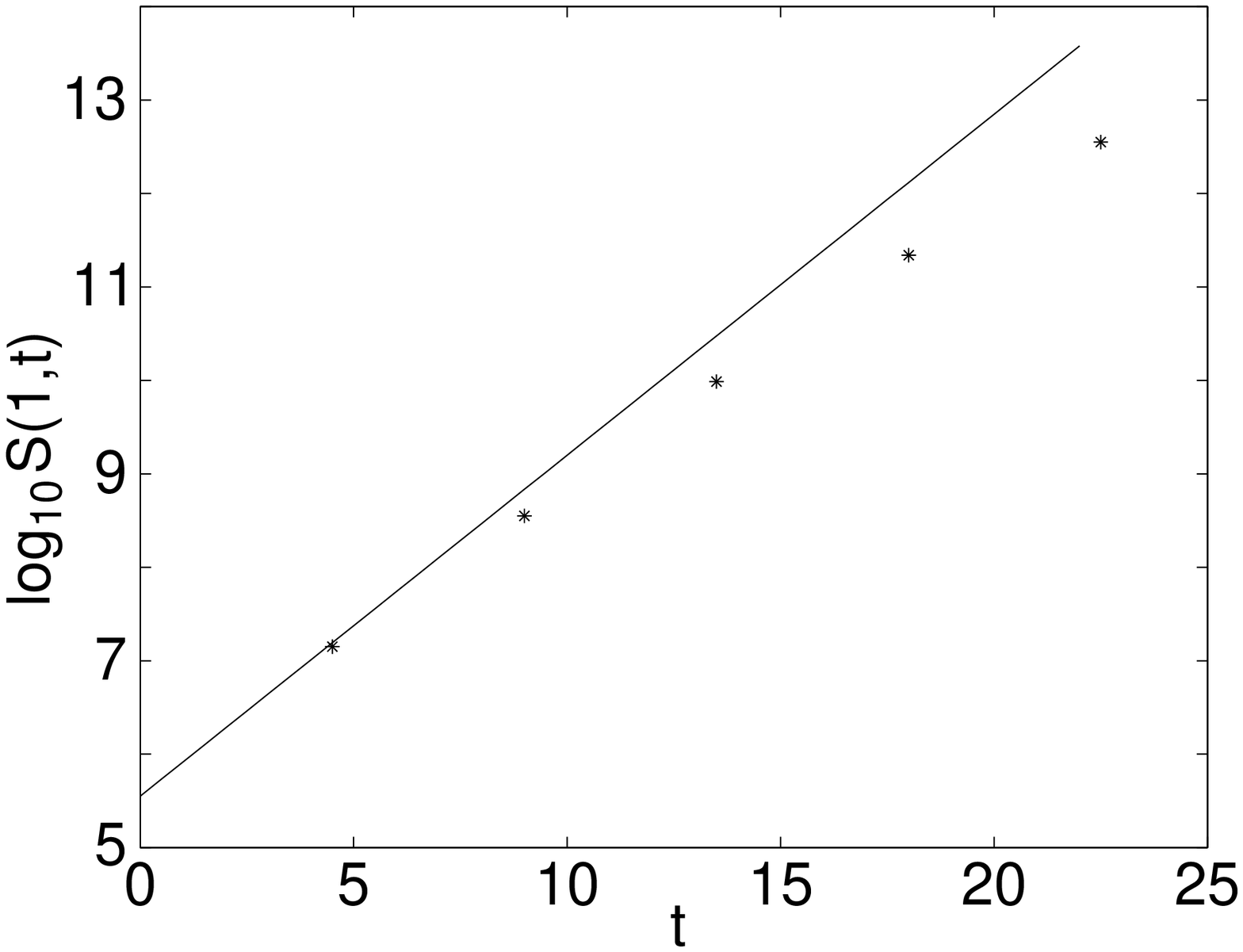}
\hfill
\includegraphics[width=.49\linewidth]{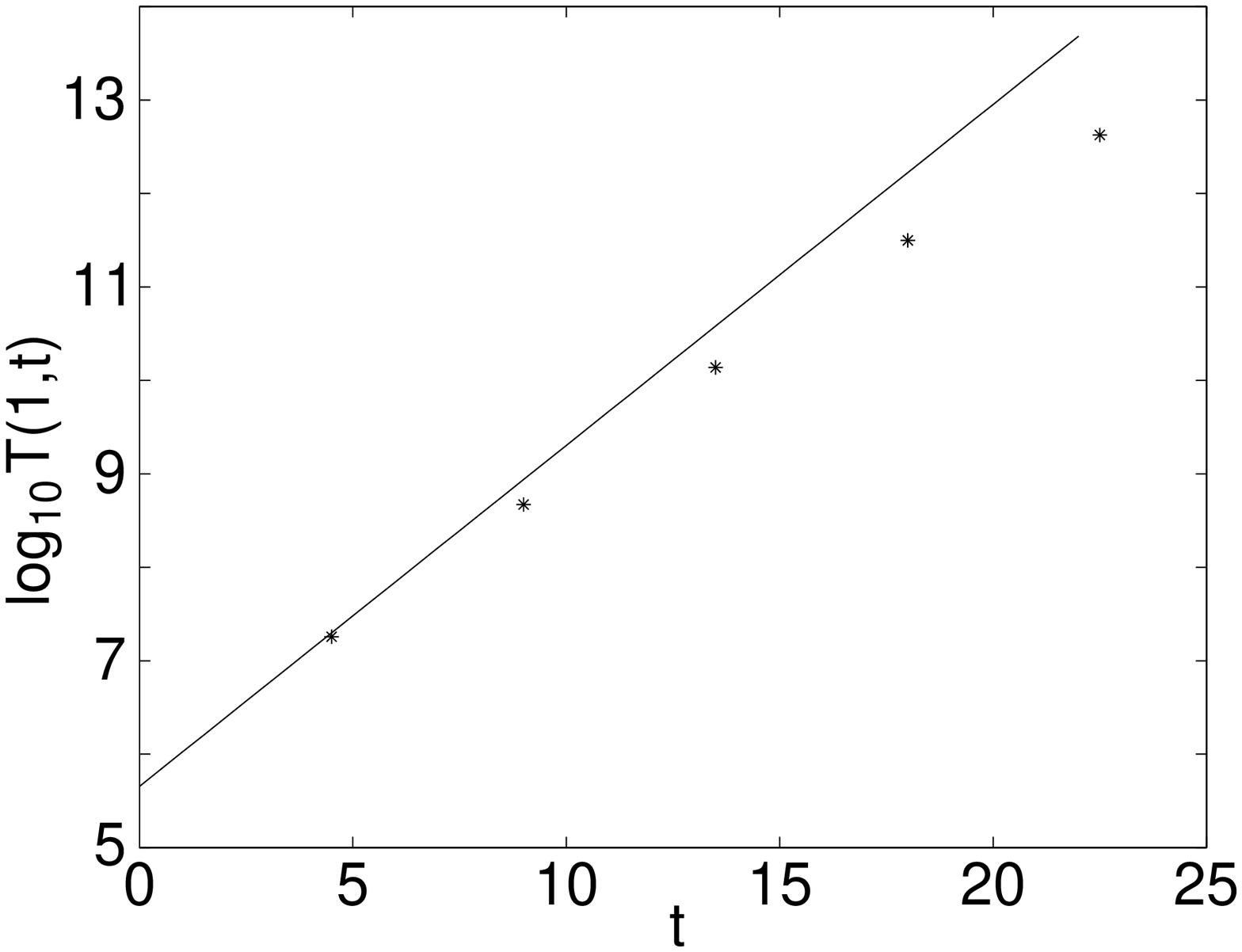}
\end{minipage}
\caption{
\noindent
Growth of $E({\bf k},t)$ (top left), $F({\bf k},t)$ (top right), $S({\bf k},t)$ 
(bottom left) and $T({\bf k},t)$ (bottom right) in time at $k=1$ for a C2 
simulation with $\Omega=0.36$. The theoretical growth for $E$, $F$, $S$ and $T$, 
(\ref{HeatSol}), (\ref{Fsoln}), {\ref{SHeatSol}} and (\ref{THeatSol}) respectively, 
have also been plotted (solid lines).}
\label{GrowRate}
\end{figure}

Before we consider the effects of finite diffusivity, we should check how the growth rates
of the spectra behave in time. That is, how rapidly they move vertically up or down in time.
By recording the value of each spectra for $k=1$, at different snapshots in time, 
we can compare this numerical growth to the theoretical solutions (\ref{HeatSol}), 
(\ref{SHeatSol}), (\ref{THeatSol}) and (\ref{Fsoln}) for $E$, $S$, $T$ and $F$ respectively. 
These results can be found in Fig. \ref{GrowRate}. Along with the numerical values (points) 
the corresponding theoretical growth has also been plotted (solid line). It should be noted that this
is only a rough check. Firstly, because we have had to ignore the $\sqrt{t}$ prefactor
in each analytic solution, and secondly because the numerical values of the spectra at $k=1$ are 
likely to fluctuate at large time as we are again in the tails of the particle distribution.
Nevertheless, all the spectra are growing in time in agreement with theory.

\subsubsection{$\kappa \neq 0$ simulations}

\begin{figure}[h]
\centering
\includegraphics[width=.49\linewidth]{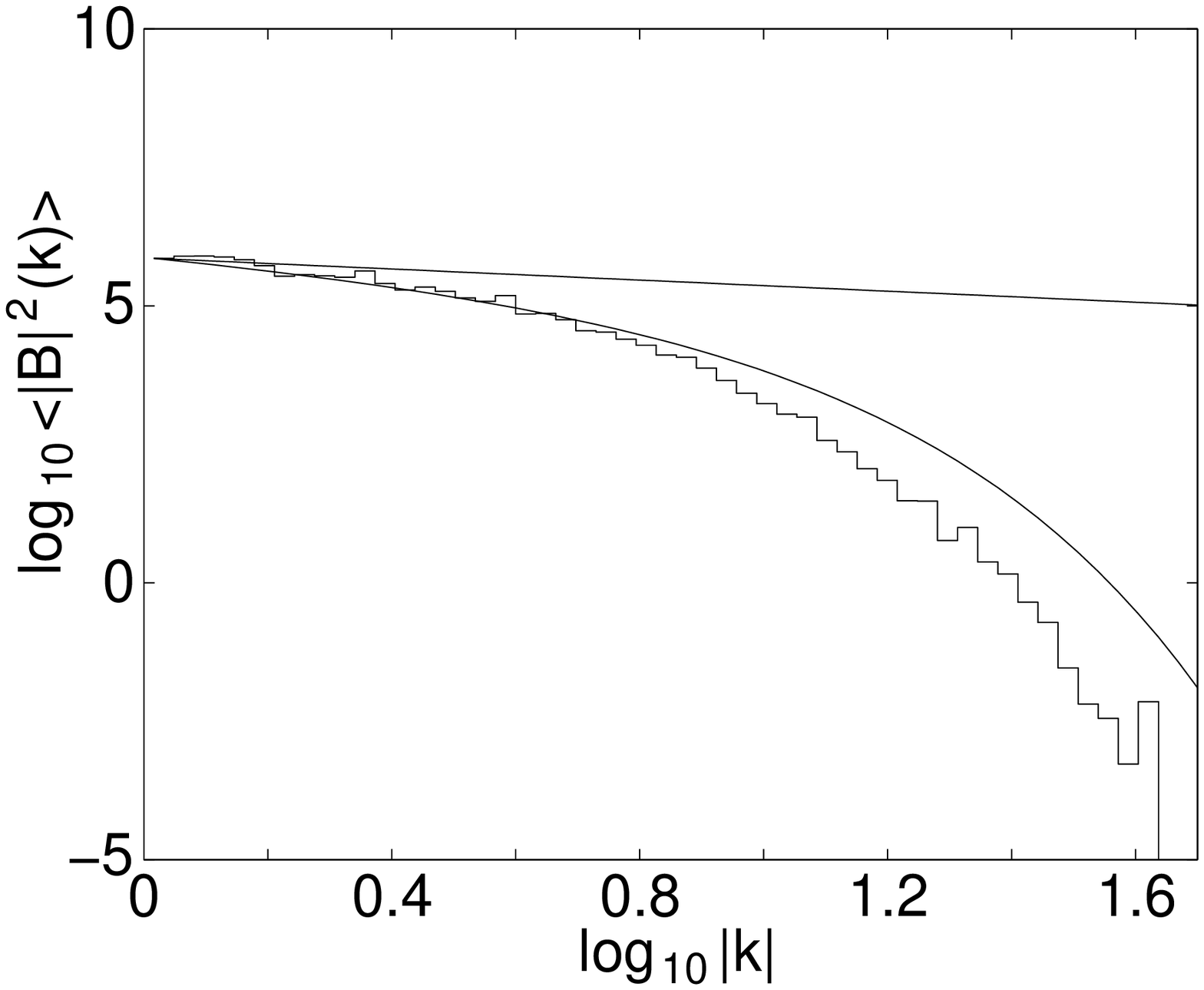}
\includegraphics[width=.49\linewidth]{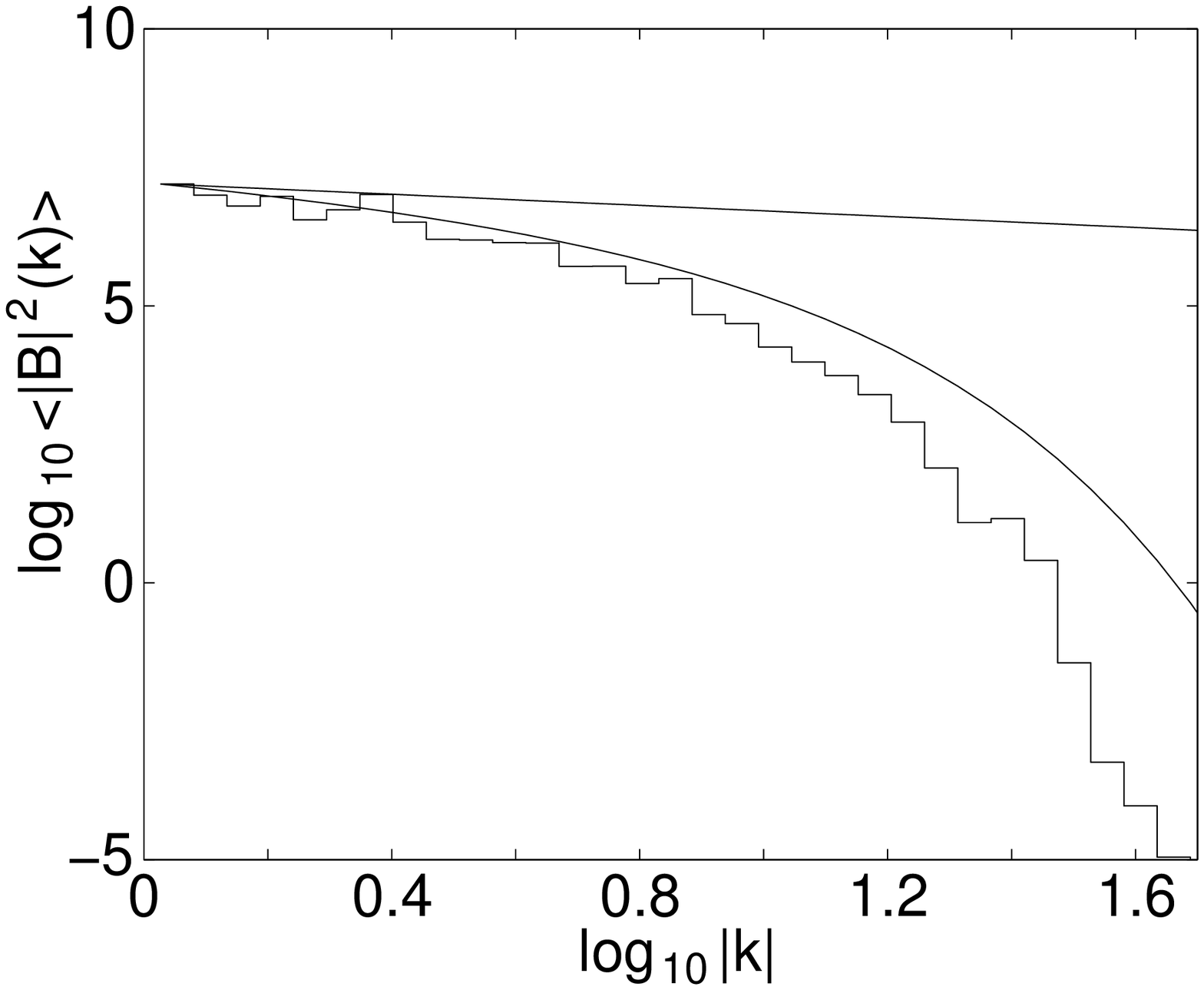}
\caption{
\noindent
$E = <{\bf B}^{2}(k,t)>$ for a C2 simulation with $\Omega = 0.36$ and $\kappa=0.005$.
Averaging has been performed over 480000 realisations. The time here is $t=6$ (left figure)
and $t=12$ (right figure). The particles have been sorted into 100 bins. The theoretical 
solution (\ref{Esoly}) (curve) and a slope of $k^{-1/2}$ (straight line) have also been 
plotted.}
\label{woo2}
\end{figure}

In this section we will discuss the results of a set of C2 simulations with 
a finite diffusivity of $\kappa = 0.005$. 
Figure \ref{woo2}
shows snapshots of $<{\bf B}^2({\bf k},t)>$ at the times $t=6$ and $12$. As with the C1 simulations, 
the finite
diffusivity produces a spectral cut-off. As expected the spectrum moves
vertically upwards in time, with the second histogram getting noiser as the density of particles in
the interval $k<k_{c+}$ is reduced. Comparing these energy spectra to their equivalent 
$C1$ counterparts, Fig. \ref{Esp36}, we should note that the $C2$ numerics appear
to be slightly hyper-diffusive. This effect is most probably numerical, originating
from the integration of the $k^2$ integral found in (\ref{fourier}). 

\begin{figure}[h]
\centering
\includegraphics[width=.49\linewidth]{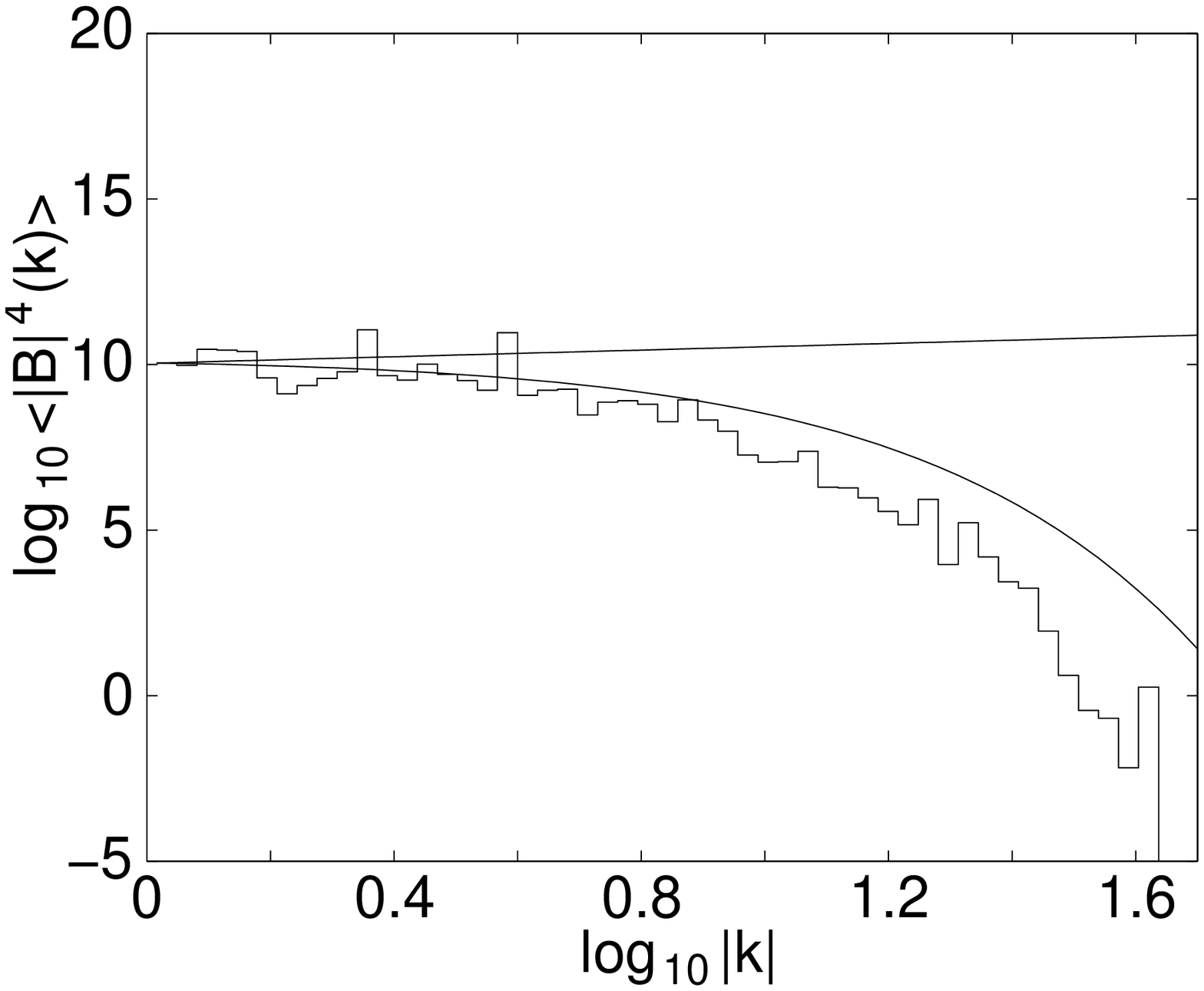}
\includegraphics[width=.49\linewidth]{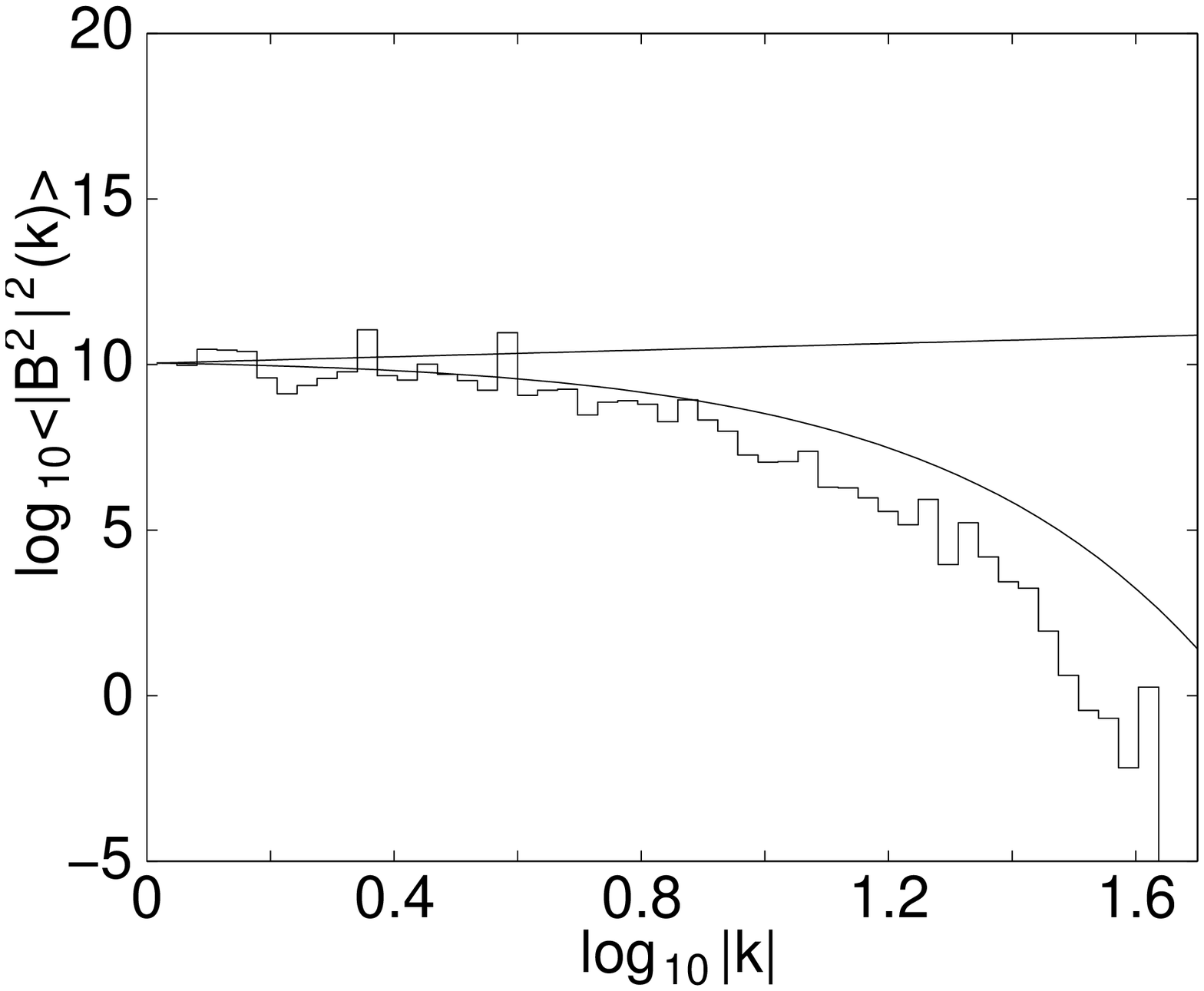}
\caption{
\noindent
$S = <|{\bf B}|^{4}(k,t)>$ (left figure) and $T = <|{\bf B}^2|^{2}(k,t)>$ (right figure) 
for a C2 simulation with $\Omega = 0.36$ and $\kappa=0.005$. The time here is $t=6$. 
Averaging has been performed over 480000 realisations. The particles have been sorted into 
100 bins, (but the axes have been re-scaled). 
The theoretical curves (\ref{Ssol2}) (left figure) and (\ref{Tsol2}) (right figure) 
as well as a slope of $k^{+1/2}$ (straight line) have also been plotted.}
\label{woo4}
\end{figure}

\begin{figure}[h]
\centering
\includegraphics[width=.49\linewidth]{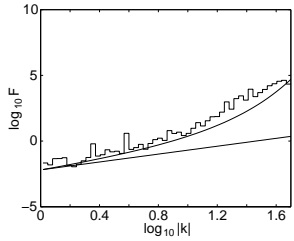}
\caption{
\noindent
$F = S/E^2$ for a C2 simulation with $\Omega = 0.36$, $\kappa=0.005$ and $t=6$. 
Averaging has been performed over 480000 realisations. The particles have been sorted 
into 100 bins, (but the axis have been re-scaled). The theoretical curve (\ref{Fsol2}) 
and slope $k^{+3/2}$ (straight line) have also been plotted.}
\label{woo5}
\end{figure}

\begin{figure}[h]
\centering
\includegraphics[width=.49\linewidth]{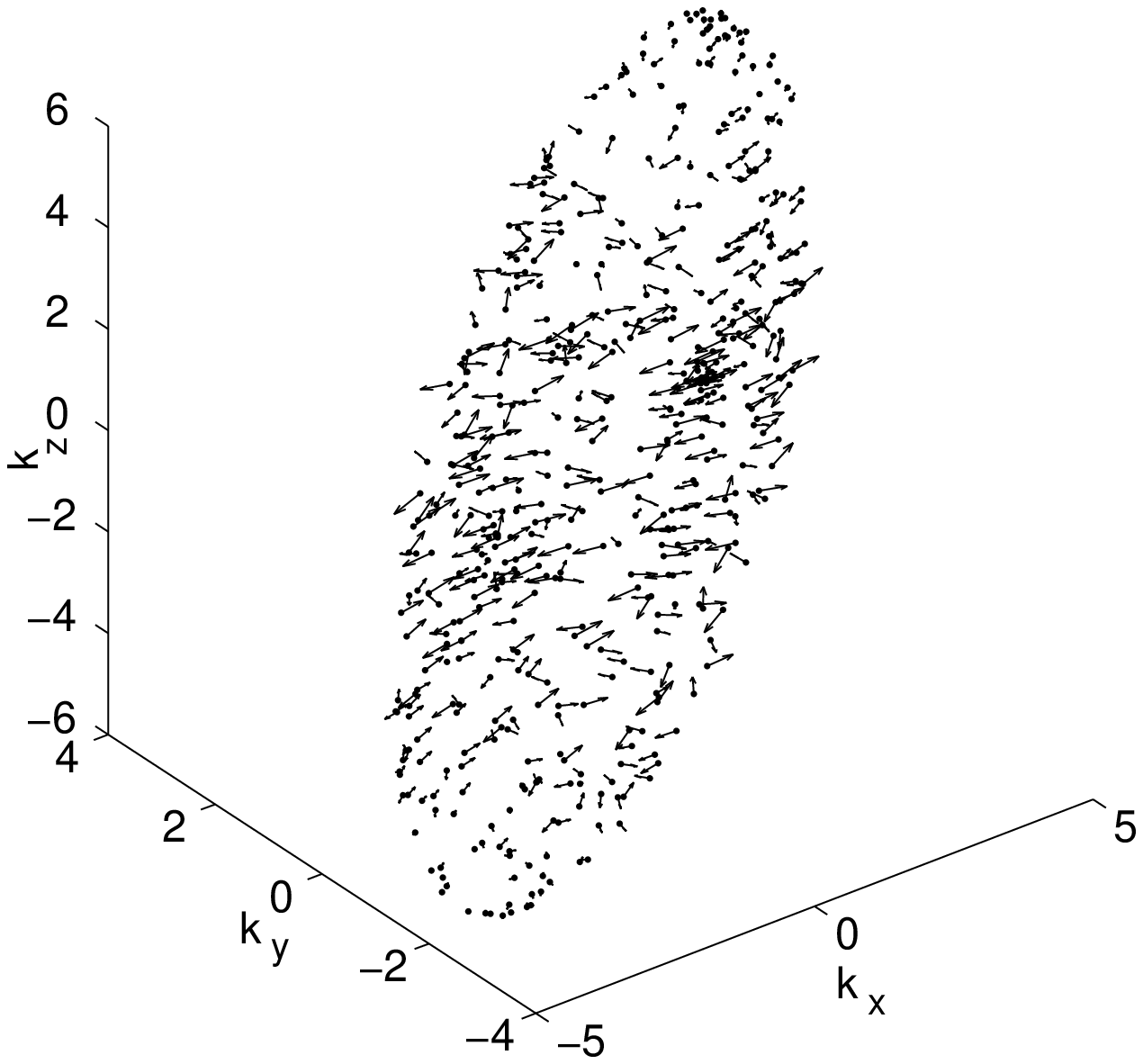}
\includegraphics[width=.49\linewidth]{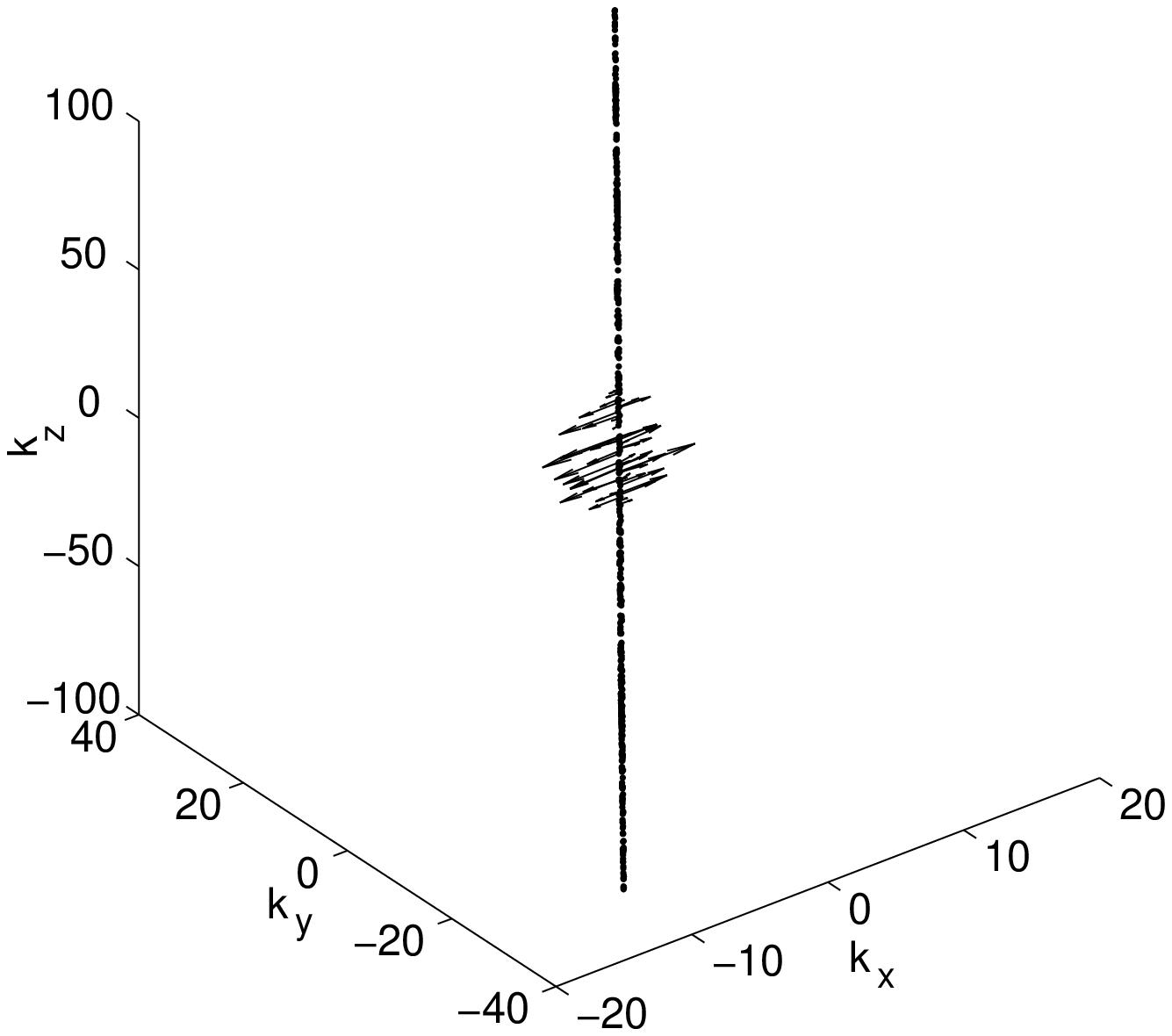}
\caption{
\noindent
The magnetic field of 500 wavepackets 
in a C2 simulation with $\Omega = 0.36$ and 
$\kappa=0.005$. The figures show the particle's positions in $k$-space
at $t=6$ and their corresponding real magnetic fields for 
two different realisations of the strain matrix. The left figure
corresponds to the same strain field as the left-hand graph of 
figure \ref{Ellipsoids1} and similarly for the right-hand graphs.}
\label{EllipDiff}
\end{figure}

The higher order correlators $S$ and $T$ have been plotted in Fig. \ref{woo4}
for a time $t=6$. We should note that the particles here have been sorted into 100 bins. However, the axis
in each graph has been re-scaled to take into account the spectral cut-off. For small $k$, 
qualitatively these figures
are in agreement with the large time asymptotic solutions. However, as with the energy spectrum, 
these results seem to slightly hyper-diffusive. The final figure we will consider is \ref{woo5} 
which shows flatness $F$ at $t=6$. Again we observe good qualitative agreement with theory. 
In particular, one can observe a $k^{3/2}$ scaling region for small $k$ and a heightened 
flatness at larger $k$ in agreement with equation (\ref{Fsol2}). 
This coincides with the analytical prediction that the inclusion 
of a finite diffusivity increases the flatness  (and hence small-scale intermittency) 
at large $k$. The corresponding normalised mean polarisation spectrum has not been included here 
as it is again flat, much like the perfect conductor regime (see Fig. \ref{Polpic})
but with a spectral cut-off. As with the non-diffusive case, on average the normalised 
polarisation appears Gaussian with $P=0.32$. Again, it is interesting to consider the magnetic field 
of an ensemble of 
particles that have been subjected to the same realisation of the strain matrix. Figure \ref{EllipDiff}
shows two such snapshots of an ensemble of $500$ magnetic particles. These two figures should be compared
with Fig. \ref{Ellipsoids1} which show the same realisations but without diffusion.
The right-hand figure in particular demonstrates why, in the diffusive regime, an ellipsoid will cover
$k$-space more sparsely due to the decay of its magnetic field at its tips. This is the reason why 
spectral flatness increases in the diffusive regime.

\section{Numerical Investigations Beyond the KKM}

The purpose of the stochastic codes outlined above was to investigate the small-scale dynamo
system for the case of a Gaussian white strain field. In contrast, the following two codes 
were developed to test the universality of the analytical KKM results summarised in section 
\ref{Fsec}, when more realistic representations of the velocity field are chosen. These 
simulations are based on the integration of the induction equation (\ref{induct}),
which when written in a Lagrangian frame of reference in $k$-space\footnote{Strictly 
speaking, the Fourier transforms used in deriving this
equation must be taken over a box. The box size is greater than the 
characteristic length-scale of magnetic turbulence, but less than the 
characteristic length-scale of the velocity field. In this case,
model (\ref{straindef}) corresponds to ignoring the quadratic (in $x$) terms 
in the scale separation parameter. The strain here is measured along the fluid path.}
takes the form
\begin{equation}\label{Amp}
    \frac{d}{dt} B_m  = \sigma_{mi} B_i - \kappa k^2 B_m,
\end{equation}
where \mbox{$ d/dt \equiv \partial_t + \dot{k}_{j}\partial_{k_j}$} and
\begin{equation}\label{Vec}
   \dot{k}_j = - \sigma_{ij} k_i.
\end{equation}
Numerically we consider an ensemble of particles (wavepackets), whose individual magnetic
fields evolve according to (\ref{Amp}) and wavevectors according to 
(\ref{Vec}).

\subsection{Code 3}

In code 3 (C3), we consider a synthetic Gaussian strain field with a finite correlation time.
The r.m.s. of the different strain components in this numerical experiment ranged from
$2.9$ to $3.6$ and the correlation time was $0.02$. Thus the correlation time was about
sixteen times less than the characteristic strain distortion time. If we choose a timestep 
which is much smaller than the correlation time, we no longer need to consider a stochastic scheme, 
but can instead use a more traditional second-order Runge-Kutta method to integrate 
(\ref{Amp}) and (\ref{Vec}). This experiment consisted of 
2048 strain realisations
with  2048 magnetic wavepackets and the timestep was chosen to be twenty times smaller than
the strain correlation time. Initially, the particles are randomly distributed 
within a ball of radius $2$ in the $k$-space. Each particle is given a randomly orientated magnetic field 
(with random phase) that is transverse to its initial wavevectors ${\bf q}$ and which has a 
random amplitude less than or equal to $2\times 10^{-4}$.

\begin{figure}[h]
\centering
\includegraphics[width=.49\linewidth]{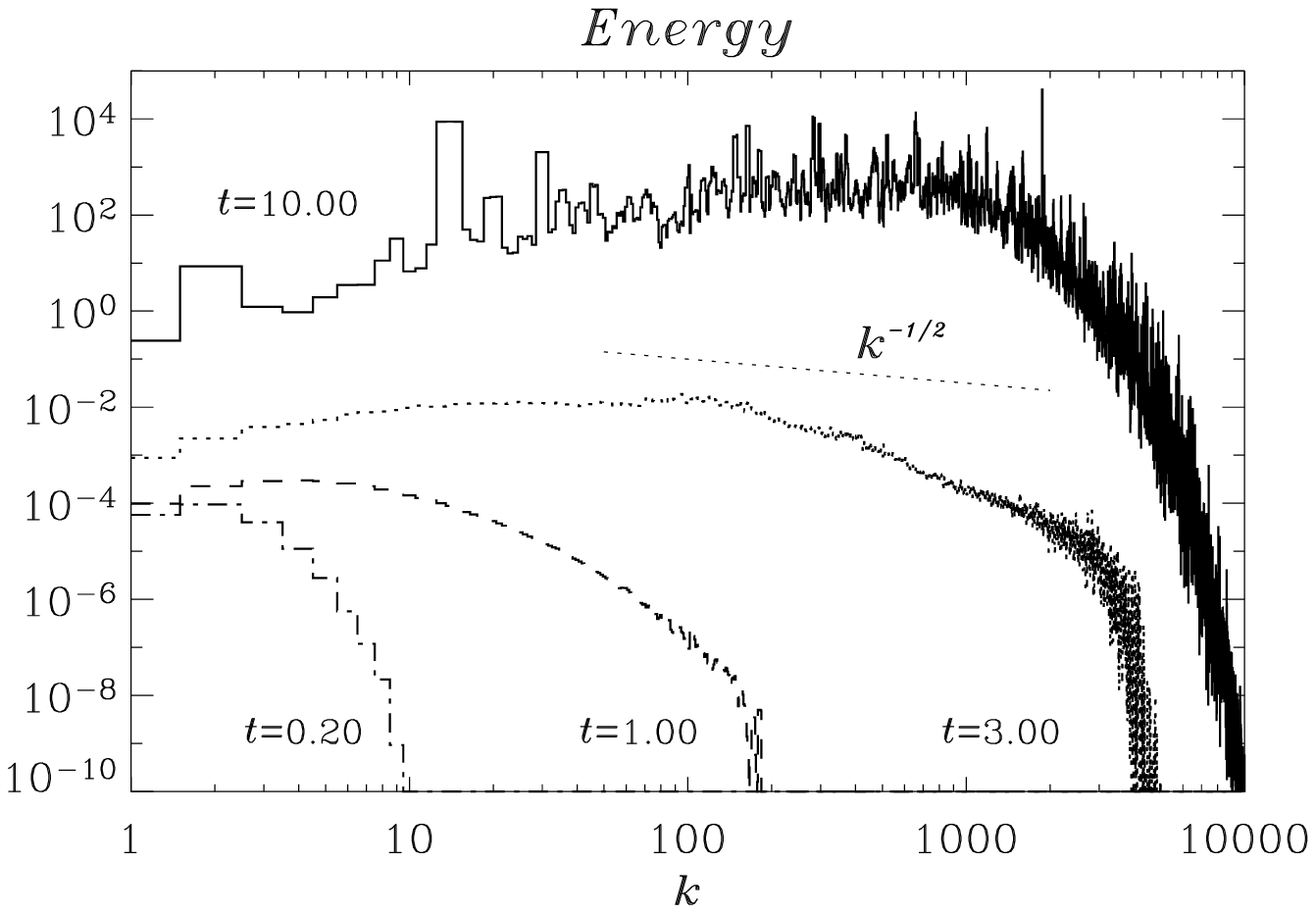}
\includegraphics[width=.49\linewidth]{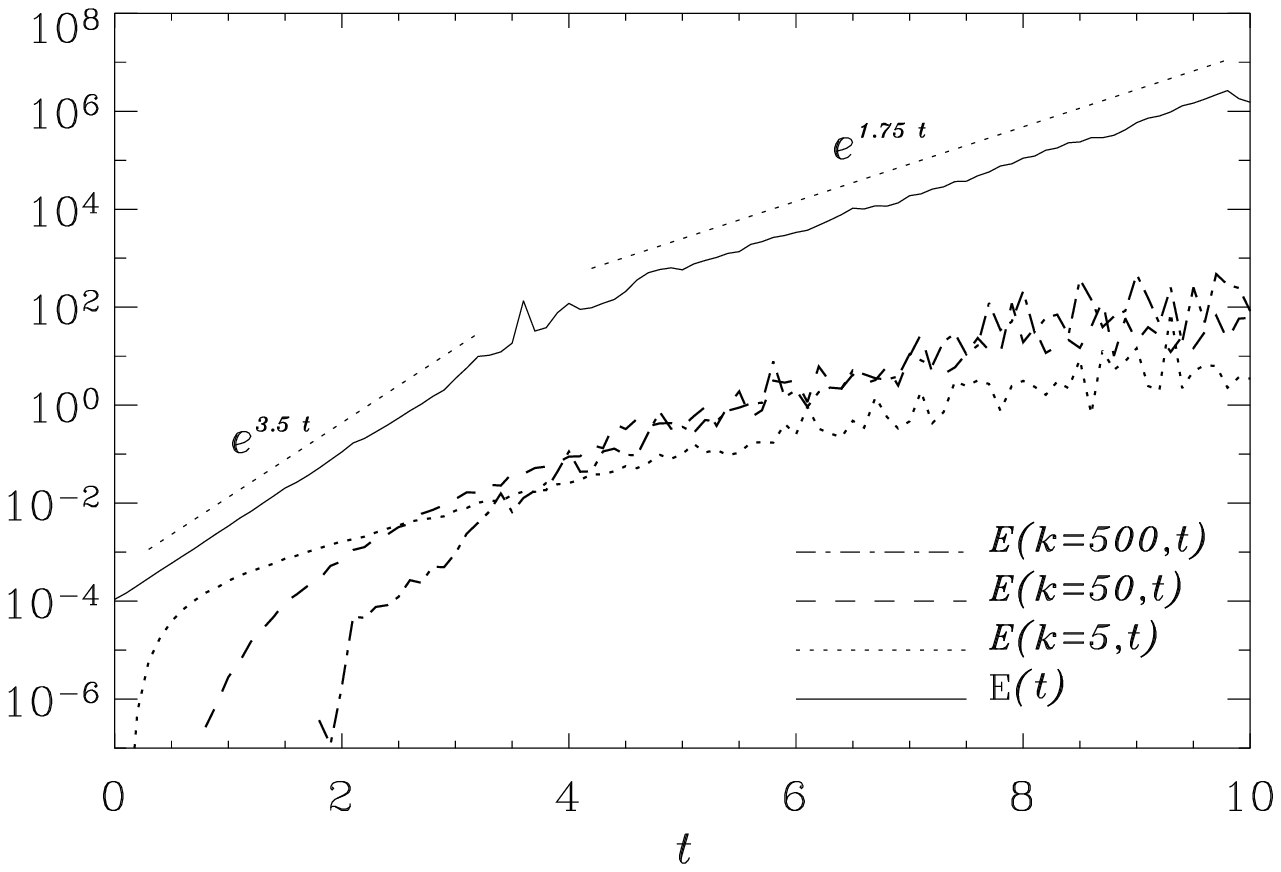}
\caption{
\noindent
The magnetic energy spectrum $E(k,t)$ (left figure) and growth of the 
total magnetic energy $<{\bf B}^2(t)>$ (right figure) 
for the case of a synthetic Gaussian strain field with 
finite correlation time. The right-hand figure also shows 
the growth of the energy for a selection of individual $k$-modes.}
\label{EGM}
\end{figure}

\begin{figure}[h]
\centering
\includegraphics[width=.49\linewidth]{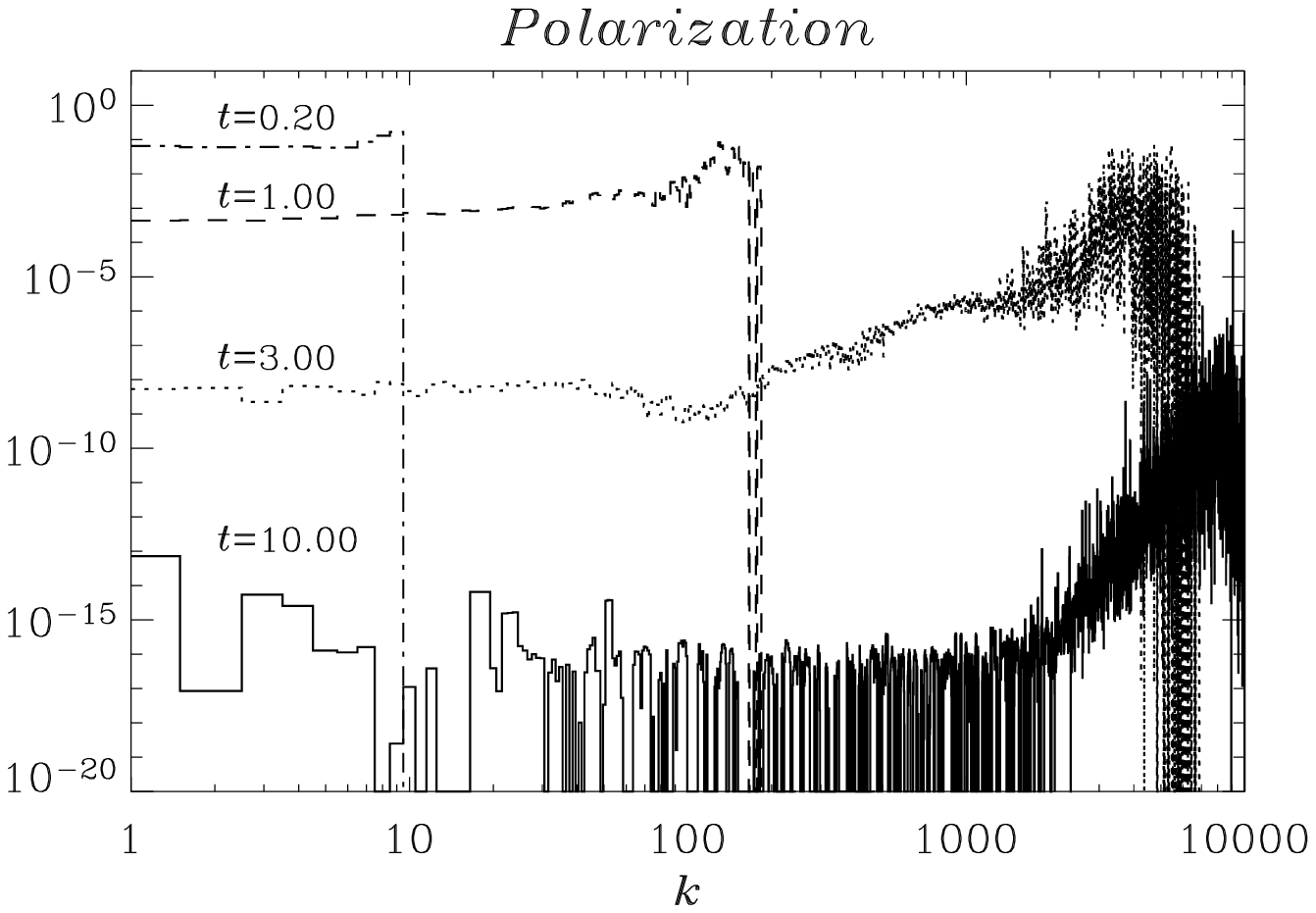}
\includegraphics[width=.49\linewidth]{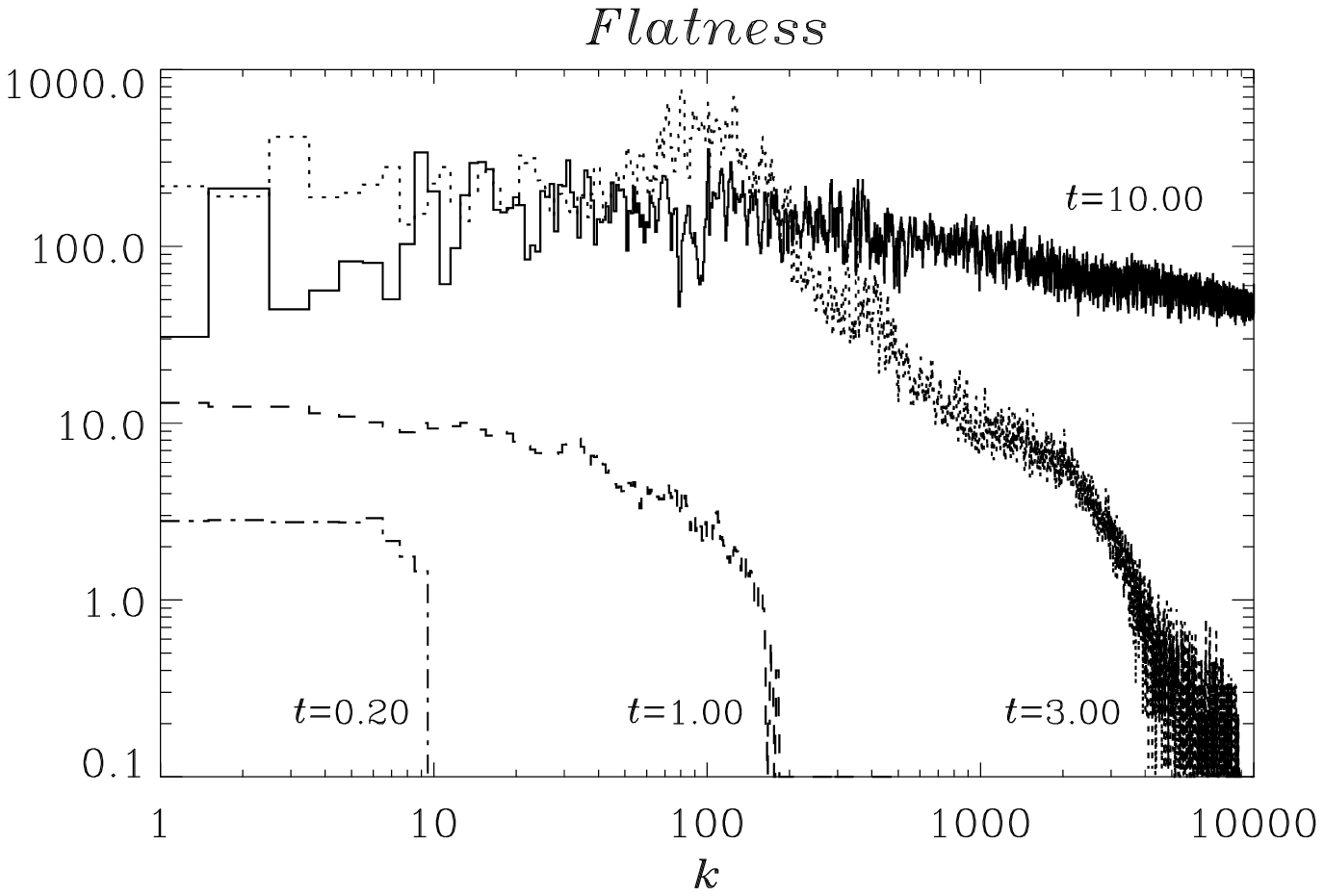}
\caption{
\noindent
The normalised mean polarisation (left figure) and spectral flatness (right figure) 
for the case of a synthetic Gaussian strain field with finite correlation time.}
\label{PFGM}
\end{figure}

Figure \ref{EGM} shows the energy spectrum (left figure) and the growth of the total energy (right figure).
The energy spectrum is consistent with theory, although there is no clear $k^{-1/2}$ range. At later
time (in the diffusive regime) one can see that the spectrum acquires a cut-off.
In the right-hand figure, the total energy grows exponentially
with two different scaling regimes, as expected. The time at which
diffusion becomes important is around $4$. In agreement with theory, in the diffusive regime the
growth rate of the total energy is approximately equal to the growth of the individual $k$-modes.
Remarkably, the ratio of the growth rate in the perfect conductor regime to the growth rate
in the diffusive regime is in good agreement with theoretical predictions.
In Fig. \ref{PFGM} one can see the corresponding mean polarisation (left figure) and 
spectral flatness (right-figure) from this simulation. One can see that the polarisation
spectrum is flat and it rapidly decreases in time. Both of these features are
in agreement with the theoretical KKM results. Thus, the KKM based 
analytical prediction that the magnetic turbulence becomes plane polarised is
much better represented in this simulation than in the previously discussed stochastic simulations.
The spectral flatness rapidly grows in time  which is also in agreement with the KKM theory.
However, there is no obvious region
of the theoretical $k^{3/2}$ scaling.

\subsection{Code 4}

Code 4 (C4) is almost identical in implementation to C3. However, in this 
numerical experiment the strain matrix components were obtained from a 
$512^{3}$ spectral DNS of the Navier-Stokes equations
 at a Taylor Reynolds number of approximately 200.
The strain time series  calculated via recording
\begin{equation}
   \sigma_{ij} = \frac{\partial v_i}{\partial x_j},
\end{equation}
along $512$ fluid paths. At each of these fluid particles we put
 4096 magnetic wavepackets.
The r.m.s. of the different strain 
components in this experiment ranged from $6.6$ to $9.3$ and the correlation 
time was approximately $0.08$. Thus, the correlation time in this case is 
of the same order as the inverse strain rate which is natural for
real Navier-Stokes turbulence where both values are of the order of the 
eddy turnover time at the Kolmogorov scale. The timestep in this simulation
was taken to be two hundred times smaller than the strain correlation time.

\begin{figure}[h]
\centering
\includegraphics[width=.49\linewidth]{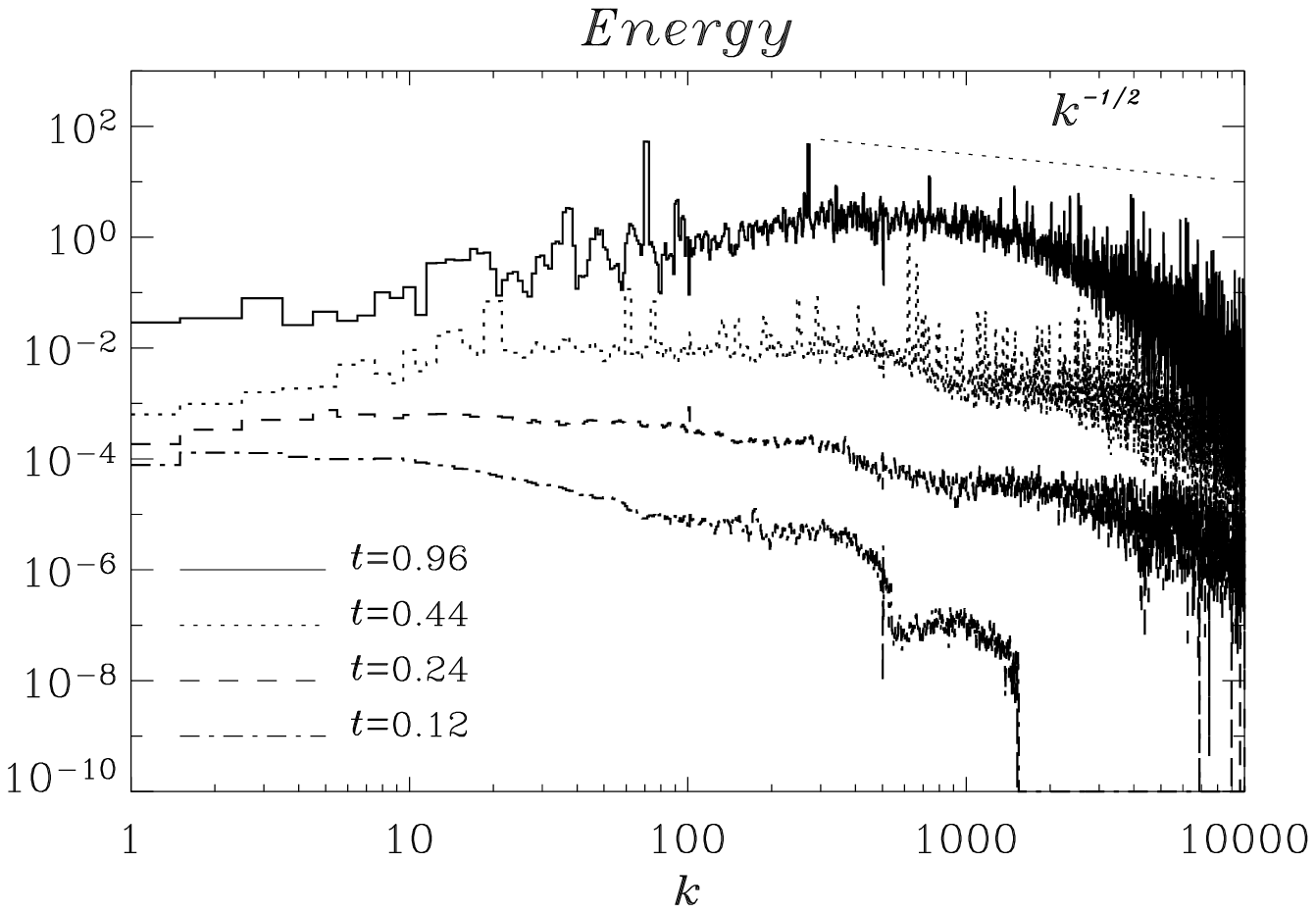}
\includegraphics[width=.49\linewidth]{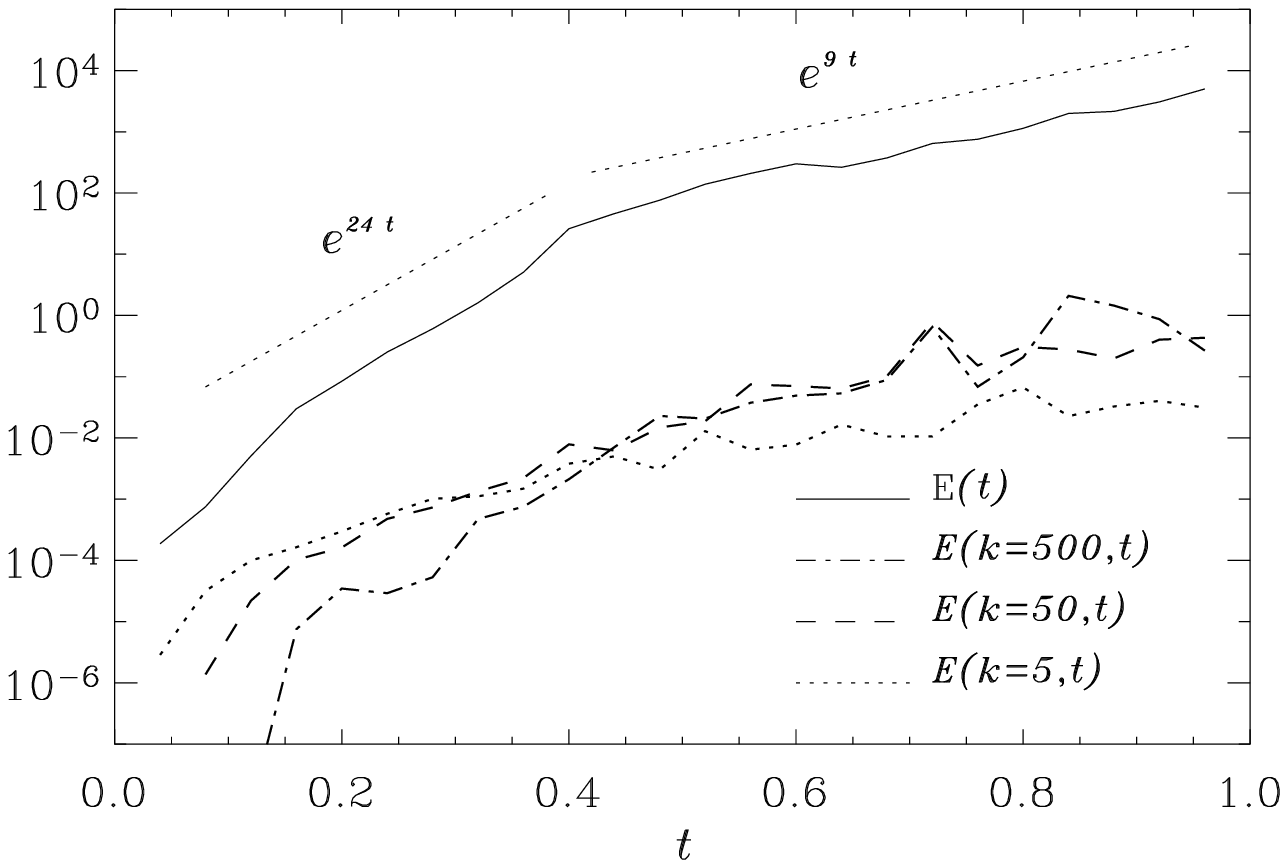}
\caption{
\noindent
The magnetic energy spectrum $E(k,t)$ (left figure) and growth of the 
total magnetic energy $<{\bf B}^2(t)>$ (right figure) 
for the case of a strain obtained from a $512^3$ DNS.
The right-hand figure also shows the growth of the energy for a 
selection of individual $k$-modes.}
\label{EDNS}
\end{figure}

In Fig. \ref{EDNS} we consider the energy spectrum (left figure) and growth of
the total energy (right figure). There is a good agreement with the 
theoretical behaviour of the energy spectrum. One can see a region of $k^{1/2}$ scaling and a
log correction in the diffusive regime.  Note that the 
pure  $k^{1/2}$  scaling changes to the log-corrected spectrum at the time when the high-$k$ tail
hits the dissipative scale which approximately corresponds to the third curve in Fig.  \ref{EDNS}
measured at $t=0.44$.
The growth of the total energy is again well represented. 
The growth rate exponent changes at around $t=0.4$
which marks the transition between the perfect conductor  and the diffusive regimes.
In agreement with the theory, the growth of the total energy
the diffusive regime is approximately equal to the growth of the energy found in
individual $k$ modes. Again the ratio of the growth rates in the perfect conductor and the diffusive regimes is 
consistent with theoretical predictions. 

\begin{figure}[h]
\centering
\includegraphics[width=.49\linewidth]{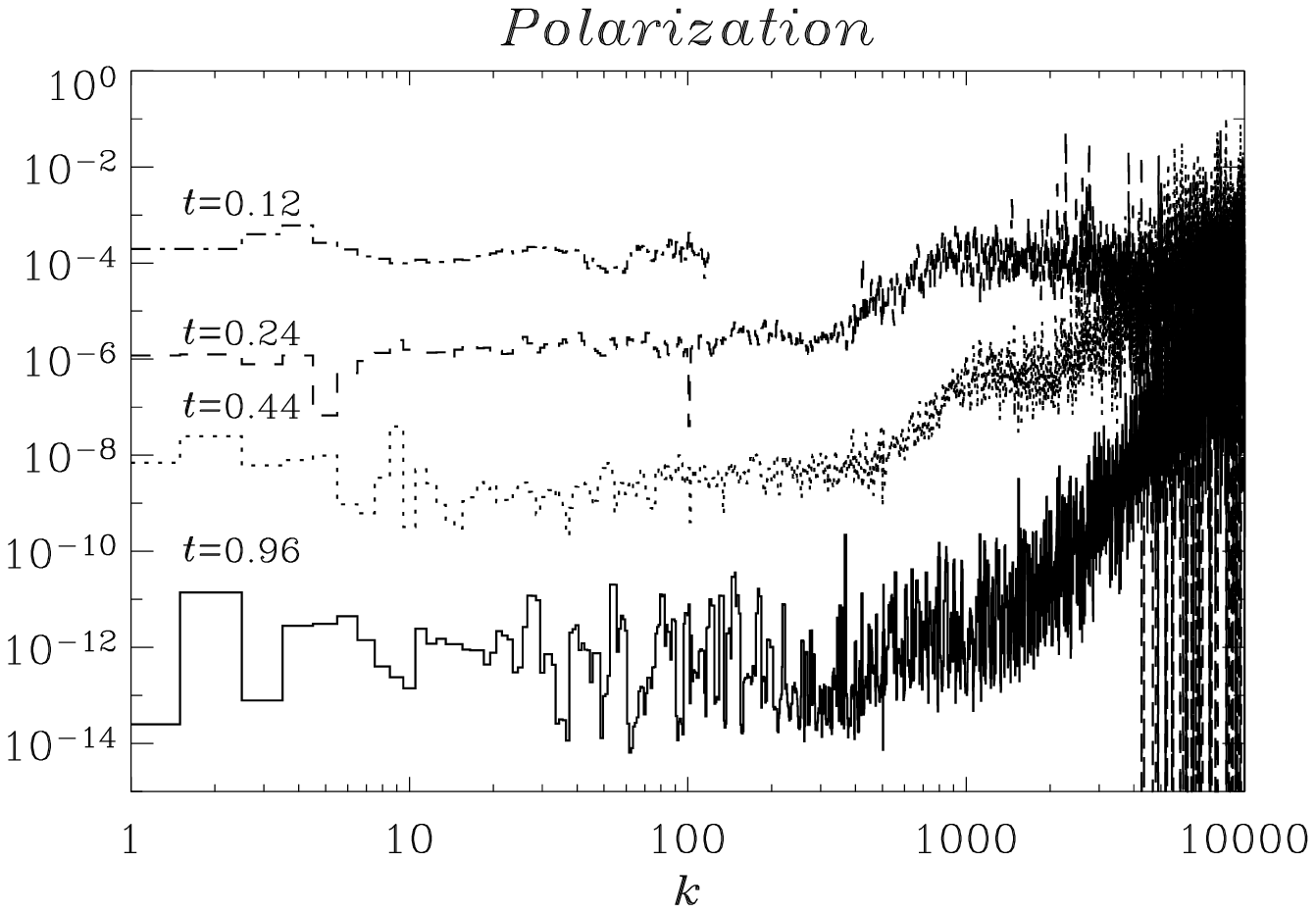}
\includegraphics[width=.49\linewidth]{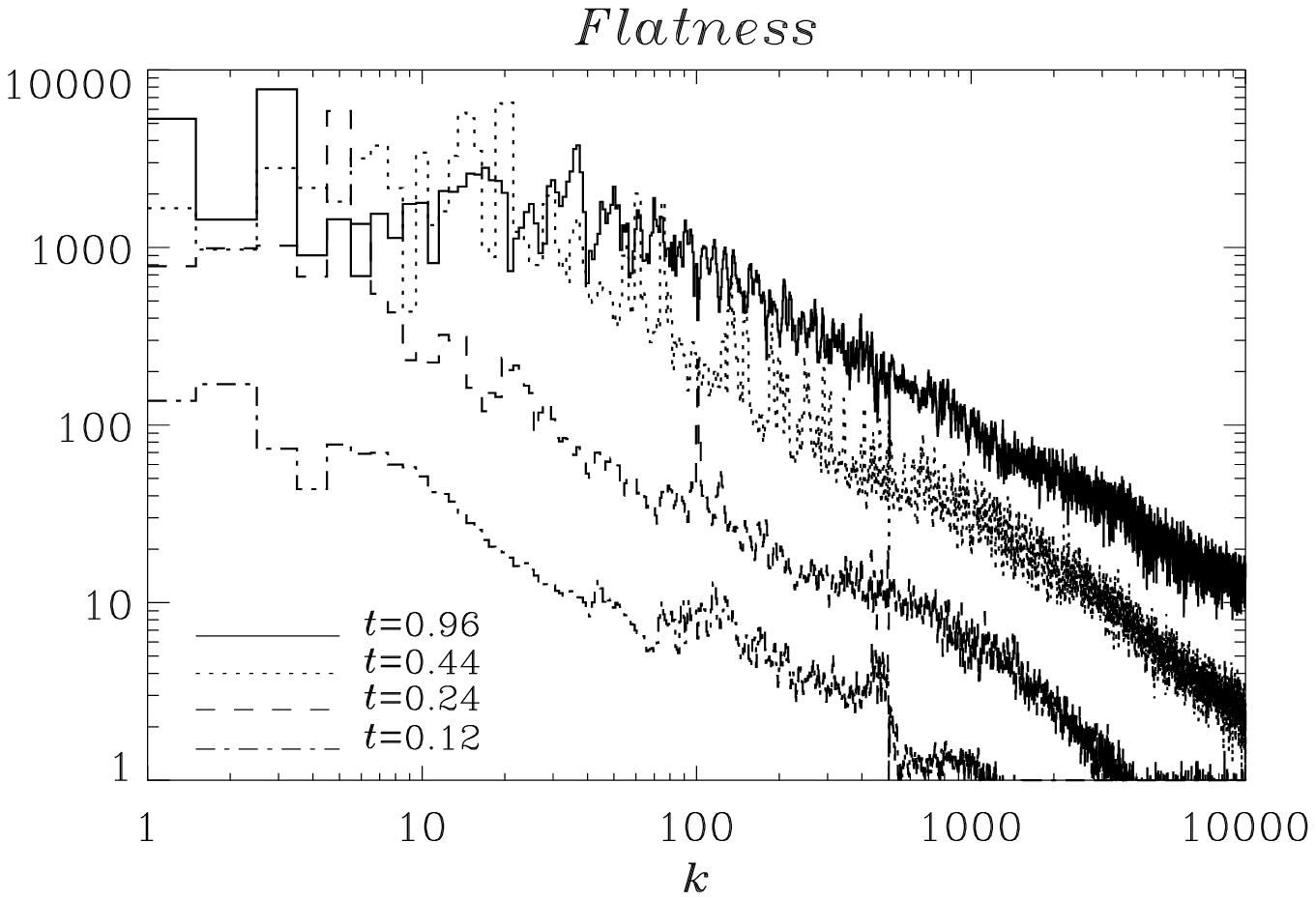}
\caption{
\noindent
The normalised mean polarisation (left figure) and spectral flatness 
(right figure) for the case of a strain obtained from a $512^3$ DNS.}
\label{PFDNS}
\end{figure}

The C4 results for the mean polarisation and the flatness spectra are shown in Fig. \ref{PFDNS}.
One can see that the mean polarisation (left figure)
decays in time and is spectrally flat which is in agreement with the KKM theory. The spectral 
flatness (right figure) grows in time, but the shape of the spectrum is
different from the theoretical result of $k^{3/2}$. This is similar to the corresponding flatness results
obtained in C3 simulations.
In fact, the deviations of the flatness from theory,
in both the C3 and C4 simulations, might not be due to changes in
strain statistics. The correlation time of the strain in the C3 simulations
is short, and therefore one would expect that its numerical results would be 
closer in behaviour to KKM than C4. However, the degree to which the
flatness differs from the KKM results is similar in both the C3 and C4 simulations.
Thus, we may conclude that our algorithm is failing to reproduce
some features of the higher order correlators.

\section{Discussion}

In this numerical investigation we have used a range of numerical models, based on Lagrangian 
particle dynamics, to investigate the small-scale turbulent dynamo problem. For the KKM based stochastic
models, using a delta-correlated Gaussian strain, we firstly considered the scaling of the 
even order moments $<{\bf B}^{2n}(t)>$ and found good agreement with the theoretical 
predictions (\ref{ourchert1}) and (\ref{ourchert2}) for the perfect conductor and 
diffusive regimes respectively.
Next we considered various spectral objects including the magnetic energy, polarisation 
and flatness, and made comparisons to their large time asymptotic analytical solutions. 
Because of the rich analytical results available, the KKM is a good testing ground for
our stochastic numerical models. The behaviour of the energy spectrum in both
diffusive and non-diffusive simulations was found to grow in time and fit the large-time 
asymptotic forms predicted by theory, with evidence of a $k^{-1/2}$ scaling region in each 
case. The fourth order correlators were also considered. Although noisier than 
the energy spectrum, these spectra also appeared consistent with the analytical 
results. However, there was evidence of some hyper-diffusion in these numerical results
are high wavenumbers. The new analytical objects, the mean polarisation and 
spectral flatness, were investigated. The flatness was well behaved with an observed 
$k^{3/2}$ scaling and the growth rate in time
in agreement with the analytical predictions. 
In contrast, the polarisation of the magnetic turbulence in these simulations 
was mis-represented. Although it was spectrally flat, in agreement with the theory, it appeared to saturate
at its Gaussian value of 1/3 rather 
than decaying in time. However, on considering individual realisations 
it was apparent that at least some of the predicted decaying behaviour was observed
qualitatively in these simulations. This also provided us with a 
useful visualisation of the dynamo system in terms of a cactus leaf in $k$-space.
The linear polarisation corresponds to the $k$-space alignment of the
magnetic field ``thorns'' along the direction of the smallest Lyapunov exponent, i.e. transverse to
the cactus leaf pad. The flatness growth with $k$ is directly linked to the fact that 
at larger $k$ the cactus leaves will cover the $k$-space more sparsely.
We also computed the Lyapunov exponents and found that the largest 
exponent was well represented which ensures the correct 
growth of the total magnetic energy. The behaviour of the 
smallest exponent was badly reproduced, which may be connected to the 
observed behaviour of the polarisation.

A second set of experiments was also performed to test the universality of the KKM based
theoretical results with respect to changes in the strain statistics. These two simulations, 
used a finite-correlated Gaussian and a DNS generated strain fields respectively.
For the finite-correlated Gaussian strain based model, the energy spectrum was found to be 
consistent with theory, although no $k^{-1/2}$ scaling region could clearly be
distinguished. For the DNS strain based model there was an agreement with 
theory for both the growth rate of the total energy and the  $k^{-1/2}$ scaling. Remarkably,
these two numerical experiments demonstrated excellent agreement with the
theoretical behaviour of the polarisation, in each case being both spectrally
flat and decaying in time. Also in agreement with the theory is the rapid time growth 
of the flatness. In contrast, the spectral shape of the flatness in these simulations
was not as well represented as in the stochastic based results. For both the 
finite-correlated Gaussian and DNS strain based simulations the flatness spectrum
was far from the predicted $k^{3/2}$ scaling. We believe this
difference from theory is a numerical artifact from our algorithm, rather than
directly due to changes in the form of the velocity field. We conclude that the gross features
predicted by the KKM theory seem to be robust and insensitive to changes in the strain 
statistics.

Interestingly, \cite{schek2,schek3} recently 
investigated the behaviour of the real space curvature of the field 
${\bf C}={\bf b}\cdot\nabla{\bf b}$ where ${\bf b}={\bf B}/|{\bf B}|$. 
They found that the curvature of the magnetic field is anti-correlated 
with its strength, which corresponds to folded and strongly 
stretched structures. This agrees with our result that the Fourier modes of the magnetic field
tend to a state of plane polarisation. However, it should be noted that the Fourier polarisation 
gives more information than the curvature statistics. Indeed, the zero curvature 
requirement allows both layered and filamentary configurations. Further, layered magnetic 
field may vary its direction as one passes from one layer to another. 
Our analysis narrows down the choice and indicates that the magnetic field structures are 
layers in coordinate space, thus ruling out the existence of any filament structures. 
Further, the plane polarization result also eliminates the possibility of any ``twists'' 
in the magnetic field between the individual layers.
That is, the sheet magnetic field is aligned along the same direction. Although 
the field in successive sheets can be parallel or anti-parallel to each other.
In fact, the presence of one neutral direction in the Lagrangian deformations tells us 
that the layers have a finite width in one direction and thus look, in coordinate space, 
like ribbons with the magnetic field lying along these ribbons.

\bigskip

\begin{acknowledgements}{}
We would like to thank Warwick University's Fluid Dynamic Research 
Centre for the use of their computer facilities.
\end{acknowledgements}

\section*{Apppendix}

In general, the strain matrix can be defined as
\begin{equation}\label{GeneralS}
<\sigma_{ij}(t)\sigma_{kl}(0)> = 
\end{equation}
$$
C_1 (\delta_{ik}\delta_{jl} + C_2\delta_{ij}\delta_{kl} 
- (1+C_2)\delta_{il}\delta_{jk}) \delta(t),
$$
where $C_{i}$ are constants (see for example \cite{McComb}). Written in this way the 
strain statistics automatically satisfy the incompressibility condition.

For our particular choice (\ref{OurStrain}), we have,
\begin{equation}\label{OurSFull}
      <\sigma_{ij}(t)\sigma_{kl}(0)> = \Omega \left(\delta_{ik}\delta_{jl} 
                                     -\frac{1}{3}\delta_{ij}\delta_{kl}\right) \delta(t),
\end{equation}
giving $C_1=\Omega$ and $C_2=-1/3$. In contrast, the covariance chosen by \cite{Chertkov}
and \cite{Falkovich} is 
\begin{equation}\label{ChertSFull}
      <\sigma_{ij}(t)\sigma_{kl}(0)> = \frac{\lambda_1}{3} (4 \delta_{ik}\delta_{jl} 
                                     - \delta_{ij}\delta_{kl} 
                                     - \delta_{il}\delta_{jk}) \delta(t),
\end{equation}
corresponding to $C_1 = 4\lambda_1/3$ and $C_2=-1/4$, where $\lambda_1$ is the growth 
rate of a material line element in the flow.
Setting $i=k=\alpha$ and $j=l=\beta$ in both (\ref{OurSFull}) and (\ref{ChertSFull}) we find,
\be
   <\sigma_{\alpha\beta}(t)\sigma_{\alpha\beta}(0)> = 8 \Omega \delta(t),
\label{Omst}
\ee
\be
   <\sigma_{\alpha\beta}(t)\sigma_{\alpha\beta}(0)> = 10 \lambda_1 \delta(t).\label{Lmst}
\ee
The latter expression is the previous stated \cite{Chertkov} formalism 
(\ref{ChertStrain}) from section \ref{Chert}.

\newpage


\newpage


\newpage

\newpage

\newpage

\newpage

\newpage

\newpage

\newpage

\newpage

\newpage

\begin{thebibliography}{}
\bibitem[Balkovsky \& Fouxon(1999)]{Balkovsky} 
Balkovsky, E., \& Fouxon, A. 1999, Phys. Rev. E, 60(4), 4164

\bibitem[Batchelor(1950)]{Batch1} 
Batchelor, G.K. 1950, Proc. Roy. Soc, 202A, 405

\bibitem[Batchelor(1954)]{Batch2} 
Batchelor, G.K. 1954, Proc. Roy. Soc, 2

\bibitem[Chandran(1998)]{chandran} 
Chandran, B.D.G. 1998, \apj, 492, 179

\bibitem[Chertkov et al.(1999)]{Chertkov} 
Chertkov, M., Falkovich, G., Kolokolov, I., \& Vergassola, M. 1999, 
Phys. Rev. Lett., 83(20), 4065

\bibitem[Childress \& Gilbert(1995)]{childress}
Childress, S., \& Gilbert, A. 1995 
Stretch, Twist, Fold: The Fast Dynamo (Berlin: Springer-Verlag)

\bibitem[Falkovich et al.(2001)]{Falkovich} 
Falkovich, G., Gawedzki, K., \& Vergassola, M. 2001, Rev. Mod. Phys., 73, 913

\bibitem[Goldhirsch et al.(1987)]{Goldhirsch} 
Goldhirsch, I., Sulem, P.-L., \& Orszag, S.A. 1987, Physica D, 27, 311

\bibitem[Kazantsev(1968)]{Kazantsev} 
Kazantsev, A.P. 1968, Sov. Phys. JETP, 26, 1031

\bibitem[Kloeden et al.(1997)]{Kloeden} 
Kloeden, P.E., Platen, E., \& Schurz, H. 1997, 
Numerical Solution of SDE Through Computer Experiments (Berlin: Springer-Verlag)

\bibitem[Kraichnan \& Nagarajan(1967)]{Kraichnan} 
Kraichnan, R.H., \& Nagarajan, S. 1967, Phys. Fluids, 10, 859

\bibitem[Kraichnan(1974)]{Kraichnan74} 
Kraichnan, R.H. 1974, J. Fluid Mech., 64, 737

\bibitem[Kulsrud \& Anderson(1992)]{Kulsrud} 
Kulsrud, R., \& Anderson, S. 1992, \apj, 396, 606

\bibitem[Kulsrud(1999)]{kulsrud99} 
Kulsrud, R. 1999, ARA\&A, 37, 37

\bibitem[McComb(1990)]{McComb} 
McComb, W.D. 1990, The Physics of Fluid Turbulence
(Oxford: Clarendon Press)

\bibitem[Moffatt(1978)]{moffatt}
Moffatt, H.K. 1978, Magnetic Field Generation in Electrically
Conducting Fluids (Cambridge: Cambridge University Press)

\bibitem[Nazarenko et al.(2003)]{NazWesZab} 
Nazarenko, S.V., West, R.J., \& Zaboronski, O. 2003, 
submitted to Phys. Rev. E

\bibitem[Parker(1979)]{parker}
Parker, E.N. 1979, 
Cosmic Magnetic Field, Their Origin and Activity (Oxford: Clarendon Press)

\bibitem[Press et al.(1993)]{numrec} 
Press, W.H., Falnnery, B.P., Teukolsky, S.A., \& Vetterling, W.T. 1993,
Numerical Recipes in C (Cambridge: Cambridge University Press)

\bibitem[Soward(1994)]{Soward} 
Soward, A.M. 1994, Fast Dynamos, 
Lectures on Solar and Planetary Dynamos, Chap. 6, 
Proctor, M.R.E., \& Gilbert, A.D. Eds (Cambridge: Cambridge University Press)

\bibitem[Schekochihin et al.(2001)]{schek2} 
Schekochihin, A.A., Cowley, S., \& Maron, J.L. 2001, Phys. Rev. E., 65, 016305

\bibitem[Schekochihin et al.(2002a)]{Schekochihin} 
Schekochihin, A.A., Boldyrev, S.A., \& Kulsrud, R.M. 2002a, \apj, 567, 828

\bibitem[Schekochihin et al.(2002b)]{schek3}  
Schekochihin, A.A., Maron, J.L., Cowley, S., \& McWilliams, J.C. 2002b, \apj, 576, 806

\bibitem[Zel'dovich et al.(1984)]{Zeldovich} 
Zel'dovich, Y.B., Ruzmaikan, A.A, Molchanov, S.A., \& Sokoloff, D.D. 1984,
J. Fluid Mech., 144, 1
\end{thebibliography}
\end{document}